\pdfoutput=1

\documentclass[11pt,twoside,a4paper,cmspaper,final,collab]{cms-tdr}
\def\svnVersion{9210ad8-D}\def\svnDate{2020/03/03}\def\cmsCernNoTag{CERN-EP-2019-238}\def\cmsCernDate{\today}\def\cmsMessage{"Published in the Journal of Instrumentation as \href{http://dx.doi.org/10.1088/1748-0221/15/02/P02027}{\doi{10.1088/1748-0221/15/02/P02027}.}"}

\begin{document}\cmsNoteHeader{MUO-17-001}

\hyphenation{had-ron-i-za-tion}
\hyphenation{cal-or-i-me-ter}
\hyphenation{de-vices}
\RCS$HeadURL$
\RCS$Id$

\newcommand{\pp}{\Pp{}\Pp}

\newcommand{\hiptid}{high-\pt ID\xspace}
\newcommand{\hiptidcaps}{High-\pt ID\xspace}

\newcommand{\Nseg}{\ensuremath{N_{\text{seg}}}\xspace}
\newcommand{\Nshower}{\ensuremath{N_{\text{shower}}}\xspace}
\newcommand{\Pshower}{\ensuremath{P_{\text{shower}}}\xspace}
\newcommand{\TuneP}{TuneP\xspace}

\newcommand{\residgen}{\ensuremath{R_{\text{reco-gen}}}\xspace}
\newcommand{\residcosmic}{\ensuremath{R_{\text{cosmic}}}\xspace}

\newcommand{\TnP}{tag-and-probe\xspace}
\newcommand{\Eironc}{\ensuremath{E^{\text{iron}}_{\mathrm{c}}}\xspace}

\cmsNoteHeader{MUO-17-001}
\title{Performance of the reconstruction and identification of high-momentum muons in proton-proton collisions at $\sqrt{s} = 13\TeV$}

\date{\today}

\abstract{
The CMS detector at the LHC has recorded events from proton-proton collisions, with muon momenta reaching up to 1.8\TeV in the collected dimuon samples. These high-momentum muons allow direct access to new regimes in physics beyond the standard model.
Because the physics and reconstruction of these muons are different from those of their lower-momentum counterparts,
this paper presents for the first time dedicated studies of efficiencies, momentum assignment, resolution, scale, and showering of very high momentum muons produced at the LHC. These studies are performed using the 2016 and 2017 data sets of proton-proton collisions at $\sqrt{s} = 13\TeV$ with integrated luminosities of 36.3 and 42.1\fbinv, respectively.
}

\hypersetup{%
pdfauthor={CMS Collaboration},%
pdftitle={Performance of the reconstruction and identification of high-momentum muons in proton-proton collisions at sqrt(s)=13 TeV},%
pdfsubject={CMS},%
pdfkeywords={CMS, physics, muons, detector performance}}

\maketitle

\tableofcontents  \clearpage

\section{Introduction} \label{sec:introduction}

One of the main tasks of the CMS experiment is to search for new phenomena
in proton-proton (\pp) collisions delivered by the CERN LHC.
Good identification and precise measurement of muons,
electrons, photons, and jets over a large energy range and at high instantaneous luminosities
are necessary for these searches to be effective.
In particular, searches for heavy gauge bosons such as the $\PZpr$~\cite{Leike:1998wr, Langacker:2008yv} and $\PWpr$~\cite{Hsieh:2010zr} rely on
precise reconstruction of muons up to very high momentum.
With the data recorded from \pp collisions in Run 2 at
$\sqrt{s}=13\TeV$, corresponding to integrated luminosities of 36.3\fbinv in 2016 and 42.1\fbinv in 2017,
the CMS detector has recorded a sufficiently large sample of higher-energy muons to allow the first detailed studies of such muons at the LHC, presented here. For some analyses that require an independent data set with all CMS subdetectors activated, the luminosities recorded are slightly lower with 35.9\fbinv in 2016 and 41.5\fbinv in 2017.

Previously published studies of the CMS muon detectors~\cite{Chatrchyan:2013sba} and muon reconstruction~\cite{Chatrchyan:2012xi}
were based on data from \pp collisions recorded during Run 1 in 2010 at $\sqrt{s}=7\TeV$, as well as on
data recorded in 2015 and 2016 at 13\TeV~\cite{Sirunyan:2018fpa}.
An extensive description of the performance of the muon detector and the muon reconstruction software is given in Refs.~\cite{Chatrchyan:2013sba,Chatrchyan:2012xi},
while Ref.~\cite{Sirunyan:2018fpa} focuses on significant improvements made to the muon system during the long shutdown period in 2013--2014 between LHC Runs 1 and 2.
These changes resulted in reconstruction software and the high-level trigger (HLT) that were shown to have similar or better performance than in 2010, despite the higher instantaneous luminosity.

In this paper, we present performance measurements of the muon triggering, reconstruction, identification, and momentum assignment, for muons with high transverse momentum $\pt>200\GeV$. Above this threshold, the effects of radiative energy losses in the steel flux-return yoke of the solenoid due to pair production, bremsstrahlung, and photonuclear interactions, as well as detector alignment, become significant enough to motivate dedicated studies.

Various sources of high-momentum muons are used to ensure significant and meaningful results.
We include muons from the decays of high-mass off-shell standard model (SM) vector bosons, denoted as high-mass Drell--Yan events (DY),
and muons from the decay of on-shell \PW or \PZ bosons recoiling against jets, denoted as \PZ(\PW)+jets events.
In addition, we study high-momentum muons originating from cosmic rays, recorded during both the \pp collisions and dedicated periods with no beam.

\section{The CMS detector}
\label{sec:cmsdetector}

The central feature of the CMS apparatus is a superconducting solenoid of 6\unit{m} internal diameter, providing a magnetic field of 3.8\unit{T}.
Within the solenoid volume are a silicon pixel and strip tracker, a lead tungstate crystal electromagnetic calorimeter (ECAL),
and a brass and scintillator hadron calorimeter (HCAL), each composed of a barrel and two endcap sections.
Forward calorimeters extend the coverage in pseudorapidity $\eta$ provided by the barrel and endcap detectors.
Muons are detected in gas-ionization chambers embedded in the steel flux-return yoke outside the solenoid.

Events of interest are selected using a two-tiered trigger system~\cite{Khachatryan:2016bia}. The first level (L1), composed of custom hardware processors, uses information from the calorimeters and muon detectors to select events at a rate of around 100\unit{kHz} within a fixed time interval of less than 4\mus. The second, high-level trigger (HLT) consists of a farm of processors running a version of the full event reconstruction software optimized for fast processing, and reduces the event rate to around 1\unit{kHz} before data storage.

Muons are measured in the range $\abs{\eta} < 2.4$ with detection planes made using three technologies: drift tubes (DTs), cathode strip chambers (CSCs), and resistive plate chambers (RPCs). The single-muon trigger efficiency with respect to reconstructed muons exceeds 90\% over the full $\eta$ range with respect to reconstructed muons, and the efficiency to reconstruct and identify muons that pass the trigger requirements is greater than 96\%. Matching muons to tracks measured in the silicon tracker results in a relative \pt resolution of 1\% in the barrel and 3\% in the endcaps, for muons with \pt up to 100\GeV. The \pt resolution in the barrel is better than 7\% for muons with \pt up to 1\TeV~\cite{Sirunyan:2018fpa}.

At the end of the 2016 LHC running period, an additional pixel layer was added to the tracker; the HLT sequences were modified to sustain a higher rate due to the increase of the number of \pp interactions in the same or adjacent bunch crossings, referred to as pileup; and the detector was opened and the alignment conditions were consequently changed. These modifications could impact several studies performed in this paper; whenever it appears to be the case, it is explicitly mentioned.

A more detailed description of the CMS detector,
together with a definition of the coordinate system used and the relevant kinematic variables, can be found in Ref.~\cite{Chatrchyan:2008zzk}.

\section{Data samples and simulation} \label{sec:samples}

The studies described in this paper are mostly
based on data recorded using single-muon triggers.
In addition, for the trigger studies, we use data samples
recorded with single-electron triggers and missing transverse momentum (\ptmiss) triggers, referred to as independent data sets, since they provide unbiased samples of muons suitable
for studies of muon triggers. (We follow common usage in defining \ptmiss  as the magnitude of
the projection onto the plane perpendicular to the beam axis of the
vector sum of the momenta of all reconstructed
objects in an event.)
To maximize the sample size at high
momentum, the muon data sets from 2016 and 2017 are merged when the performance under
study is independent of the detector and software changes from one year to another; otherwise, the results are
presented for the two years separately. The results in this paper
are obtained from selected data samples consisting of events with a
pair of reconstructed muons, or with a single reconstructed muon for the trigger
study using independent data sets;  throughout, muon $\pt>53\GeV$ is required, in order to be above trigger turn-on effects at the trigger threshold of 50\GeV. Further event criteria are applied, depending on the study, and
are described in detail below.
Cosmic ray muon data, recorded in the absence of LHC beams or in gaps between
\pp collisions, are used to provide complementary
studies on the muon momentum resolution and charge
assignment.

The selected data events are compared with simulations from several
event generators that use the Monte Carlo (MC) method. The DY $\PZ/\gamma^{\star}
\to \mu^+ \mu^- $ signal samples are generated with \POWHEG
v2~\cite{Nason:2004rx, Frixione:2007vw, Alioli:2010xd} at
next-to-leading order (NLO) in both QCD and electroweak corrections,
and cover a mass range
from 50\GeV up to 5\TeV. For the studies that use exclusively the \PZ
peak ($60 < m_{\mu\mu} < 120\GeV$) and explore the high-momentum muons produced from
boosted bosons, we use samples enriched in \PZ$+$jets generated with
\MGvATNLO v2.2.2~\cite{Alwall:2014hca}. Finally, the $\PW^{*} \to \mu \nu$ signal samples, used to
validate the single-muon trigger efficiency, are generated at
leading-order (LO) with \PYTHIA 8.212 (8.230)~\cite{Sjostrand:2014zea} for
2016 (2017) studies.

The dominant backgrounds over the full dimuon mass
range are, in order of importance, \ttbar, t\PW, and \PW\PW; they
are simulated at NLO with \POWHEG. The \ttbar cross section is
calculated at next-to-NLO (NNLO) with Top$++$ v2.0~\cite{Czakon:2011xx}. Other electroweak backgrounds, such as
\PW\PZ and \PZ\PZ, are generated with \PYTHIA.

For all simulated samples mentioned above, the fragmentation and
parton showering is modeled with \PYTHIA 8.212 with the CUETP8M1~\cite{CMS-PAS-GEN-14-001}
underlying event tune for the 2016 studies or with \PYTHIA 8.230 with
CP5~\cite{CMS:2018zub} tune for 2017 studies. The NNPDF3.0~\cite{Ball:2014uwa} and NNPDF3.1~\cite{Ball:2017nwa} parton
distribution function sets are used for the 2016 and 2017 samples,
respectively. The simulation of the CMS detector response is based on \GEANTfour~\cite{Agostinelli:2002hh}; the events are then reconstructed with the
same algorithms as used for data. Pileup is also
simulated, except for studies where it is explicitly stated that this is not
the case.

\section[\texorpdfstring{High-$\pt$}{High-pT} muon reconstruction overview]{\texorpdfstring{High-$\boldsymbol{\pt}$}{High-pT} muon reconstruction overview}
\label{sec:overview}

Most of the muons produced in CMS originate in processes such
as semileptonic decays of top quarks or heavy-flavor hadrons,
or in leptonic decays of on-shell vector bosons (\PW, \PZ).
Such muons typically have $\pt< 200\GeV$, and are referred to as low-$\pt$ muons.
On the other hand, high-\pt muons are produced in rare processes such as off-shell production of high-mass or on-shell production of high-$\pt$ \PW$^{\star}$ and \PZ$^{\star} / \gamma$ bosons,
and could be produced from the decay of beyond the standard model (BSM) particles with TeV-scale mass (\eg, \cPZpr or \PWpr bosons).

Experimentally, the main differences between high- and low-\pt muons can be understood as follows.
As the muon momentum increases, the \pt resolution of the reconstructed track degrades. In the part of the orbit in near-uniform magnetic field $B$, the measurement of \pt depends on $B$, and the radius of curvature, $R$, of the track~\cite{PDG2018}:
\begin{equation}
  \pt[\text{GeV}] = \abs{0.3 B[\text{T}] R[\text{m}]}.
\label{eq:pTvsRadius}
\end{equation}
The magnetic field is monitored with high precision and is roughly uniform at 3.8\unit{T} in the tracker volume inside the solenoid. The radius of curvature is related to the arc length $L$ and sagitta $s$ of the track via
\begin{equation}
  R[\text{m}]\approx L[\text{m}]^{2}/8s[\text{m}],
\label{eq:RadiusvsSagitta}
\end{equation}
where the approximation is valid for $L/R \ll 1$. Assigning arithmetic signs consistently to $R$, $s$, and the charge $q$ (in units of proton charge) yields
\begin{equation}
  s[\text{m}]\approx (0.3 B [\text{T}] L[\text{m}]^{2}/8) (q/\pt[\GeVns]) =  (0.3 BL^{2}/8) \kappa,
\label{eq:SagittavsPt}
\end{equation}
where $\kappa=q/\pt$ is referred to as the (signed) curvature of the muon track.  Because $s$ is linearly related to the measurement of hit positions in the detector (which have approximately symmetric uncertainties), the uncertainty in $\kappa$ (rather than in $\pt$) from the cumulative effect of hit uncertainties is (approximately) Gaussian. Hence $\kappa$ is the more natural variable for use in muon momentum resolution and scale studies, as discussed in Section~\ref{sec:Momentum}. As the \pt increases and the sagitta in the tracker decreases, the muon momentum measurement can be improved by using the large $BL^2$ between the tracker and the muon system (and within the muon system), if the $\pt$ is large enough that multiple Coulomb scattering in the calorimeters and in the steel flux-return yoke of the solenoid does not spoil the measurement. Thus high-$\pt$ muon track reconstruction and muon momentum measurement rely on matching tracks reconstructed in the inner tracker and the muon system, separated by more than 3~meters and forming a global track, as explained in Section~\ref{subsecsec:reconstruction}. However, because of the smallness of the sagitta (or more precisely, the generalizations of sagitta in nonuniform $B$) in the TeV regime, the muon \pt resolution is sensitive to alignment of the hits used to reconstruct the muon track. The impact of the detector alignment on the momenta resolution is discussed in Section~\ref{sec:Momentum}.

If a muon traveling through the steel of the magnet flux-return yoke has sufficiently large momentum, radiative energy losses (bremsstrahlung with inner and outer e$^+$e$^-$ pair production, photonuclear interactions) are no longer negligible compared to ionization energy losses. The muon critical energy for iron, \Eironc, at which the ionization energy losses are equal to the sum of all radiative losses, is around 300\GeV\cite{PDG2018}. As a consequence, the main source of energy loss for muons above \Eironc propagating through the steel between the muon subdetectors is radiative energy losses. This radiation creates cascades of particles (electromagnetic showers) and can lead to extra hits being reconstructed in the muon detectors. These showers can have a strong impact on the muon performance (\ie, triggering, reconstruction, or \pt measurement). The muon showering primarily depends on the total muon momentum, as opposed to the transverse component that is commonly used in physics analyses. Depending on the longitudinal component of momentum, muons with $\pt>200\GeV$ can have energies above \Eironc. The potential presence of showers around the muon track is what motivates the choice of $\pt>200\GeV$ to define a high-\pt muon in the paper. Dedicated algorithms for momentum assignment have been developed and are discussed in Section~\ref{subsecsec:reconstruction}. In addition, in order to understand the behavior of high-\pt muons and the impact of showering along the CMS detection sequence, we parameterize the showering and then confront simulation with data on the various muon performance aspects. This shower tagging is discussed in Section~\ref{subsecsec:showering}, whereas the results of the muon performance as a function of muon showering are shown in Sections~\ref{sec:Efficiency} and~\ref{sec:Momentum}.

Some BSM searches, involving high-\pt muons, probe processes with small cross sections for which negligible backgrounds from SM processes are expected. High efficiency for measuring TeV muons is particularly important for obtaining a high sensitivity in such searches. For example, the current upper limit~\cite{Sirunyan:2018exx} on the product of
production cross section and branching fraction for a \zp boson with a mass of 2\TeV, $\sigma(\zp){\mathcal{B}}(\zp\to \mu\mu)$, is ${\mathcal{B}}(10^{-7})$ smaller than that of the SM \PZ boson, $\sigma(\PZ){\mathcal{B}}(\PZ\to \mu\mu)$. In such analyses, the signal efficiencies are derived with simulated samples. While the simulations can be validated in some kinematic regions using \PZ boson events in data, the lack of signal at higher masses forces the analysis strategy to extrapolate into the highest \pt regions. Therefore, it is important to have uniform reconstruction, identification, and triggering efficiencies as a function of the muon $p$ and $\pt$, and to ensure that any sensitivity to muon showering is understood. Dedicated high-\pt muon identification criteria have been developed and further improved during LHC Run 2 in order to provide robustness with increasing muon \pt; they are detailed in Section~\ref{subsecsec:identification}. The level of agreement between the performance in data and simulation is quantified in terms of data-to-simulation efficiency ratios called scale factors (SF).

\subsection{Reconstruction} \label{subsecsec:reconstruction}

In the standard CMS reconstruction procedure for \pp collisions,
muon tracks are first reconstructed independently in the inner tracker and in
the muon systems~\cite{Adam:934067}. In the latter, tracks called ``standalone muons'' are
reconstructed by using information from DT, CSC, and RPC
detectors along a muon trajectory using the Kalman filter
technique~\cite{Fruhwirth:1987fm}.
In both the barrel and endcap regions, the muon detectors
reside in four ``stations'', which are typically separated by 23 to 63\cm of steel. The steel thickness prevents an electromagnetic shower from propagating across more than one station.  Within each
station, there are multiple planes of detectors, from which ``hits''
are recorded. The hits within a station are combined into local
``segments'', which are in turn combined into standalone muons.

Matching standalone-muon tracks with tracks reconstructed in the inner tracker yields combined tracks referred to as ``global muons''. If the momentum, direction, and position in the transverse plane of the inner and standalone tracks are compatible, then
the global track is fit by combining hits from the tracker track and standalone-muon track in a common fit.

Global muons are complemented by objects referred to as ``tracker muons'' that are built by propagating the inner tracker tracks to the muon system with loose geometrical matching to DT or CSC segments.
If at least one muon segment matches the extrapolated track, the track is qualified as a tracker muon.
Tracker muons have higher efficiency than global muons in regions of the CMS detector with less instrumentation and for muons with low-$\pt$.

The momentum of a muon reconstructed as a global muon can be extracted from the combined tracker-plus-standalone trajectory.
For high-\pt muons, however, extra particles produced in electromagnetic showers can contaminate the muon detectors, yielding extra reconstructed hits and segments.
These extra segments can be picked up by the trajectory building algorithm instead of the correct muon track segment, or even make the reconstruction of
the muon track in a chamber impossible. The high-\pt case thus requires
careful treatment of the information from the muon system.  A
set of specially developed TeV-muon track refits has been developed
to address this issue: the ``tracker-plus-first-muon-station'' (TPFMS)
fit, the ``Picky'' fit, and the ``dynamic truncation'' (DYT) fit. The momentum assignment is finally performed by the ``\TuneP'' algorithm, which chooses the
best muon reconstruction among the tracker-only track, TPFMS, DYT, and Picky fits.

The TPFMS fit is historically the
first alternative to the global muon fit (which is based on all the
trajectory measurements). It only uses hits from the tracker and the innermost
muon station containing hits, thus taking advantage of a large $BL^2$, while neglecting the stations that are farther along the muon's trajectory, thus reducing potential contamination from
showers. Even with this omission, by making a judicious track-by-track choice between the tracker-only fit and the TPFMS fit, the
resolution at high $\pt$ can be improved with respect to both the
tracker-only fit and the global fit~\cite{Sirunyan:2018fpa}.

Other strategies for improvement have also been developed.
If a shower in one muon station corrupts
the position measurement in that specific station, thus the thickness of the
steel layer will absorb the shower and prevent it from leaking into the next
station. Then, in principle, if it is possible to identify a station where a
shower occurs, then it can be discarded from the muon global
fit instead of rejecting most stations, as is done with TPFMS.
The Picky algorithm was developed with this approach in mind.
It identifies stations containing showers based on the hit multiplicity,
and for each of them, it imposes extra requirements on hit compatibility with the muon trajectory.
If hits in a station with showering fail these requirements, that station is removed from the trajectory fit.

The DYT fit approach is based on the observation that in some cases, when a muon
loses a large fraction of its energy, its orbit can change and
the segments (or hits) in subsequent stations may no longer be consistent with the initial
trajectory. In other cases, where the energy loss is less severe, only
hits in one station appear incompatible, while the rest of the
trajectory is negligibly changed and can be used in the fit. The DYT algorithm starts from the tracker track and proceeds outwards, iteratively adding to the fit muon hits compatible with the extrapolated track trajectory. When incompatible hits are found it ignores them or stops the fit entirely, depending on the degree of incompatibility.

Thus, the algorithm for choosing between the tracker-only fit and TPFMS has evolved into a more
general algorithm, known as the \TuneP algorithm, for choosing among the various refits
on a track-by-track basis. It uses the track fit $\chi^2/$dof tail probability and the relative \pt measurement uncertainty $\sigma_{\pt}/\pt$,
where $\sigma_{\pt}$ is the uncertainty in \pt, as determined by the Kalman filter.
The algorithm starts its search for the best track fit choice by initially
considering the Picky hypothesis and comparing its
$\sigma_{\pt}/\pt$ with the value estimated for the corresponding
track but refitted by the DYT algorithm. The refit with the
smallest uncertainty in \pt is then compared to the tracker-only fit, and the track
with the lower $\chi^2/$dof tail probability value is kept, to be finally compared with
the TPFMS refitter algorithm. The final best track is chosen after the last comparison
according to the $\chi^2$/dof tail probability. In the rare cases where there is
no convergence in the Picky algorithm refit, or in the other refits
tried consecutively, the global fit is kept.

In cases where the final candidate
track has a \pt lower than 200\GeV, the tracker-only fit is used.
Figure 1 presents the fractions for each choice of \TuneP among DYT,
Picky, and any of the other fits (TPFMS, global, or tracker-only),
as a function of the muon \pt, separately for the barrel and
endcap regions. The selected muons
come from dimuon events and are required to pass the high-\pt identification described in Section \ref{sec:highptIdEff}, and
to have $\pt>200\GeV$. To simulate the
data events, we add to DY simulation all the other electroweak
processes that arise in data and that mimic DY events (diboson,
\ttbar, single top quark, etc.). We do not add
the background from SM events comprised uniquely of
jets produced through the strong interaction, because this background is negligible above $200\GeV$.
The simulation reproduces well what is observed in data: similar fractions in
the choice among the refits across the full \pt spectrum, with a preference
for Picky in the  barrel (${\approx}$60\%) while similar fractions for DYT and Picky are
found in the endcaps (${\approx}$50\%). When DYT was first developed, its
performance was studied integrated over muon $\eta$ and
in consequence found to be driven by the endcap region where
most of the showering takes place. The high level of agreement between data
and simulation is an indication that the impact of showering on
momentum assignment is well reproduced by simulation.

\begin{figure}[!ht]
\centering
\includegraphics[width=0.49\textwidth]{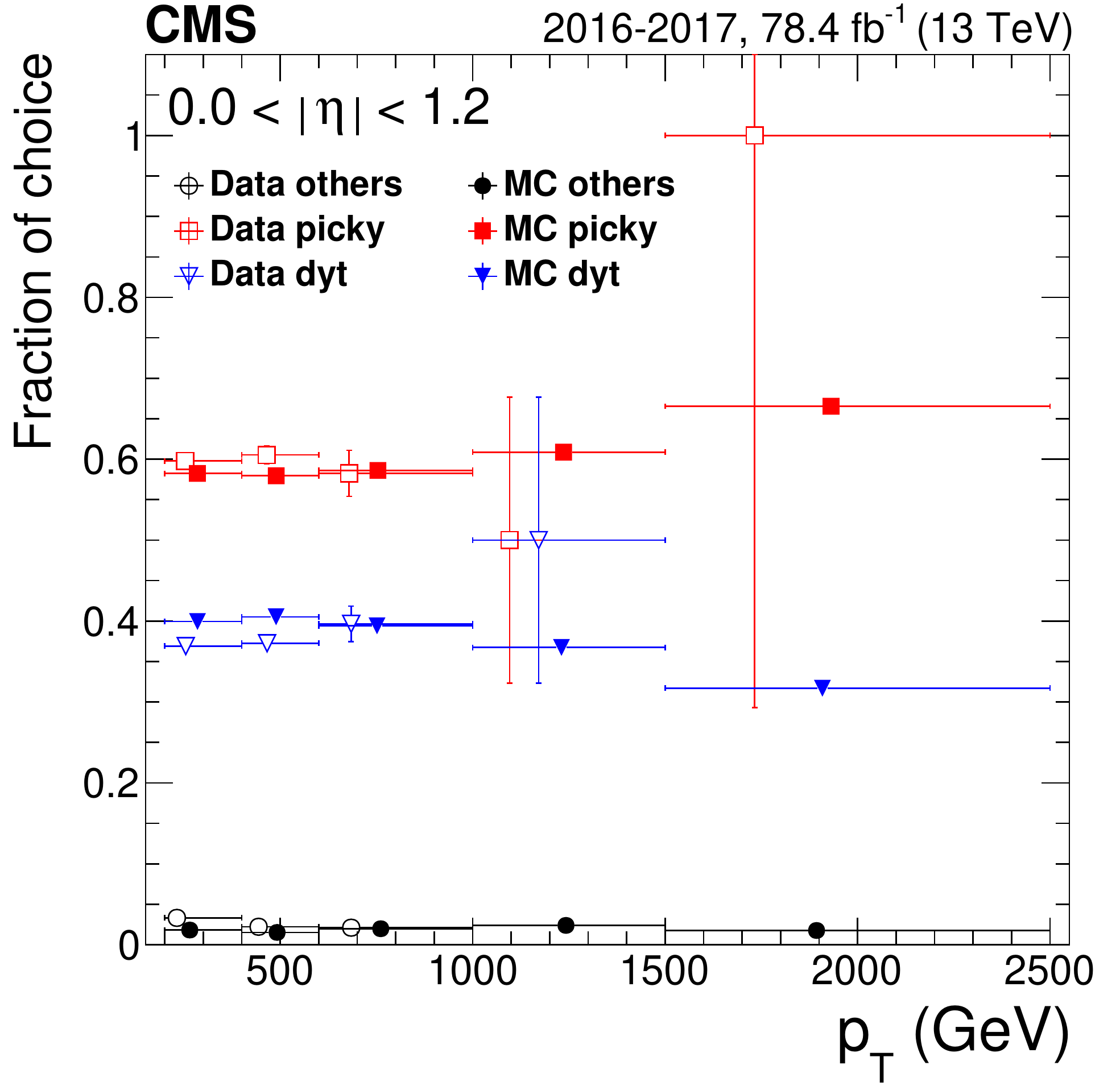}
\includegraphics[width=0.49\textwidth]{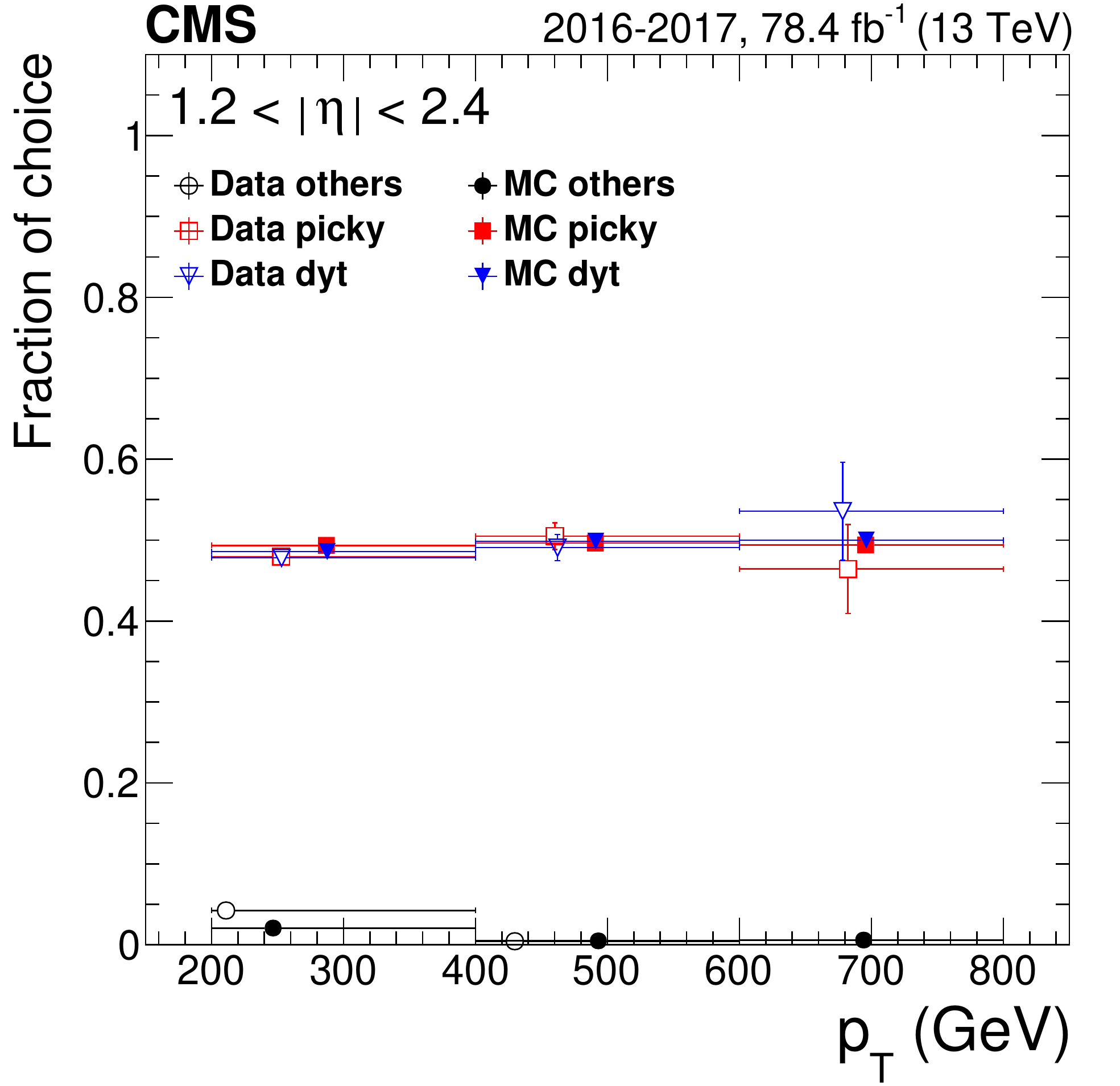}
\caption{Fraction of choices of different refit algorithms chosen by \TuneP, comparing
2016+2017 data and DY simulation for five $\pt$ ranges and for two $\eta$ categories:
(left) barrel with $\abs{\eta} < 1.2$
and (right) endcap with $1.2 < \abs{\eta} < 2.4$. The central value in each bin is obtained from the average of the distribution within the bin.}
\label{TuneP_Choice_1617}
\end{figure}

\subsection{Muon radiative energy losses: showering} \label{subsecsec:showering}

In order to understand the effect of showers on the various aspects of muon reconstruction and measurements (including triggering) we have developed empirical definitions to identify (``tag'')
and characterize showers in the muon systems.
Both data and simulation samples are used to converge on this definition of a ``shower''
and are compared to study the adequacy of the shower modeling in simulation.

The ``extended tag-and-probe'' technique (Section~\ref{sec:Efficiency}) is used to study showers in simulated high-mass DY samples and in dimuon events from the single-muon primary data sets (Section~\ref{sec:samples}).
Definitions for tags, probes, and dimuon pairs are the same as those used to measure muon reconstruction efficiency, and are described in detail in Section~\ref{sec:recoEff}.
In addition, single-muon (or antiparallel double-muon) samples uniform in $\eta$ and $p$ in the range between 5 and 2500\GeV
are generated without simulating pileup.
In this case, the muon candidates used in the analysis are required to satisfy only the selection criteria used for probes, except that the muons are not required to come from the primary vertex, since it is problematic to accurately reconstruct a vertex with only two tracks that are nearly antiparallel.

The multiplicity of segments reconstructed within a single DT or CSC station can be used as a proxy to identify showers.
The tracker track of the selected probes is extrapolated to the different station layers of the muon detectors.
Segments belonging to the chambers traversed by the propagated track are counted if they lie within $\abs{\Delta x} < 25\cm$ from the extrapolated track position,
with $\Delta x$ computed in local chamber coordinates and representing the bending direction of the track. If the extrapolated track crosses a given station layer close to the border between chambers,
or if different chambers overlap, segments satisfying the requirement on $\Delta x$ in all potentially crossed chambers are counted.
Finally, the number of track-segment matches, provided by the tracker muon identification for all the chambers involved in the computation, is also counted and subtracted from the total sum. The result of this logic is the number of \textit{extra segments} (\ie, the number of segments in addition to those belonging to a muon track),
computed independently for each station crossed by a muon. It is referred to as \Nseg.

The DT and CSC local reconstruction can generate ``ghosts'', \ie, reconstructed tracks with no corresponding genuine track, in cases of multiple track segments traversing a single chamber.
For example, in the case of DTs, the segment fitting is first performed independently in the $\phi$ and $\theta$ views of a chamber and pairs of such
``2D segments'' are then combined only at a later step of the segment reconstruction to provide a three-dimensional object.
Combinations are built out of all possible permutations of $\phi$--$\theta$ 2D segments,
leaving to the standalone track reconstruction the burden of the disambiguation.
A similar phenomenon happens for CSCs,
though with different logic due to a different approach to the segment building.

The value of \Nseg above which a station is considered to have a shower was chosen after considering several possibilities.
The probability to have at least one station with a shower increases with the muon momentum, while for very low momentum it should be close to zero.
The slope of dependence is larger for a looser requirement on \Nseg.
However, when requiring $\Nseg\ge1$, the shower probability for very low momentum is still ${\approx}$20--30\%, which suggests a large contribution from ghosts. This falls to ${\approx}$5--10\% for $\Nseg\ge2$; consequently, the requirement $\Nseg\ge2$ is chosen as the working point for shower tagging,
because this is the most sensitive definition having acceptably small mistagging of showers at low momentum.

\begin{figure*}[!t]
\centering
{\includegraphics[width=0.49\linewidth]{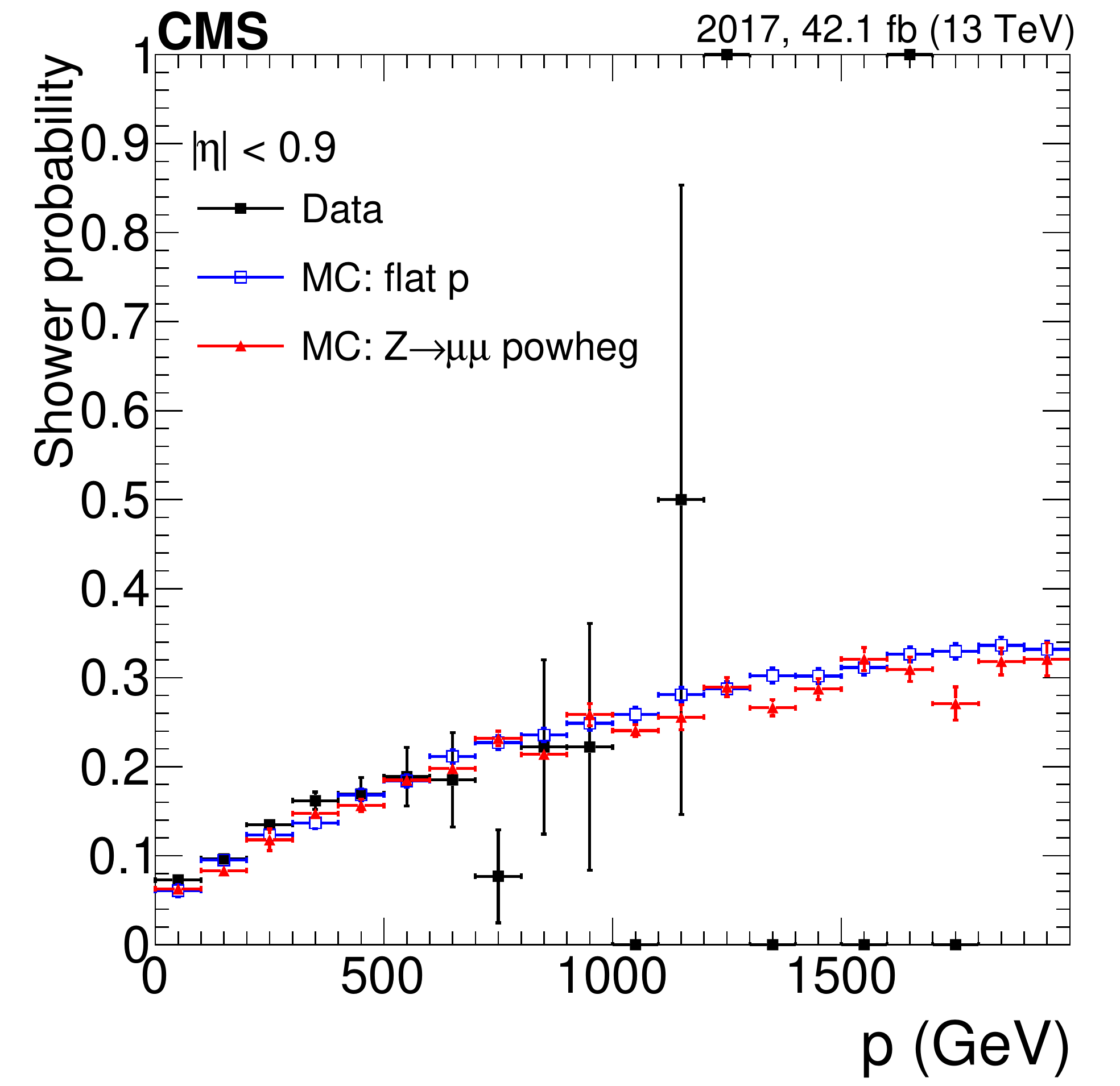}}
{\includegraphics[width=0.49\linewidth]{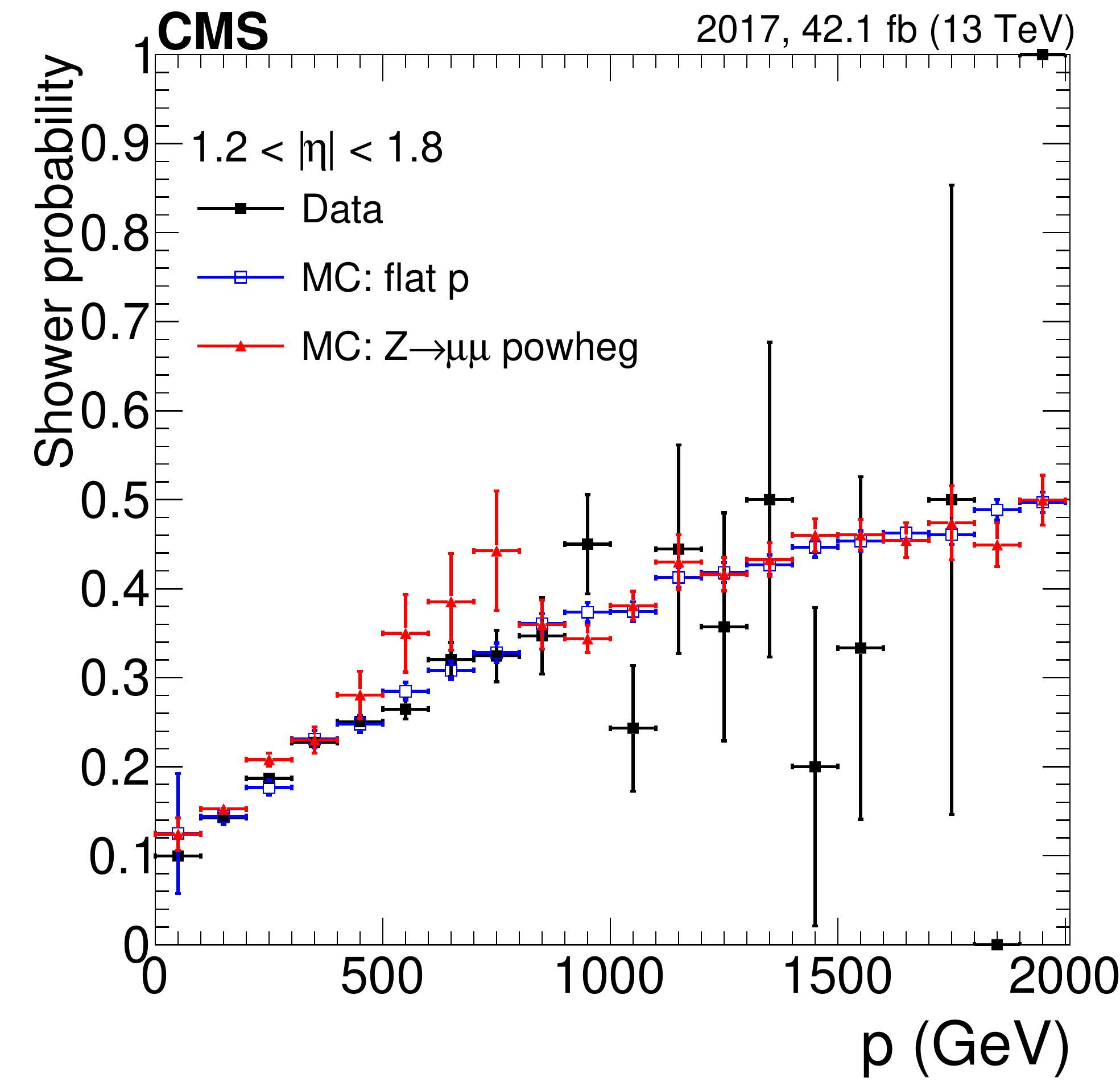}}
{\includegraphics[width=0.49\linewidth]{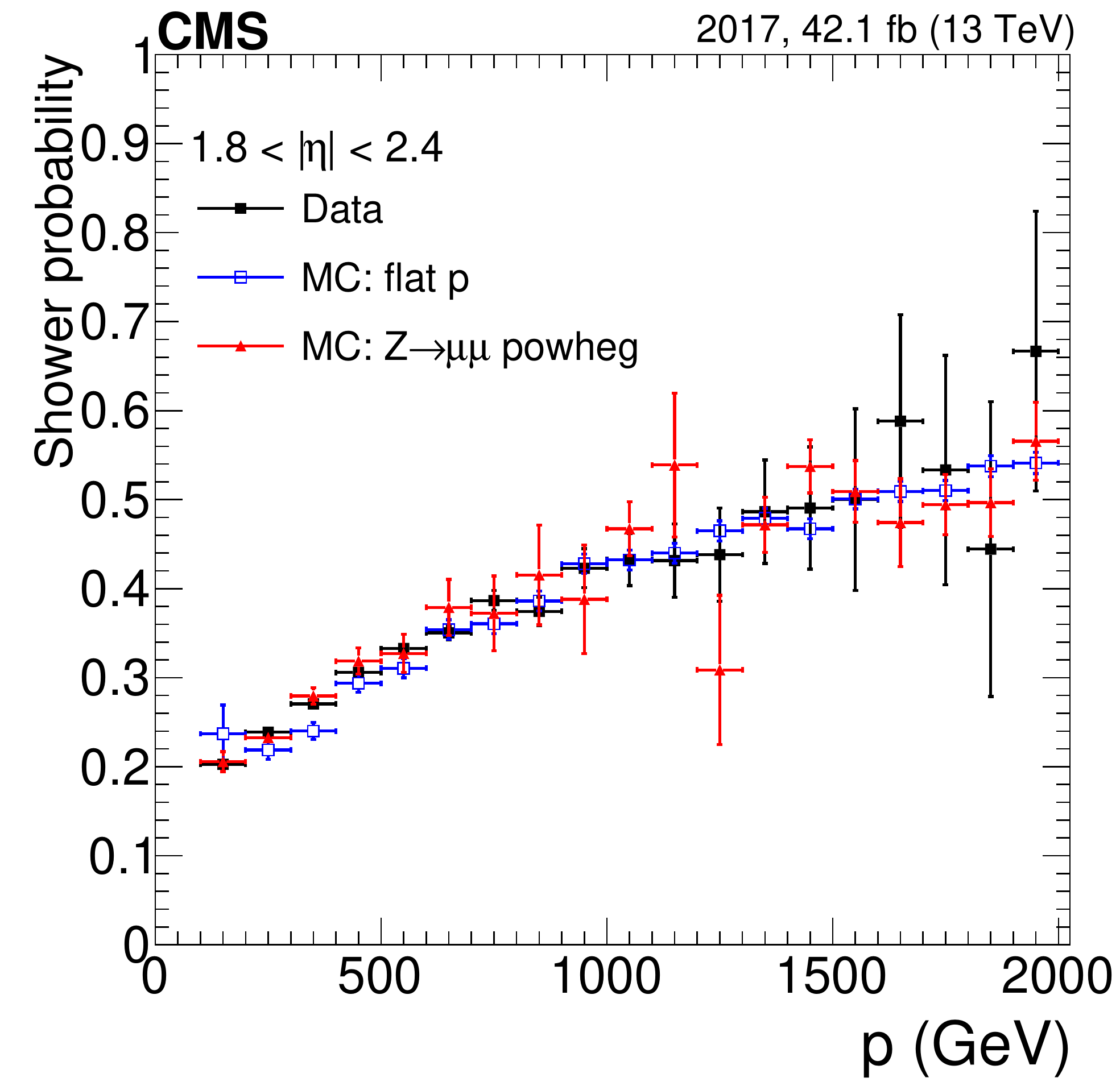}}
\caption{The probability $\Pshower(p)$ to tag at least one shower in any of the four stations, as a function of the incoming muon momentum, for (upper left) DTs; (upper right) CSCs with muon $\abs{\eta} < 1.8$; and (lower) CSCs with muon $\abs{\eta} > 1.8$ . Results are evaluated for the shower tagging definition requiring $\Nseg\ge2$. Different colors refer to: data (black), DY simulation (red), and single muons simulated with a uniform $p$ distribution (blue).}
\label{fig:showersVsP_dataVsMc_segment}
\end{figure*}

The probabilities of finding a shower in each of the four muon stations are computed separately and are compatible,
except in the first muon station in the endcap, where the shower probability
is higher than in the remaining endcap stations by ${\approx}$20\%. We attribute this to
hadronic punch-through hadrons from other collisions in the bunch crossing, wrongly tagged as muon-induced showers; this effect is not present
in the single-muon simulated sample, which does not include pileup.
For the purpose of the studies in this paper, we use a simple picture with one number characterizing the probability of tagging a shower in any station.
Figure~\ref{fig:showersVsP_dataVsMc_segment} shows the resulting probability $\Pshower(p)$ to tag a shower in at least one of the four muon stations
as a function of the muon momentum.

Results from data are compared with those from the simulated high-mass DY and single-muon samples, in the barrel and endcap regions separately. The endcaps are further split above and below $\abs{\eta} = 1.8$ to isolate the forward endcap region that has the highest shower probability.
Below 1000\GeV there is good agreement between data and simulation, thus validating the modeling of showers in simulation.

\subsection{Identification} \label{subsecsec:identification}

High-momentum muons are produced in rare processes with low cross sections and backgrounds. Often in searches the muon identification performance is measured using simulation in TeV signal regions that is validated only with extrapolations from measurements at lower momenta. In order to make this procedure more robust, the muon identification efficiency is designed to be uniformly high as a function of muon $p$ and $\pt$. For this purpose a dedicated high-\pt muon identification was designed during Run 1~\cite{Khachatryan:2014fba} (``Run 1 \hiptid''), targeting topologies involving high-\pt muons; it was further improved during Run 2 (``Run 2 \hiptid'').

In the Run~1 \hiptid, muons are required to be global muons with at least two segments reconstructed in two muon stations that match the inner track.
This selection suppresses punch-through and accidental track-to-segment matches.
The main source of inefficiency is due to the gaps between the muon chambers and is more prominent in the barrel region, where CMS has two pathways (``chimneys'') for services located around $\abs{\eta} = 0.3$.
In contrast, chambers in the endcaps overlap with each other, which provides continuous coverage.
The main update of this selection for Run 2 is to consider global muons that have only one segment matching the inner track,
but only when the extrapolation from the tracker muon to the muon system predicts that they pass through the muon system gaps.
In that case, only zero or one segment is expected to match the inner track.
This change in the Run~2 \hiptid raises the signal efficiency by 1 to 2$\%$ at high \pt and improves agreement between the data and simulation.
The efficiency gain affects high-\pt muons slightly more than lower-\pt muons because of a kinematic correlation: high-\pt muons are mostly produced from high-mass states that have low absolute rapidity and hence their muon decay products are more likely to be in the barrel region.

To guarantee that the muon system information is also used in the final momentum assignment,
the Run~1 \hiptid requires that at least one valid muon system hit be retained in the global muon fit,
which removes the outlier hits.
The global muon valid hit collection is inherited from the parent standalone muon and the hits are qualified as
valid when their addition to the global muon fit does not degrade the $\chi^2$.
However, in the presence of showers, the hit multiplicity increases and the $\chi^2$ of the standalone fit gets worse when trying to include them in the trajectory fit.
The \TuneP algorithm that has been developed to optimize the muon refit (Section~\ref{subsecsec:reconstruction}) can result in a hit collection used for the final momentum assignment that differs from the global hits collection; furthermore, if $\pt<200\GeV$, the TuneP algorithm chooses the fit using only tracker hits.
Hence, the second change from the Run~1 \hiptid to the Run~2 \hiptid consists in requiring that either
the global muon fit or the fit chosen by \TuneP use at least one valid muon system hit.
This change raises the signal efficiency by 1\% for muons with $\pt>500\GeV$,
mostly affecting the endcap region where showering (which scales with $p$, not \pt) is more abundant.

\begin{figure}[!htb]
  \begin{center}
    \includegraphics[width=.49\textwidth]{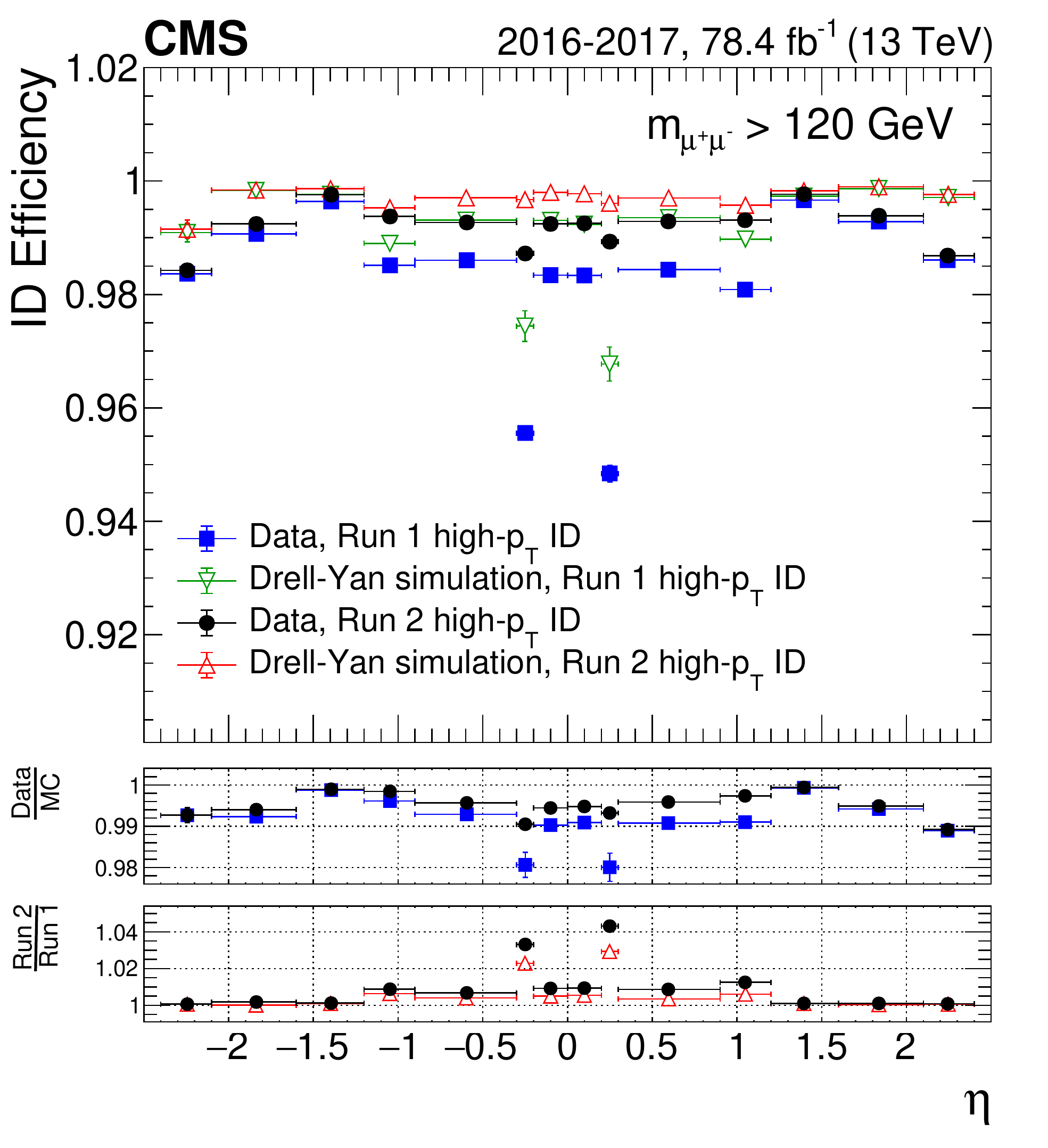}
     \includegraphics[width=.49\textwidth]{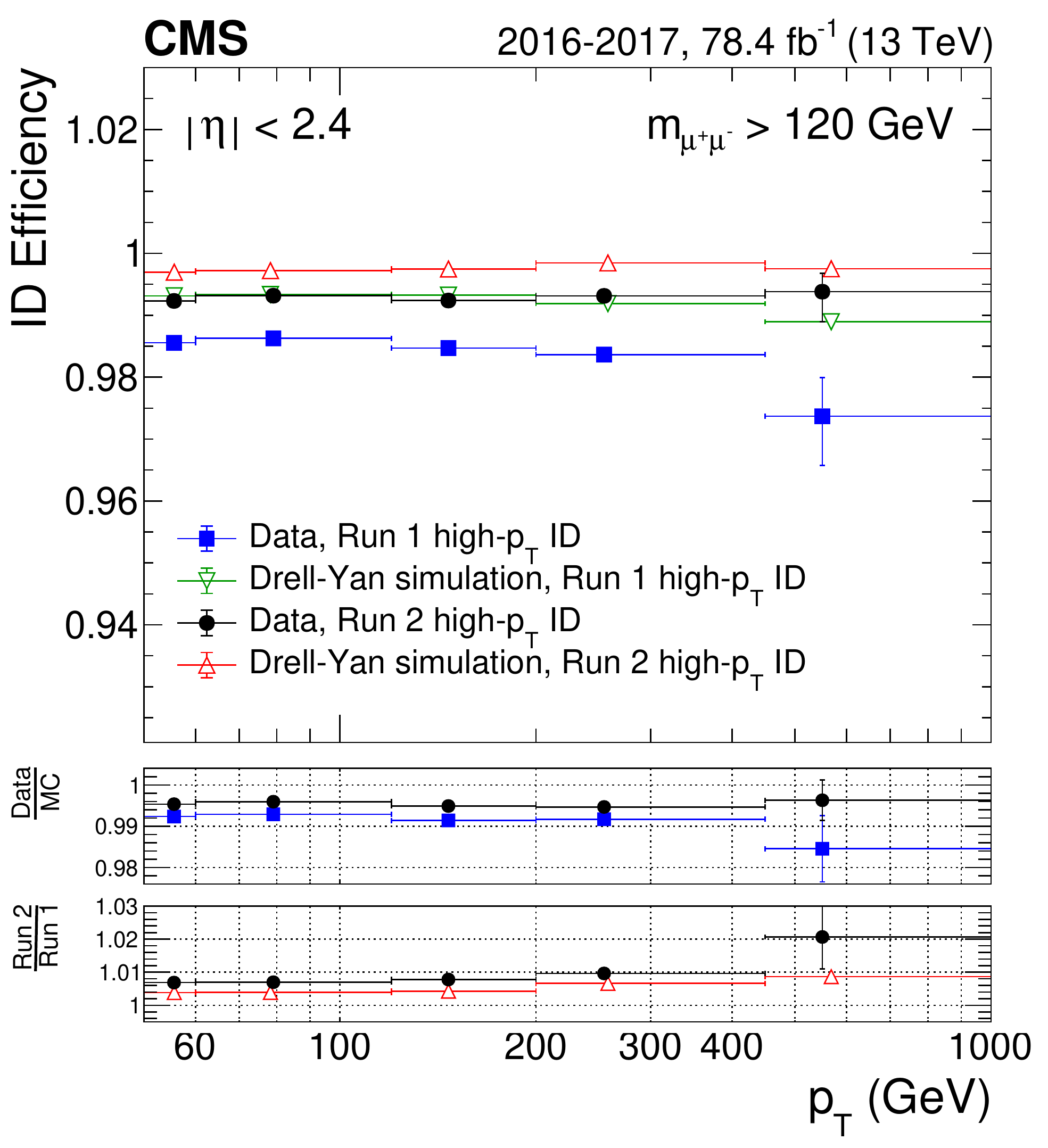}
  \end{center}
  \caption{Comparison between the efficiency of Run 2 and Run 1 \hiptid, as a function of (left) $\eta$ and (right) \pt. The efficiencies are obtained from dimuon events with a mass greater than 120\GeV to further select the high-mass DY process. The top panel shows the data to simulation efficiency ratio obtained for the Run 1 (blue squares) and for the Run 2 \hiptid (black circles). The bottom panel shows the Run 2 to Run 1 \hiptid
   efficiency ratio obtained from the data (black circles) and from simulation (red triangles). The central value in each bin is obtained from the average of the distribution within the bin.}
  \label{fig:IntroHighPtID}
\end{figure}

Figure~\ref{fig:IntroHighPtID} displays the Run~1 \hiptid efficiency as a function of muon $\eta$ and $\pt$, with comparison to the Run~2 \hiptid efficiency.
They are obtained from DY simulations and from dimuon events in data when combining the full 2016 and 2017 data sets.
The method to compute these efficiencies as well as more details and results concerning the Run~2 \hiptid efficiency are discussed in Section~\ref{sec:highptIdEff}.

The other selection criteria of the Run~2 \hiptid are the same as for the Run~1 \hiptid, with notably tight requirements on the track part of the global muon.
A minimal number of pixel hits and tracker layers is required in order to ensure that the muon originates from the center of the primary interaction,
to suppress cosmic ray muons and muons produced from meson decays in flight, and to ensure good momentum measurement resolution.
Finally, a muon is required to have a reliable $\pt$ assignment to perform the analysis; thus only global muons with a \TuneP relative \pt measurement uncertainty, $\sigma_{\pt}/\pt$, smaller than 30$\%$ are considered.

\section{Efficiency measurements}
\label{sec:Efficiency}

The tag-and-probe method~\cite{Chatrchyan:2012xi} is a standard
technique for measuring efficiencies for prompt muons coming from \cPZ boson
decays. The method provides an unbiased estimation of the total muon efficiency $\epsilon_{\mu}$
at the various stages of muon trigger, offline muon tracking
reconstruction, and
muon identification. Each component of $\epsilon_{\mu}$ is determined
individually and factorized according to:
\begin{equation}
\epsilon_{\mu} = \epsilon_{\mathrm{track}}
\epsilon_{\mathrm{ID}} \epsilon_{\mathrm{reco}} \epsilon_{\mathrm{trig}}.
\end{equation}

The efficiency $\epsilon_{\mathrm{track}}$ of the tracker track
reconstruction appears independent of the muon momentum and does not require dedicated study at
high momentum~\cite{Chatrchyan:2014fea}. All other components of
$\epsilon_{\mu}$ rely on the performance of the muon system and can
potentially be affected by muon showering as well as by the biases in the muon
system alignment. Such features would lead to a dependence of efficiency on
muon $\pt$ and
$\eta$. The individual components $\epsilon_{\mathrm{ID}}$, $\epsilon_{\mathrm{reco}}$, and
$\epsilon_{\mathrm{trig}}$ are scrutinized and computed as functions of
these kinematic variables in Sections \ref{sec:highptIdEff}-\ref{sec:hltTriggerEff}, respectively. In
addition, in order to
understand the impact of muon showering on the efficiency
and to establish if the simulation models data accurately,
the various efficiency components are studied as a function of showering,
using the shower tagging method described in
Section~\ref{subsecsec:showering}.
A slight difference with respect to the usual tag-and-probe method concerns
$\epsilon_{\mathrm{reco}}$, where the probe is a tracker muon
instead of a track. Starting from a track allows probing of
the entire muon system reconstruction, whereas for the tracker muon
requirement, there is already the assumption that at least two segments
are reconstructed in the muon chambers and that they are aligned with
the track. We have checked that this difference has a negligible impact
and no $p$ dependence.
To gain further insight into the combined L1 and HLT efficiency of
Section~\ref{sec:hltTriggerEff}, separate L1 efficiency studies are presented in Section \ref{sec:l1trigger}.

In order to compute $\epsilon_{\mu}$ up to \pt of 1\TeV, the standard tag-and-probe method has been
augmented. In this ``extended tag-and-probe'' method, we aim to collect as many prompt
high-\pt muons from the DY process as possible with maximal
suppression of backgrounds. Therefore, we do not restrict the
invariant mass of the tag and probe muons to the \PZ boson mass window. For
background rejection, we impose very tight isolation requirements on both
tag and probe muons. The isolation requirements rely exclusively on
the energy measured in the tracker, in a cone centered on the muon
track and with a radius $\Delta R = \sqrt{\smash[b]{(\Delta \eta)^2 +  (\Delta \phi)^2}}$ smaller than 0.3. No inputs from the calorimeters are
considered in the computation of isolation, to avoid including radiation emitted by the
muon that could bias the shower studies. Only muons with
total energy in the cone smaller than 30\GeV and not more than $5\%$ of their \pt are kept.
In addition to the isolation selection, kinematical criteria can be applied, such
as requiring back-to-back events in the transverse plane, or a balance between
the $\pt$ vectors of the two muons. This last set of criteria can be used to reduce the
background contribution from \ttbar
events; when they are not part of the pair selection, they are at
least used to cross-check the results. The tag muon is required to
pass the full Run~2 \hiptid described in Section~\ref{subsecsec:identification}. After applying the probe
selection, which depends on the efficiency under study, no further background subtraction is needed; the
efficiency is calculated by counting passing and failing probe
muons.

\subsection[\texorpdfstring{High-$\pt$}{High-pT} muon identification efficiency]{\texorpdfstring{High-$\boldsymbol{\pt}$}{High-pT} muon identification efficiency}

\label{sec:highptIdEff}

\begin{figure}[!t]
  \centering
  {\includegraphics[width=0.49\textwidth]{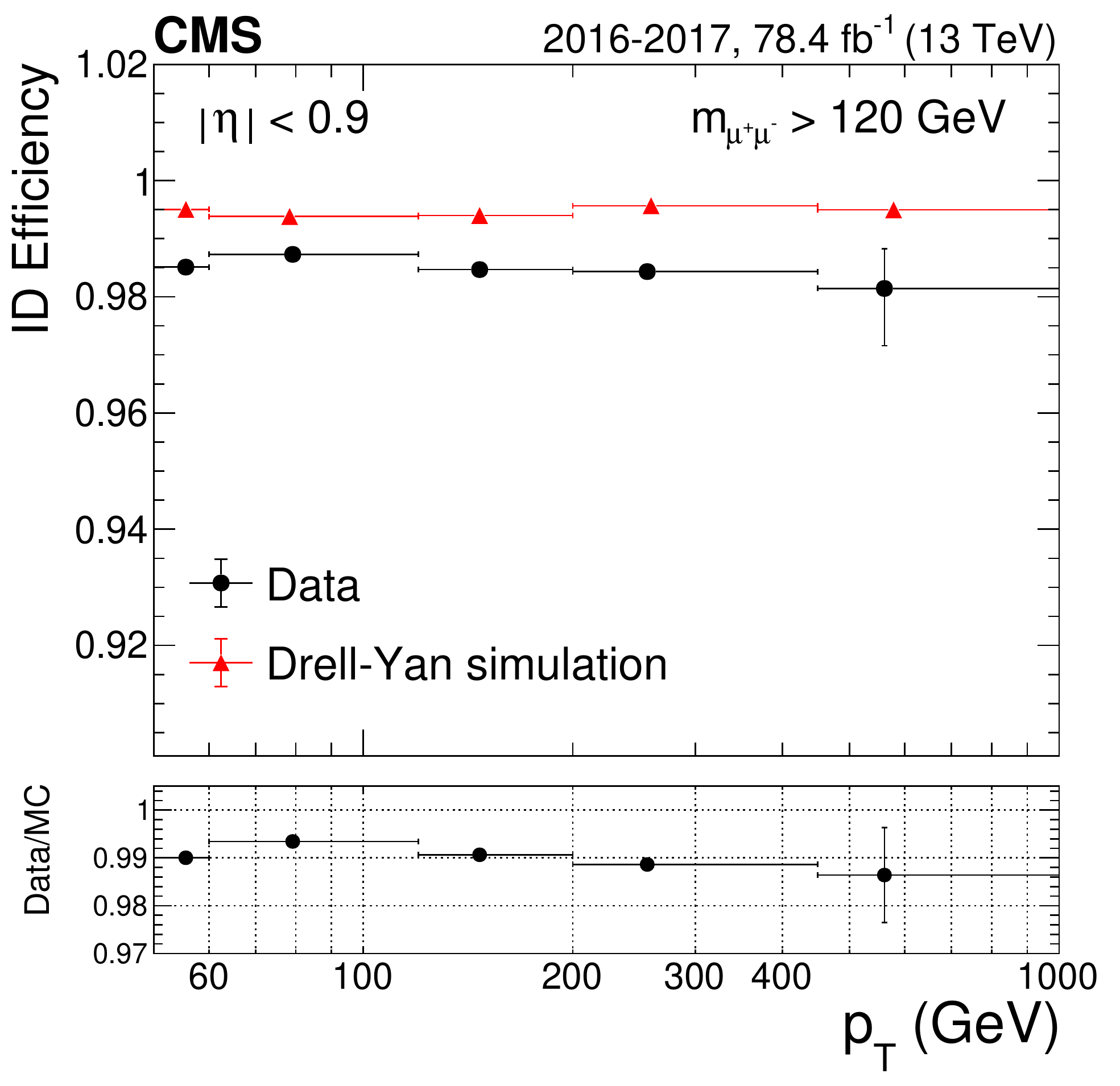}}
  {\includegraphics[width=0.49\textwidth]{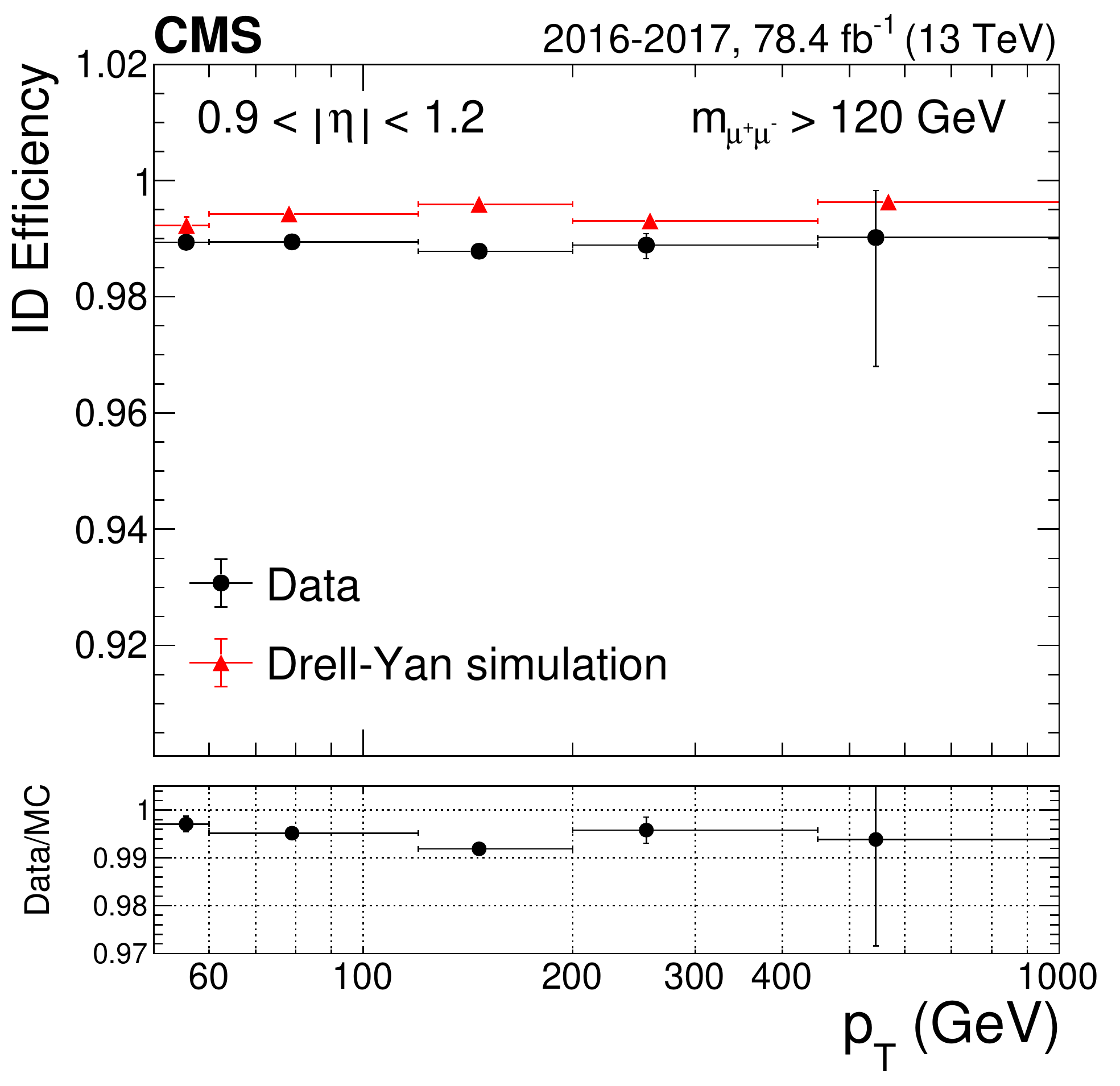}}
  {\includegraphics[width=0.49\textwidth]{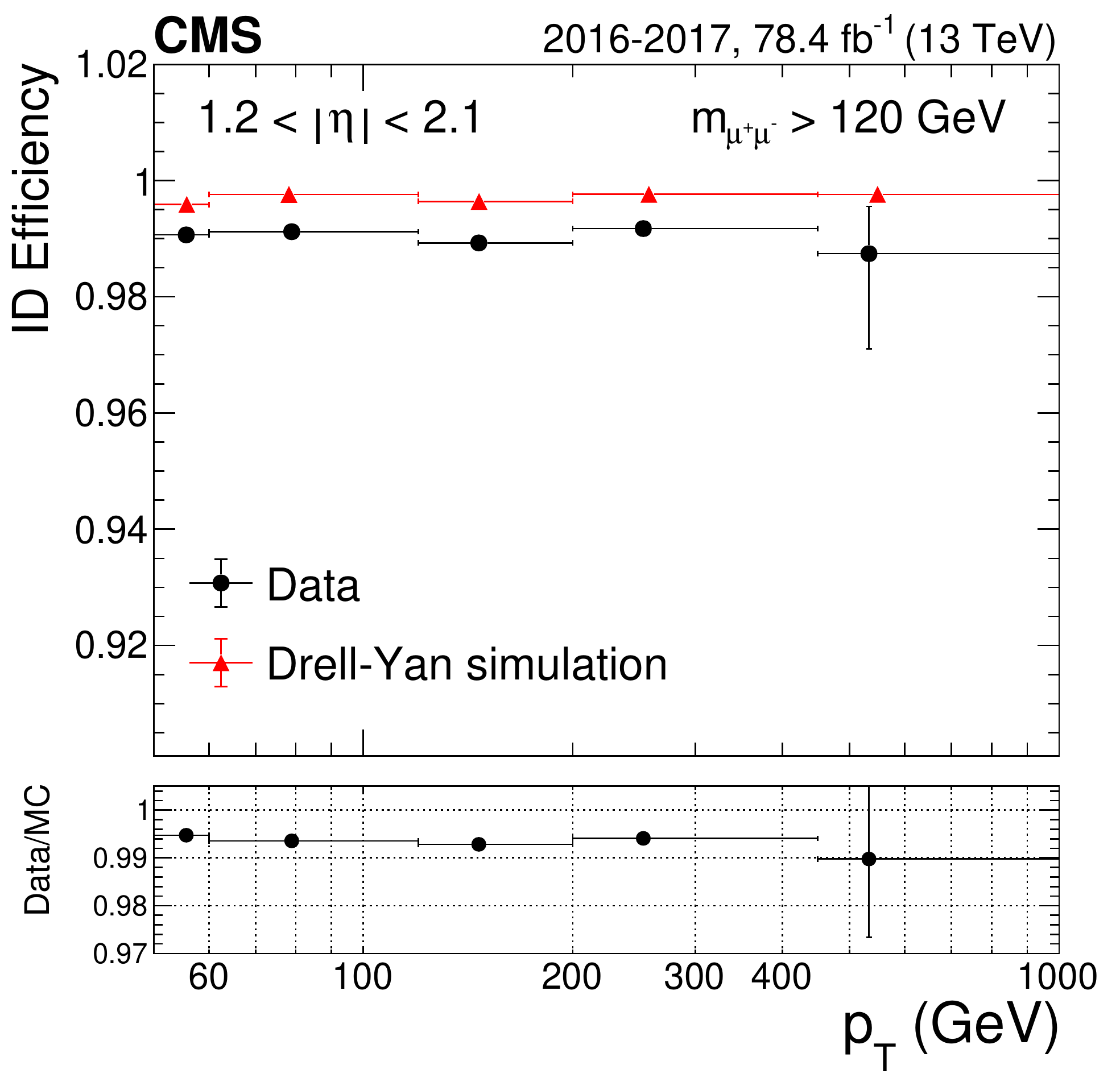}}
  {\includegraphics[width=0.49\textwidth]{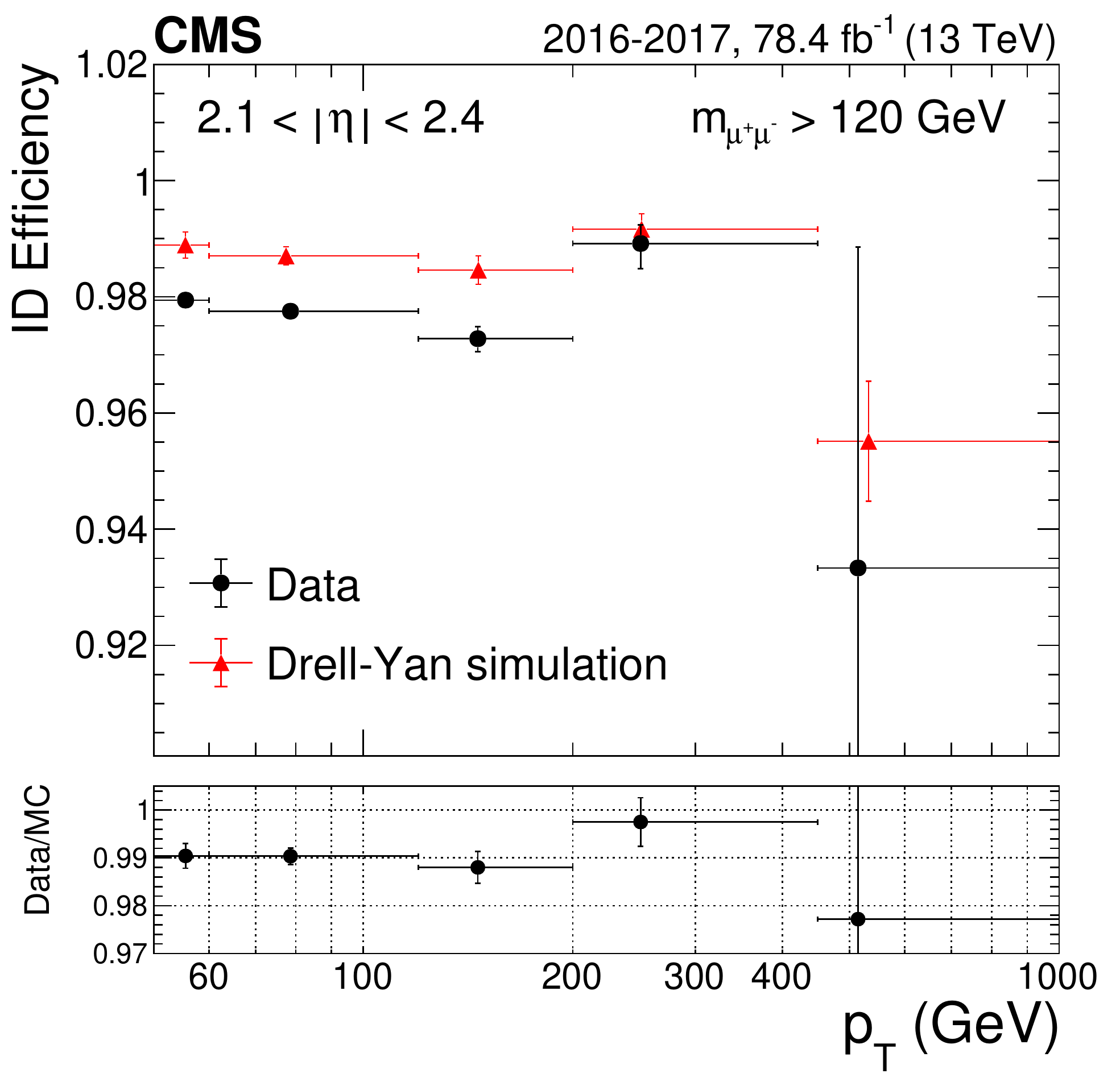}}
  \caption{\hiptidcaps efficiency for 2016 and 2017 data, and
    corresponding DY simulation, as a function of \pt for (upper left)
    $\abs{\eta}<0.9$, (upper right) $0.9<\abs{\eta}<1.2$, (lower left)
    $1.2<\abs{\eta}<2.1$, and (lower right) $2.1<\abs{\eta}<2.4$. The black
    circles represent data; the red triangles represent DY
    simulation. The data-to-simulation ratio, also called the data-to-simulation scale factor (SF), is displayed in the
    lower panels. The central value in each bin is obtained from the average of the distribution within the bin.}
  \label{fig:IDEff2016n2017}
  \end{figure}

The Run 2 \hiptid efficiency is measured using the extended tag-and-probe
method on muons that are reconstructed as global muons. The results are presented in Fig.~\ref{fig:IDEff2016n2017} for the
combined 2016 and 2017 data sets and for simulated DY samples.
The efficiency as a function of \pt is shown separately in four $\eta$ regions with different detector composition and characteristics:
$\abs{\eta}<0.9$, only composed of DTs;
$0.9<\abs{\eta}<1.2$, composed of both DTs and CSCs;
$1.2<\abs{\eta}<2.1$, only composed of CSCs;
and $2.1<\abs{\eta}<2.4$, the very forward region composed of CSCs but
very sensitive to pileup, punch through, and showering.

\begin{figure}[!th]
  \centering
  {\includegraphics[width=0.49\textwidth]{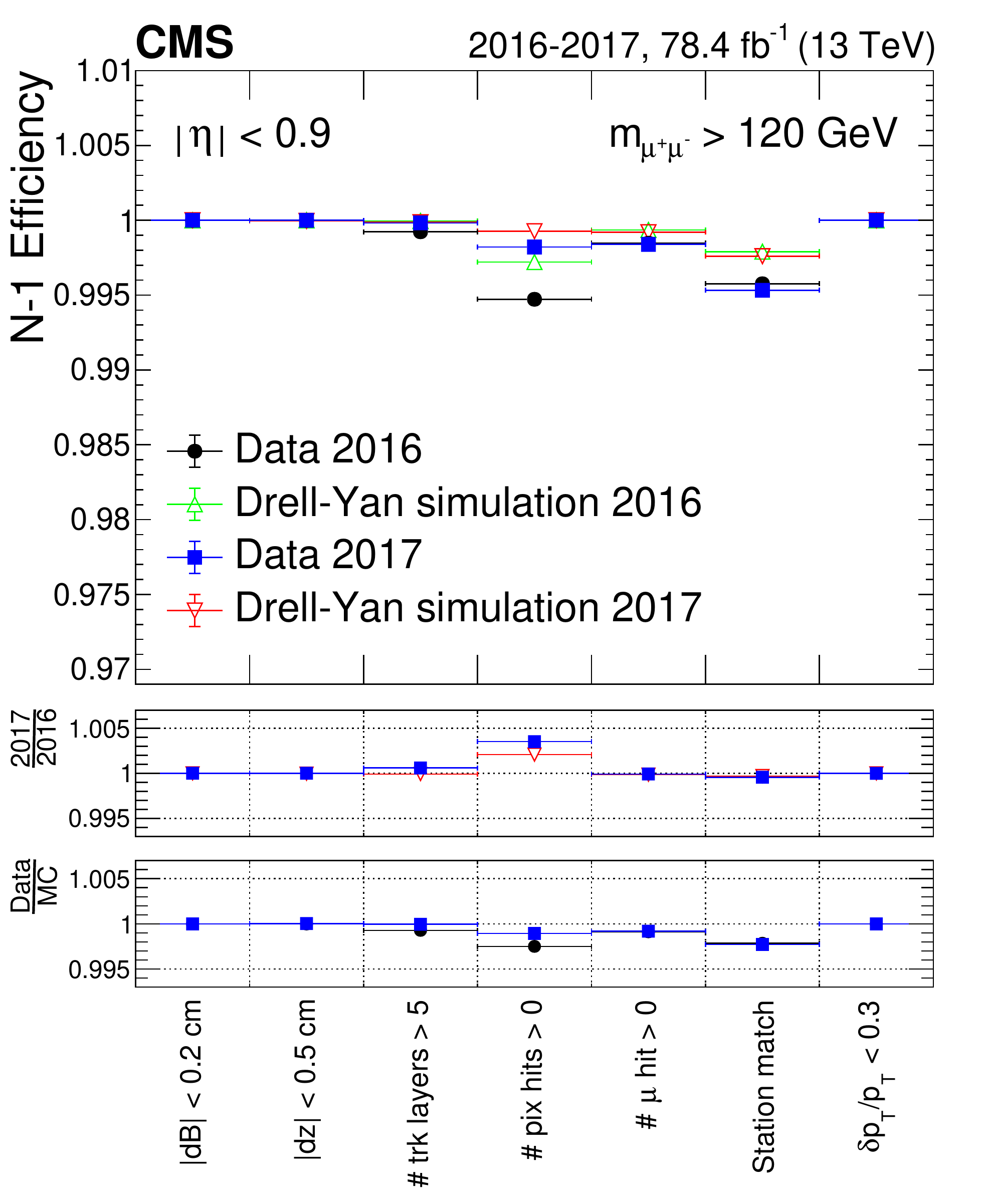}}
  {\includegraphics[width=0.49\textwidth]{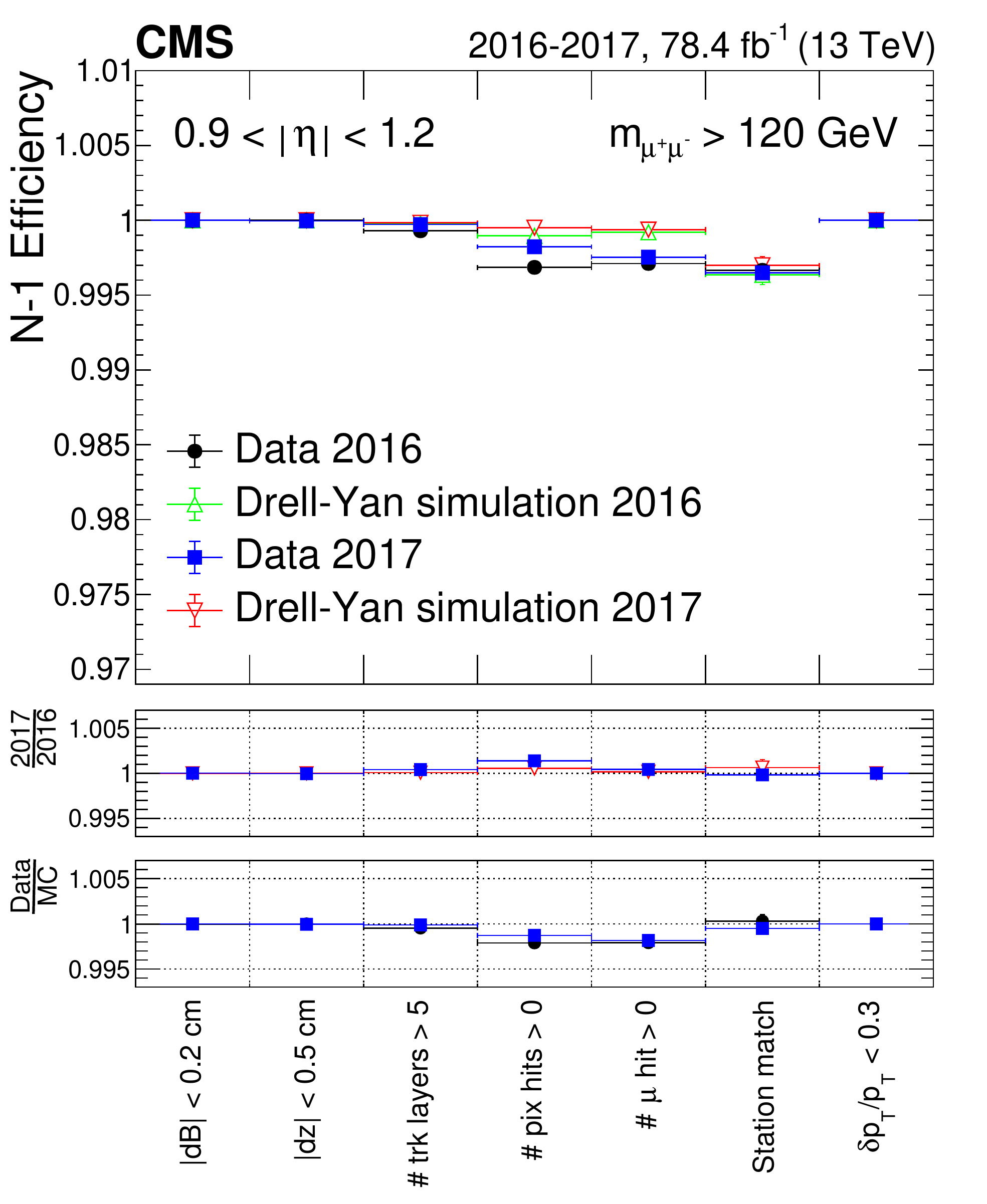}}
  {\includegraphics[width=0.49\textwidth]{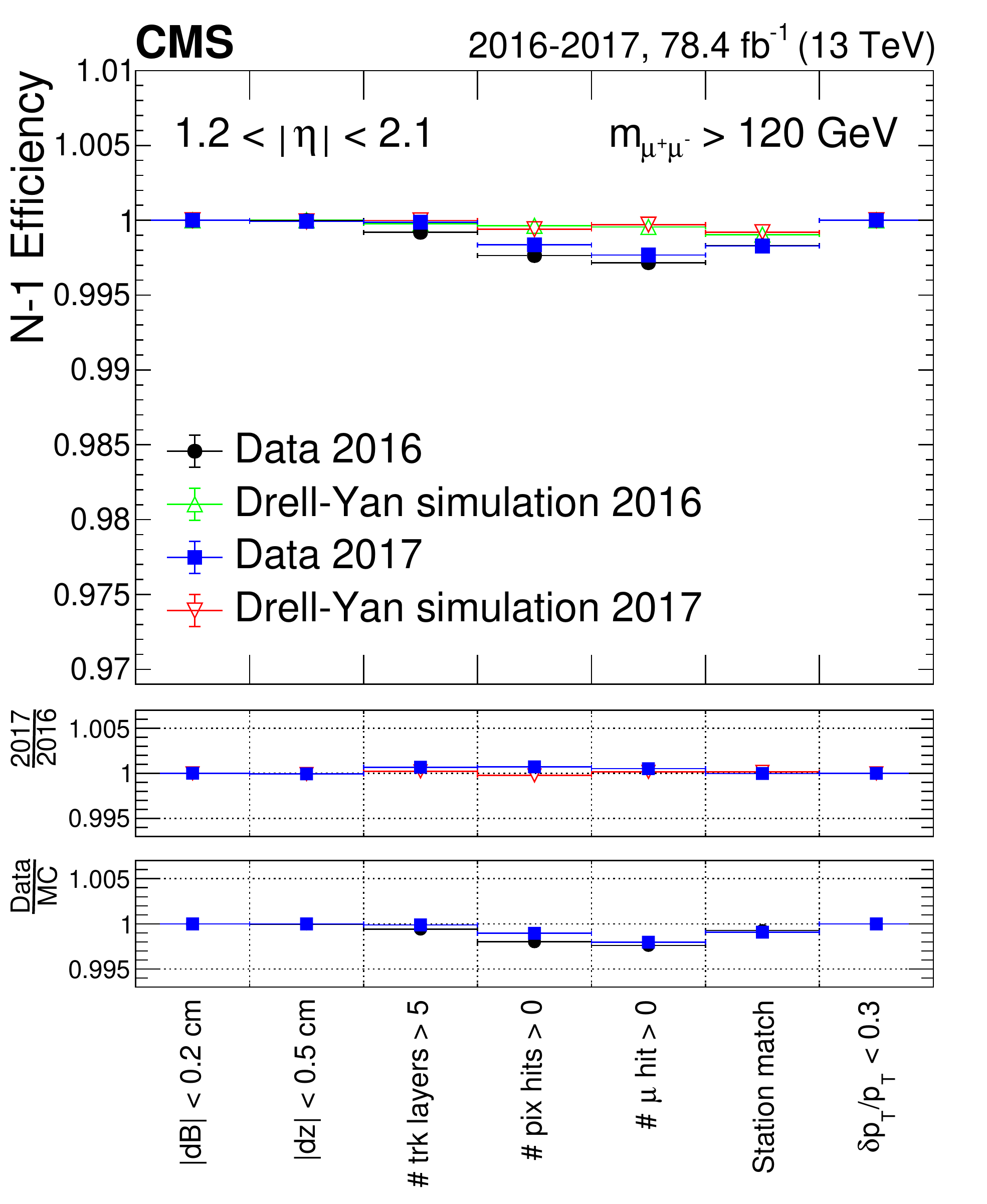}}
  {\includegraphics[width=0.49\textwidth]{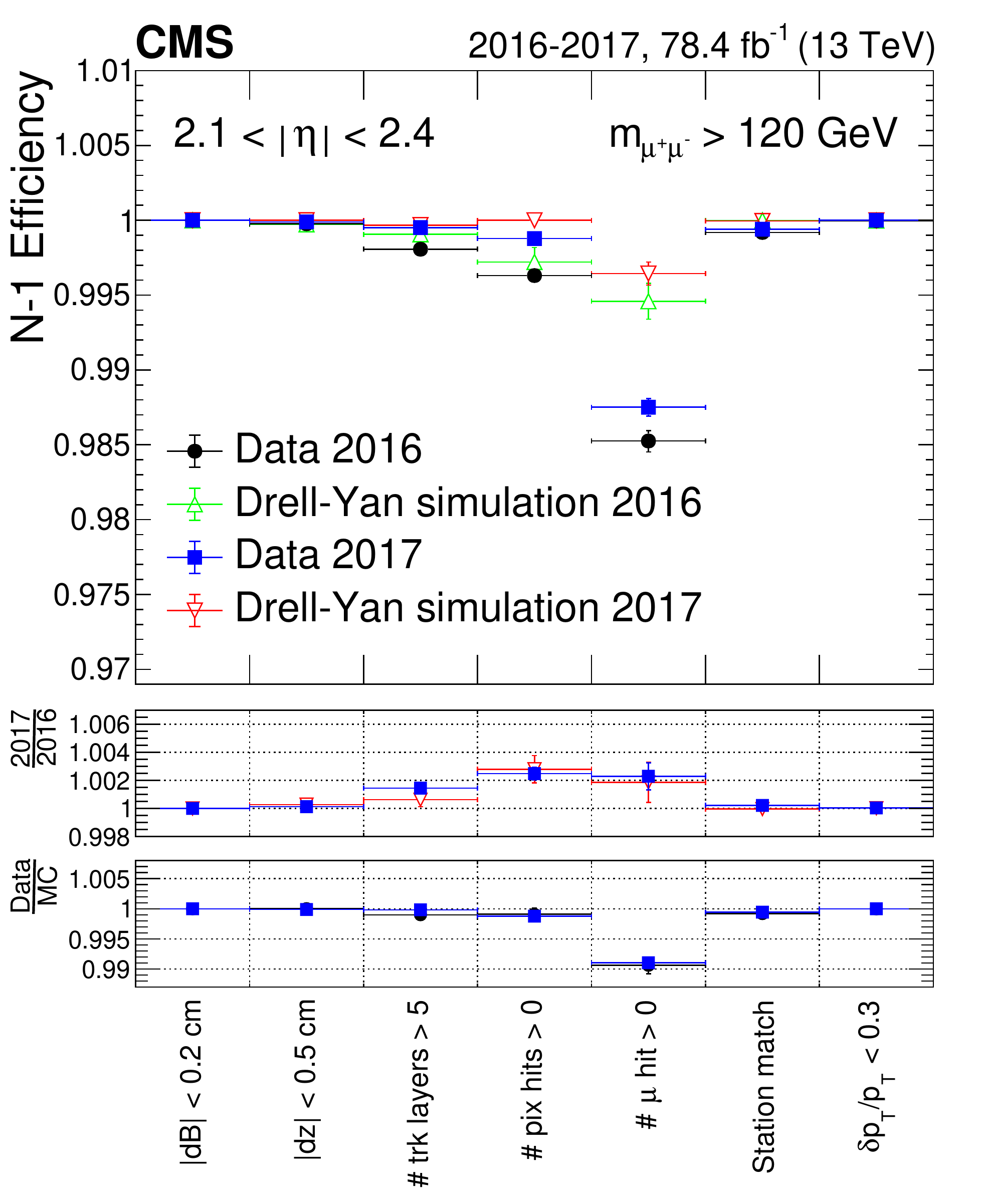}}
  \caption{The $N-1$ efficiencies, for $\pt>53\GeV$ and binned in
    $\eta$, comparison between 2016 and 2017 data sets and for the
    corresponding DY simulations, for (upper left) $\abs{\eta}<0.9$,
    (upper right) $0.9<\abs{\eta}<1.2$, (lower left) $1.2<\abs{\eta}<2.1$,
    and (lower right) $2.1<\abs{\eta}<2.4$. The black circles represent
    2016 data; the blue squares represent 2017 data. The lower panels
    display the ratio of $N-1$ efficiencies obtained for each of the criteria,
    between 2017 and 2016 data sets, and between data and their corresponding simulations for both years. }
  \label{fig:SingleValNM1YearData}
  \end{figure}

A very high identification efficiency, mostly above 98\%, is
found over the full detector acceptance. No \pt-dependent inefficiency is found
for either 2016 or 2017 data. The DY simulation predicts
slightly higher efficiency than observed in data, but the data-to-simulation agreement is uniform with
increasing \pt. The ``$N-1$ efficiencies'' for each ID requirement are individually tested by dividing
the number of probe muons passing a given selection criterion by the number of probe
muons passing all other
criteria. Figure~\ref{fig:SingleValNM1YearData} shows the results for
each criterion that are obtained for muon $\pt>53\GeV$ and binned in $\eta$.
Although the matching criteria between the muon system segments
and the inner tracker part of the global muon were updated between
Run 1 and Run 2 (Section~\ref{subsecsec:identification}), this
selection is still responsible for the slight
discrepancy between simulation and data in the barrel region.  In the
endcaps ($\abs{\eta}>1.2$), we observe a slight inefficiency in both 2016
and 2017 data with respect to the rest of the detector and to simulation, due to the
requirement of a valid muon detector hit in the final momentum fit.
Finally, we observe a small efficiency gain in 2017 (${+}$0.5\%)
with respect to 2016 in the barrel region,
which can be traced back to the tracker part of the muon Run 2 \hiptid
that links the improvement with the new pixel detector installed in CMS
between the 2016 and 2017 data taking periods.

\begin{figure}[!t]
  \centering
  {\includegraphics[width=0.49\textwidth]{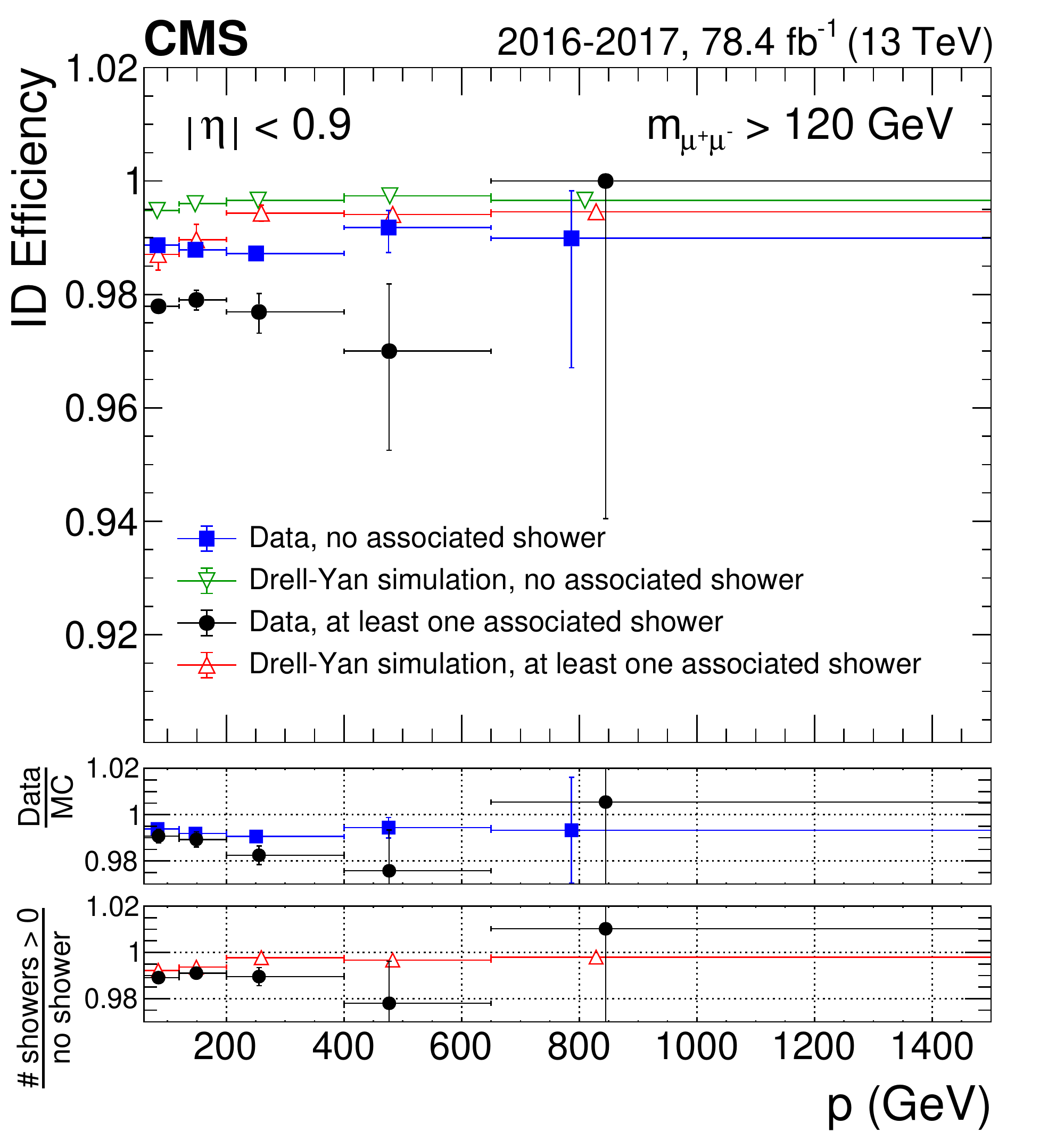}}
  {\includegraphics[width=0.49\textwidth]{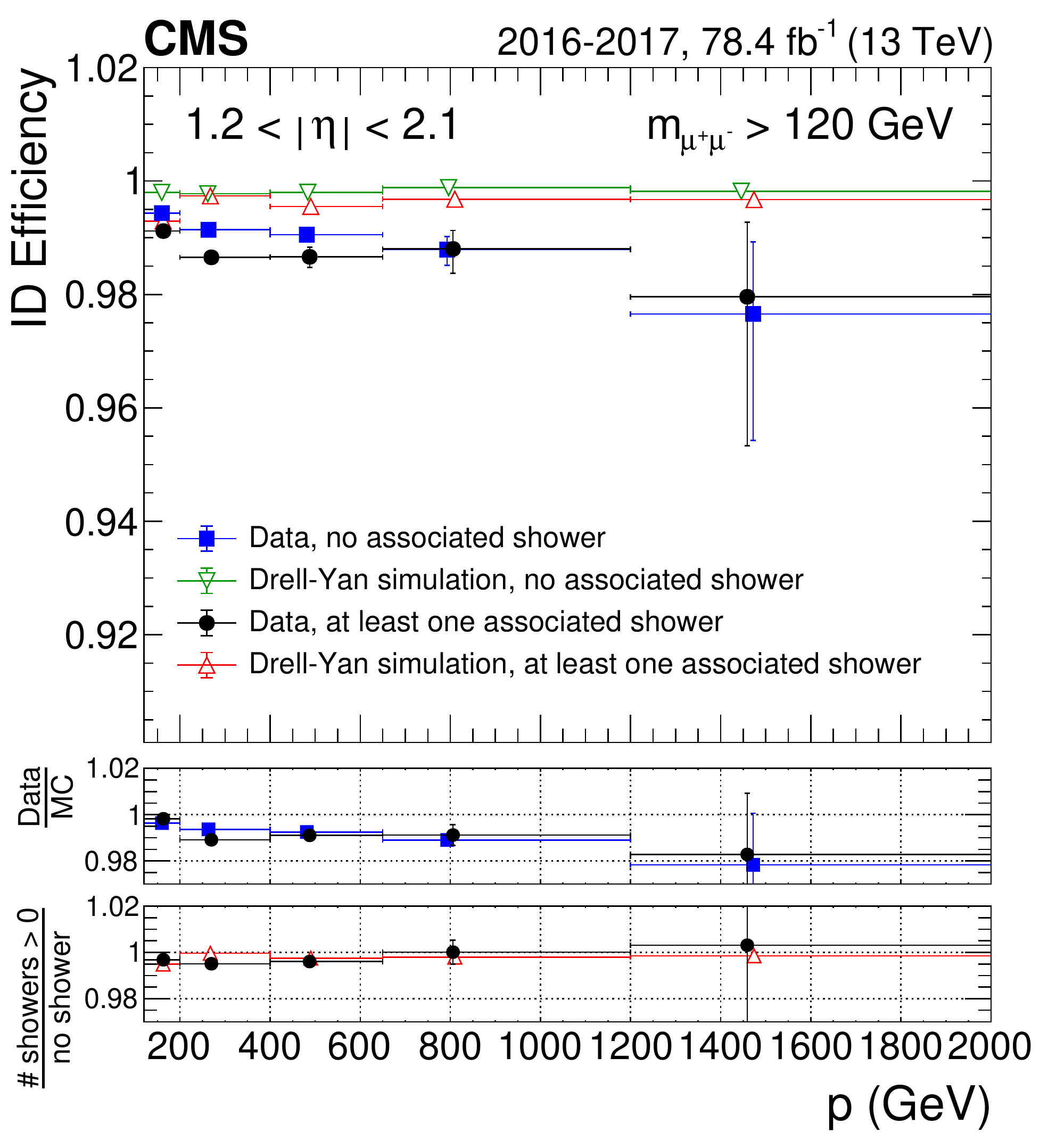}}
   {\includegraphics[width=0.49\textwidth]{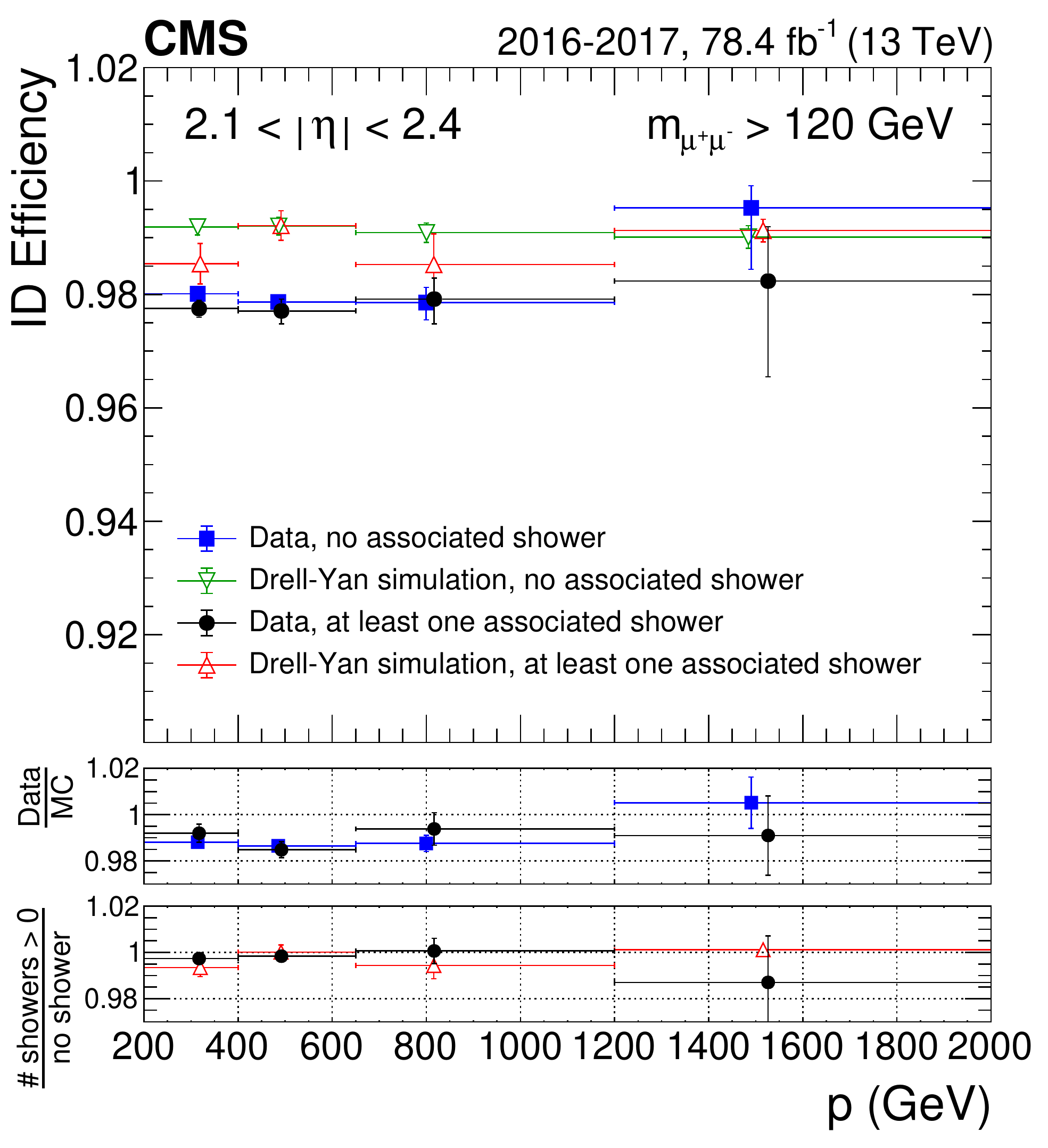}}
  \caption{\hiptidcaps efficiency for 2016+2017 data, and
    corresponding DY simulation, as a function of $p$ for (upper left)
    $\abs{\eta}<0.9$, (upper right) $1.2 < \abs{\eta} < 2.1$, and (lower) $2.1 < \abs{\eta} < 2.4$. The blue squares
    show efficiency for muons in data with no showers tagged; the green inverted triangles show
the same for muons in DY simulation.
    The black circles correspond to muons in data with at least one shower tag, while
    the red triangles are the same for muons in DY simulation. The central value in each bin is obtained from the average of the distribution within the bin.}
  \label{fig:IDVsShower}
  \end{figure}

The Run~2 \hiptid efficiency is very high and no trend is
observed with increasing $\pt$. The results are also
provided as a function of the muon $p$ in
Fig.~\ref{fig:IDVsShower}. The 2016 and 2017 data sets are combined in
order to reach higher sensitivity.
The efficiencies are further split into two
categories, whether or not a shower is tagged, given a muon.
The overlap region ($0.9<\abs{\eta}<1.2$) is not included, to avoid double counting
from CSC and DT segment-overlap that biases the shower tagging definition.
No effect due
to showering can be seen in the endcap region (upper right and lower plots), but a slight
decrease in the efficiency of 1\% is visible over the full momentum spectrum in the
barrel region (left plot) for muons with an associated shower. This
inefficiency is due to requirements on the matching of the inner track to the segments in the muon system,
which are responsible for most of the inefficiency in the barrel
region. In most of the cases, the muon is failing these identification criteria
because it fails to be reconstructed as a tracker muon, despite the
fact that the global reconstruction is successful. It appears likely that those
muons are emitting showers in the calorimeters, which cause a change in
trajectory before entering the muon system, so that the tracker-track
extrapolation does not match the segments.

\subsection{Reconstruction efficiency}
\label{sec:recoEff}

\begin{figure}[ht]
\begin{center}
  \includegraphics[width=0.48\textwidth]{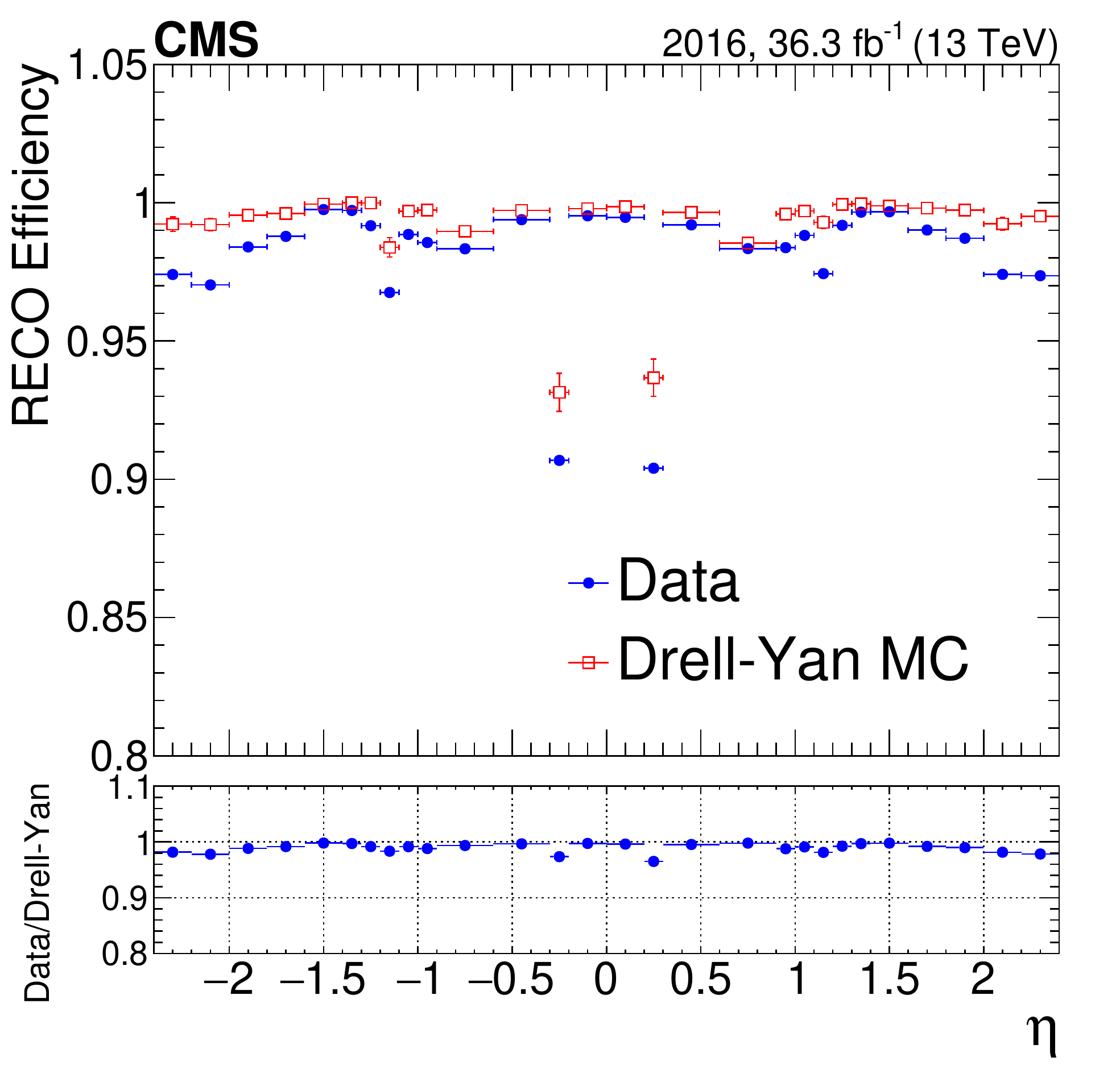}
  \includegraphics[width=0.48\textwidth]{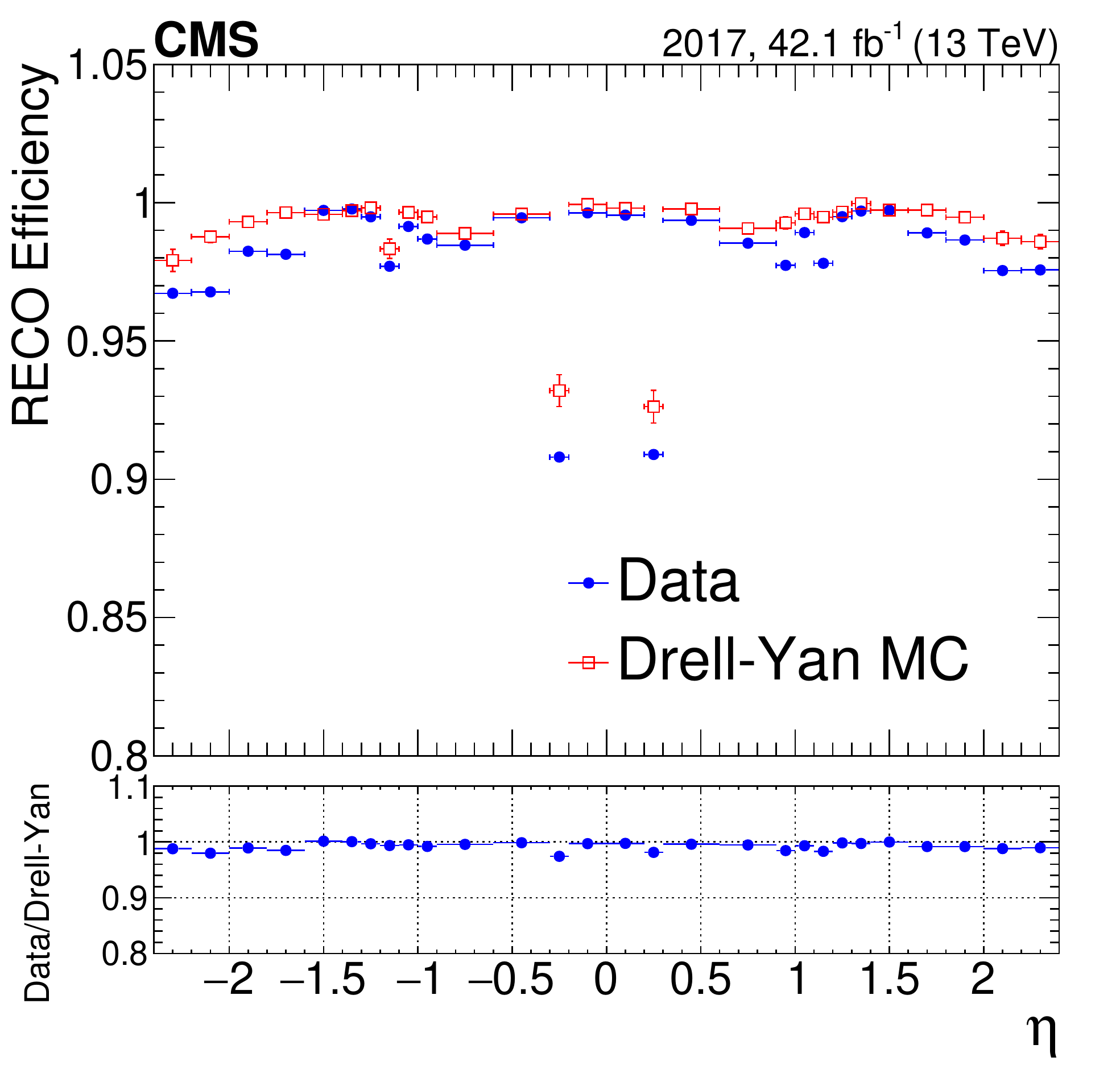}
\end{center}
\caption{Standalone muon reconstruction efficiency as a function of muon $\eta$ for the (left) 2016 and (right) 2017 data sets. The blue points represent the data, while the red empty squares represent the simulation.}
\label{fig:fig_reco_sta_eta}
\end{figure}

The standalone and global muon reconstruction efficiencies are studied as a function of muon $\eta$ and $p$ using the extended tag-and-probe
method.
The selected probe muons are required to be good quality tracker muons,
and the efficiency to reconstruct either standalone or global muons is calculated with respect to these probes.
Figure~\ref{fig:fig_reco_sta_eta} shows the 2016 and 2017 standalone muon reconstruction efficiency as a function of muon $\eta$
for muons with $\pt>53\GeV$. The efficiency is above 99\% in the barrel region and up to $\abs{\eta}=1.6$, both for data and simulation, and for both data sets.
For $\abs{\eta}>1.6$, the simulation does not reproduce the slight inefficiency observed in data.

\begin{figure}[ht]
\begin{center}
\includegraphics[width=0.48\textwidth]{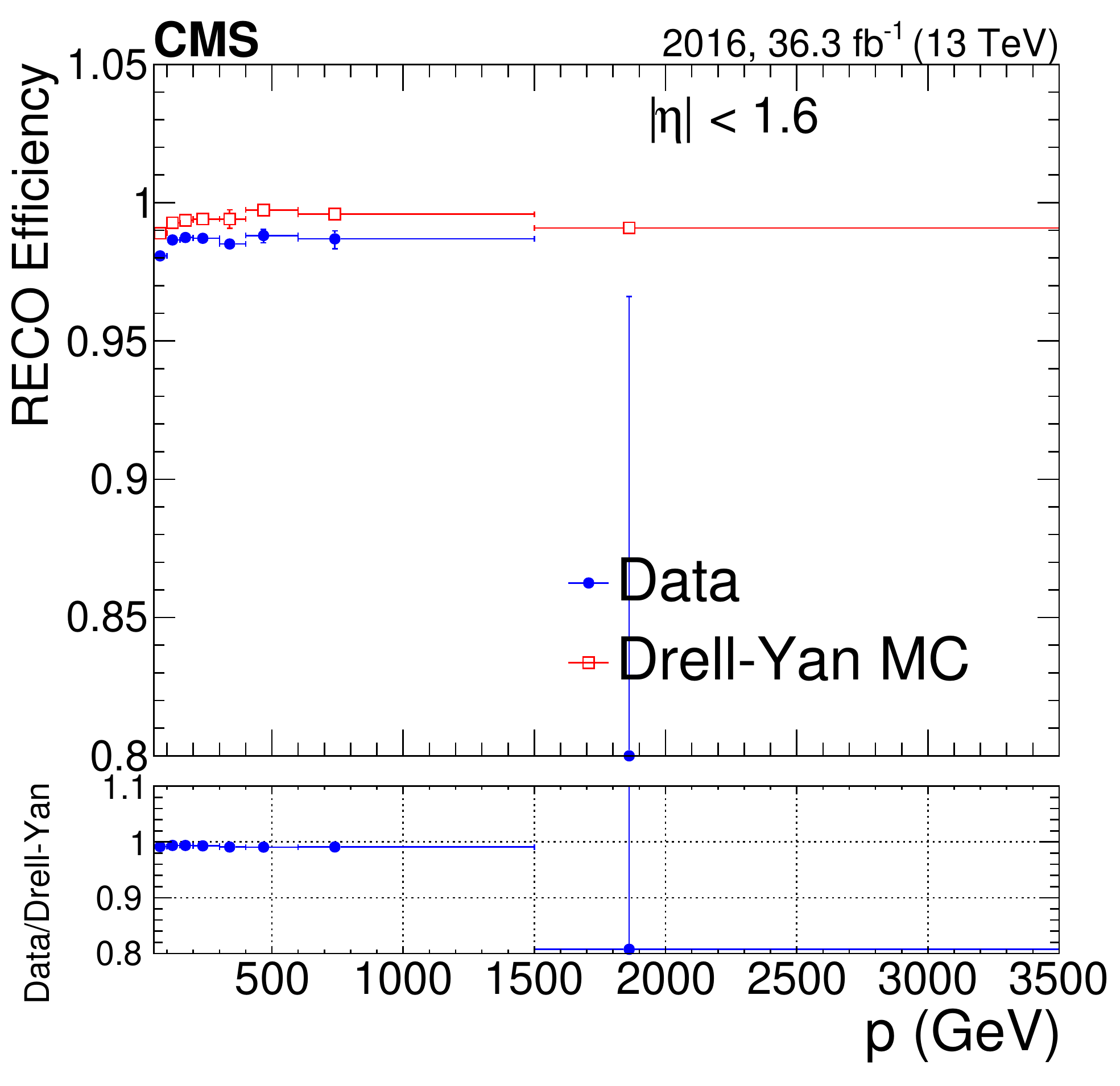}
\includegraphics[width=0.48\textwidth]{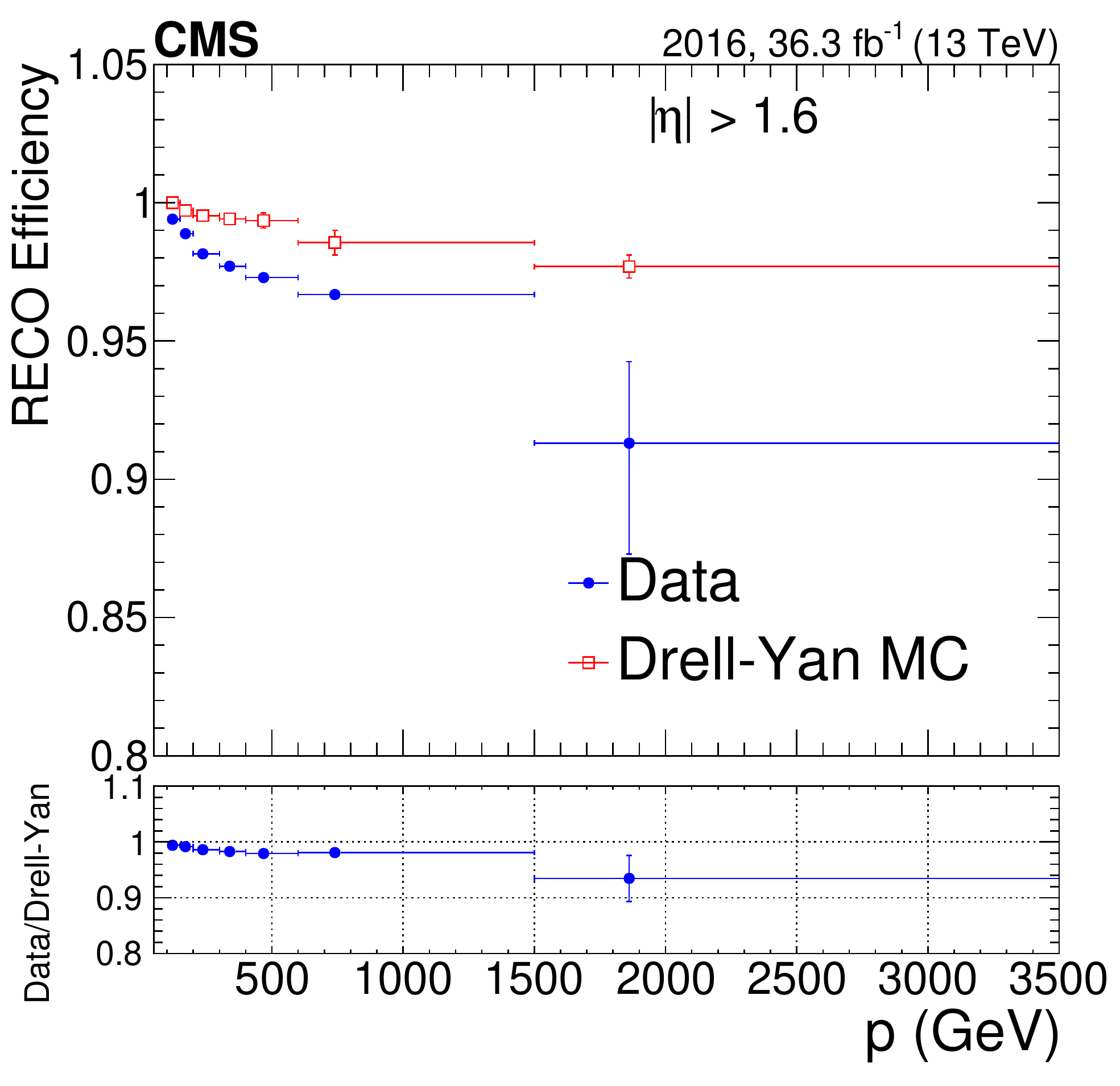}
\includegraphics[width=0.48\textwidth]{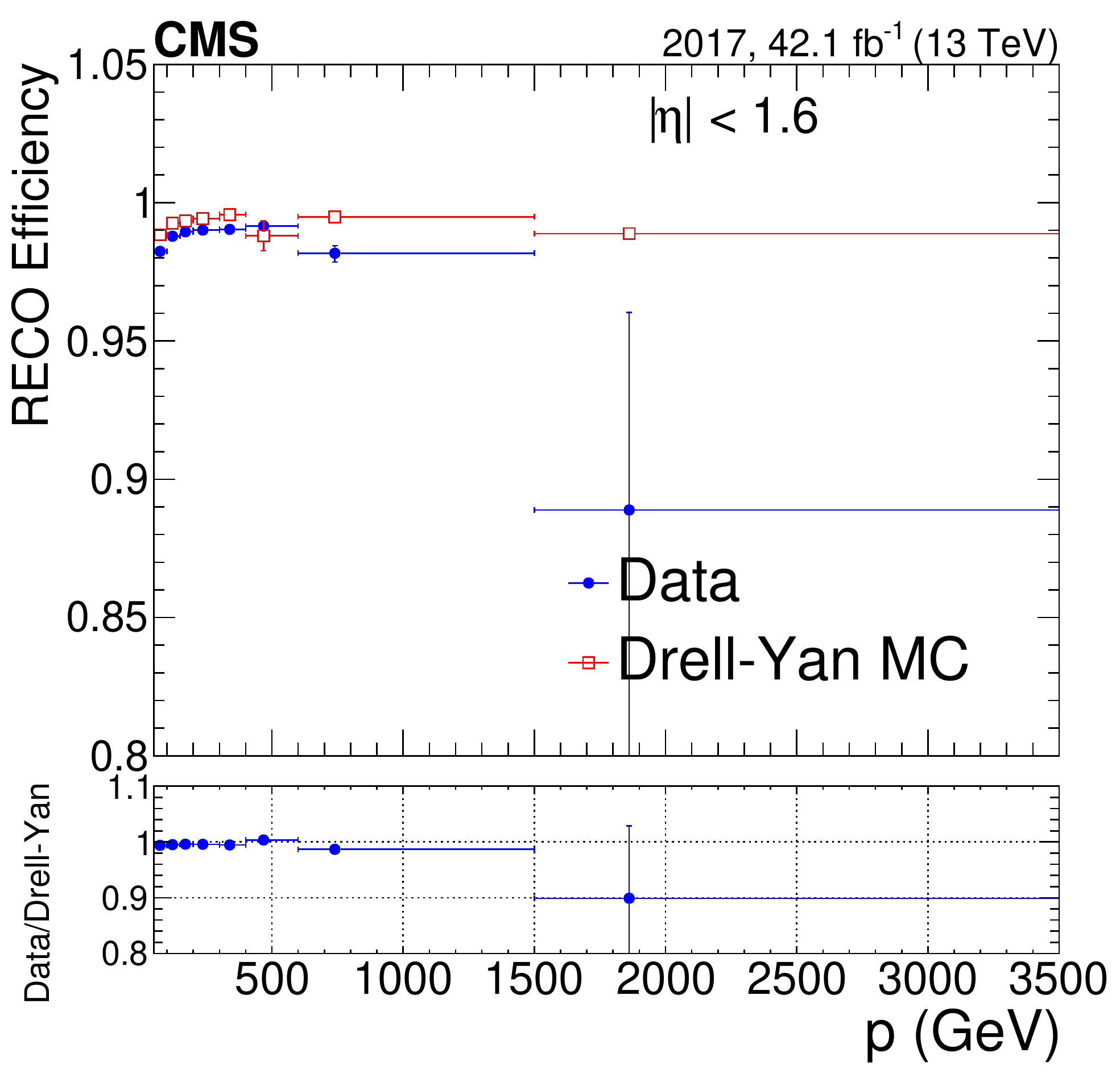}
\includegraphics[width=0.48\textwidth]{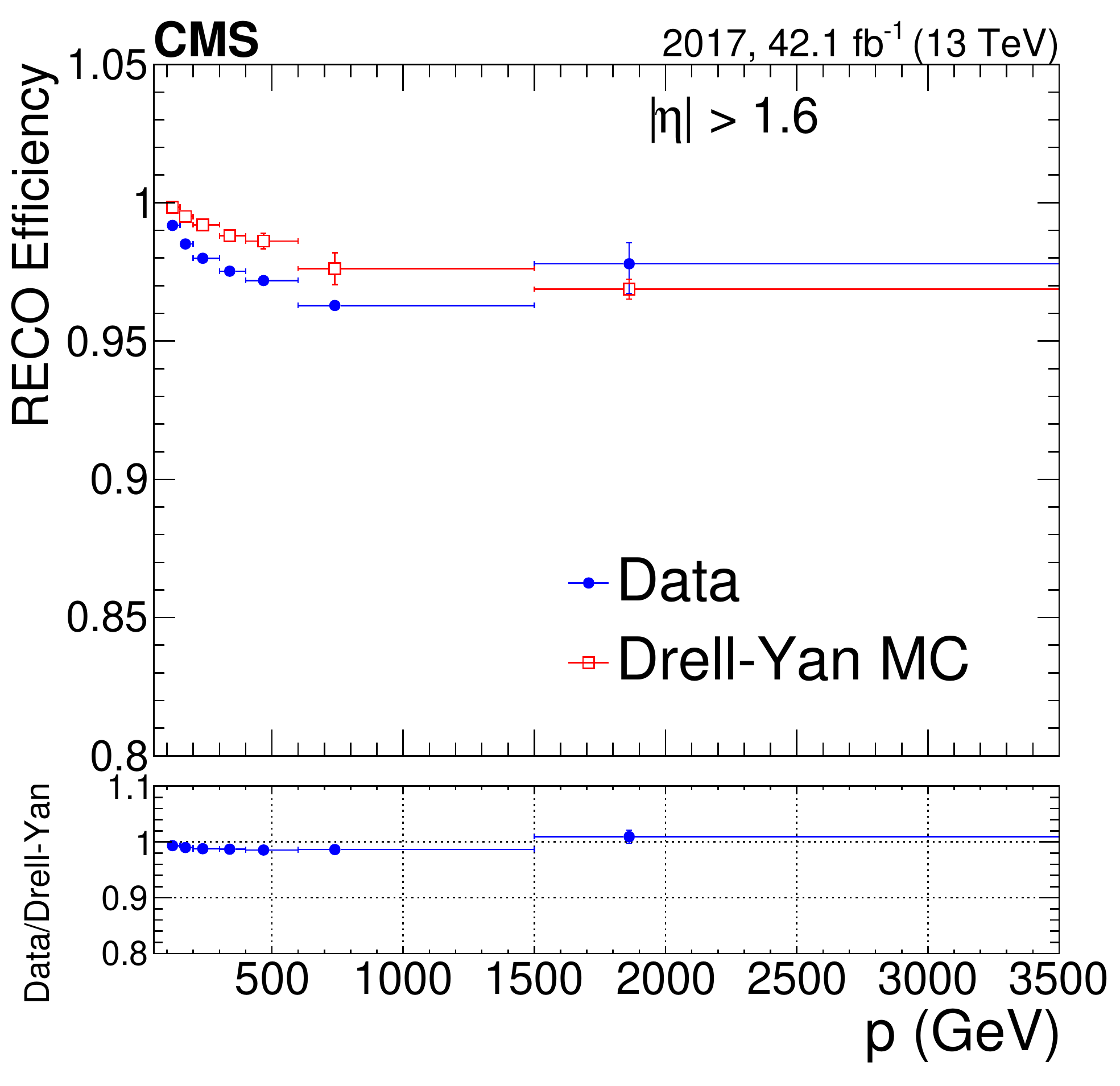}
\end{center}
\caption{Standalone muon reconstruction efficiency as a function of muon momentum in two different $\abs{\eta}$ regions: (left) $\abs{\eta} < 1.6$, and (right) forward endcaps from, $1.6 < \abs{\eta} < 2.4$. The upper row shows the 2016 results, with blue points representing data and red empty squares representing simulation. The lower row shows the 2017 results. The lower panels of the plots show the ratio of data to simulation. The central value in each bin is obtained from the average of the distribution within the bin.}
\label{fig:fig_reco_sta}
\end{figure}

\begin{figure}[!ht]
\begin{center}
\includegraphics[width=0.49\textwidth]{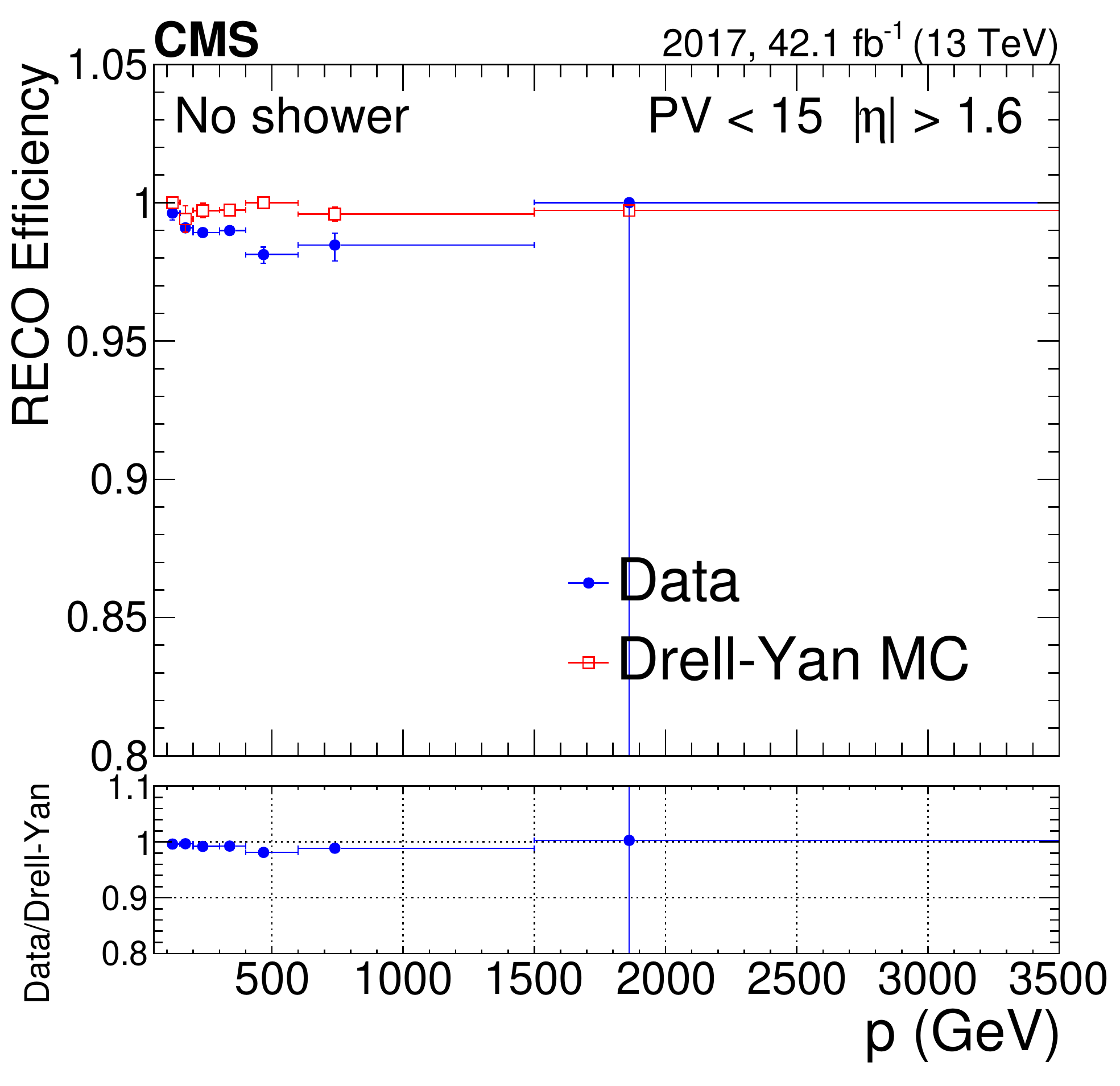}
\includegraphics[width=0.49\textwidth]{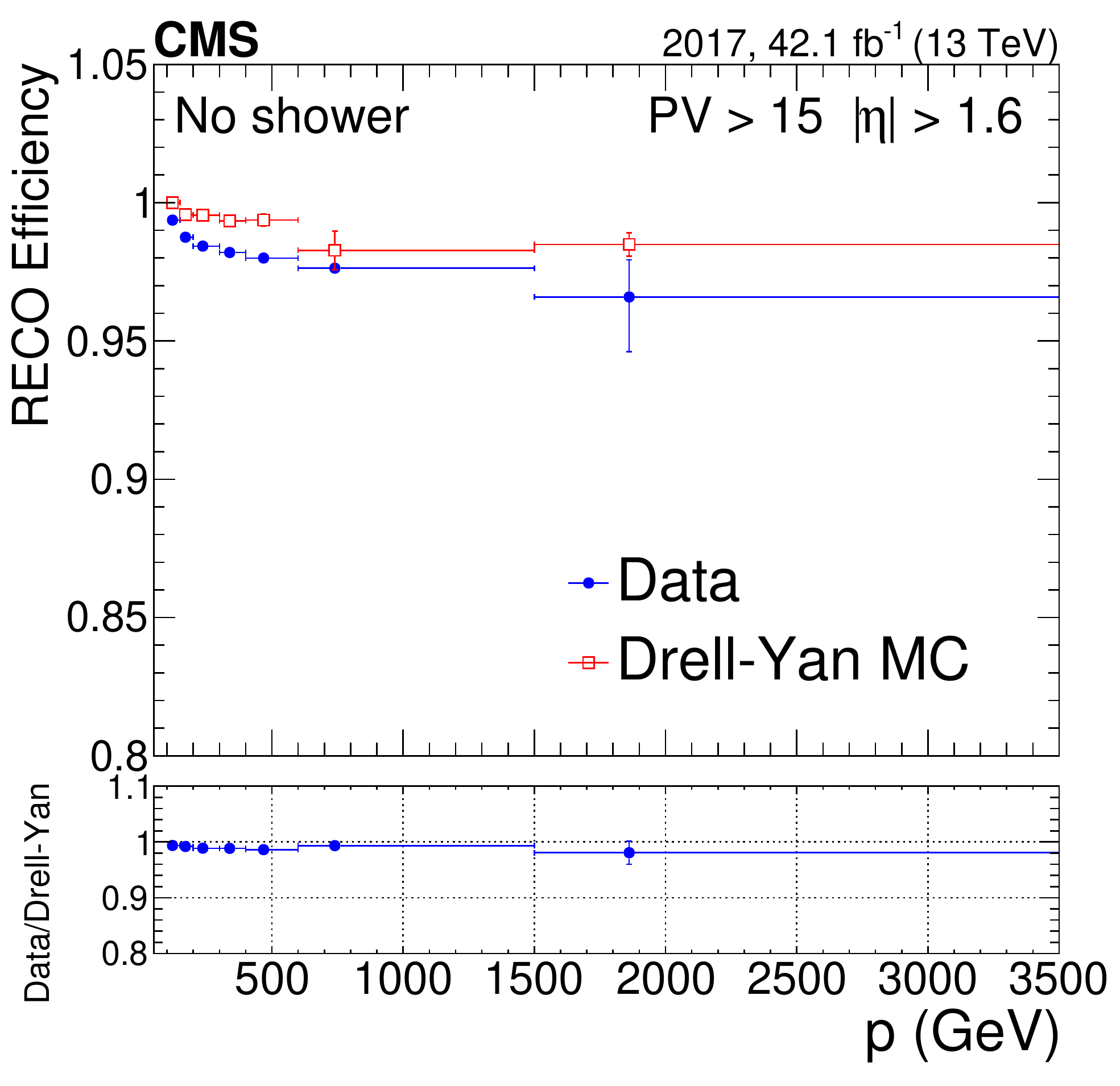}
\includegraphics[width=0.49\textwidth]{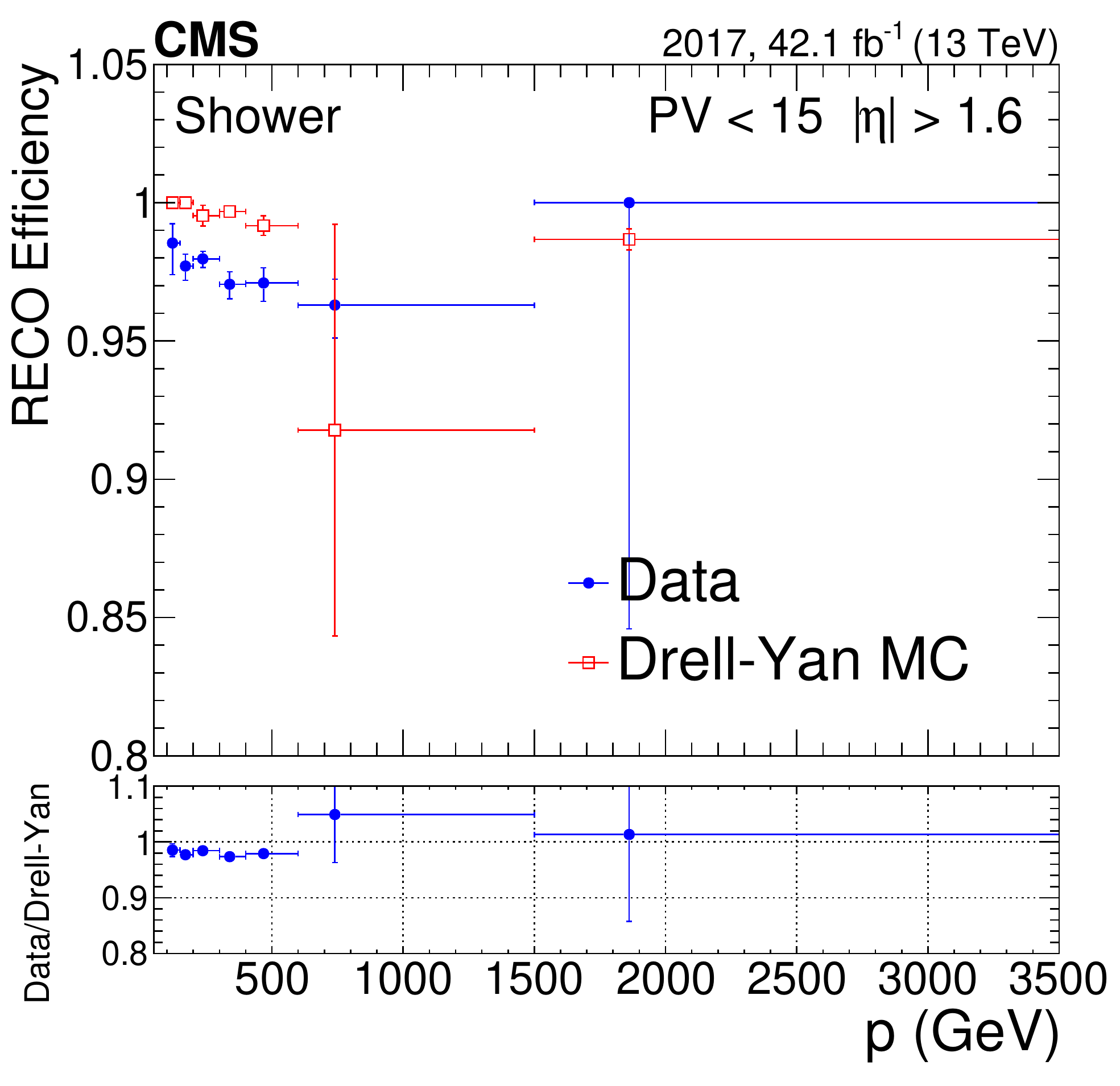}
\includegraphics[width=0.49\textwidth]{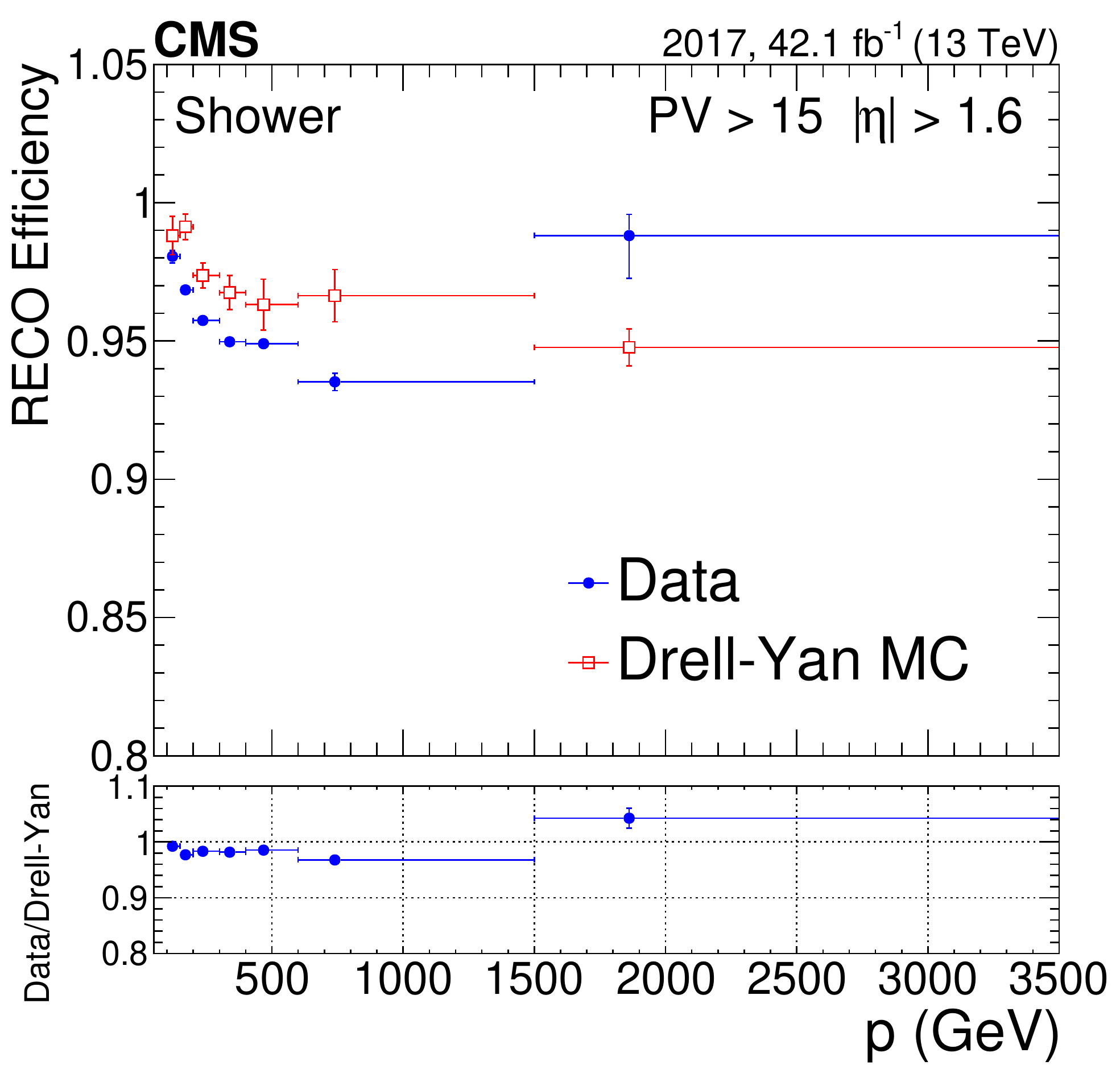}
\end{center}
\caption{Standalone muon reconstruction efficiency as a function of muon $p$ for muons with $1.6 < \abs{\eta} < 2.4$. The left plots are for low pileup (up to 15 vertices) while the right plots are for higher pileup. The upper plots are obtained with events without any showers;
the lower ones contain events with at least one shower. The blue points represent data and the red empty squares represent simulation. The lower panels of the plots show the ratio of data to simulation. The central value in each bin is obtained from the average of the distribution within the bin.}
\label{fig:fig_reco_stabis}
\end{figure}

To characterize the inefficiency seen in the forward part of the detector and in both data sets, Fig.~\ref{fig:fig_reco_sta}
shows the standalone muon reconstruction efficiency as a function of $p$ for $\abs{\eta}<1.6$
and for the forward endcaps $(1.6<\abs{\eta}<2.4)$.
The measured efficiency in the $\abs{\eta}<1.6$ region is uniform in $p$ up to approximately 2\TeV in both data and simulation. In the region $1.6<\abs{\eta}<2.4$, a decreasing trend as a function of $p$ is observed in both data and simulation, although it is more pronounced in data by approximately 2\%.
In order to separate out the possible effect of pileup (in particular since the forward part of the detector suffers from the dense track activity), Fig.~\ref{fig:fig_reco_stabis} compares the standalone reconstruction efficiency obtained in data with DY simulation for events with low pileup environment (defined as having less than 15 reconstructed primary vertices) and for events with higher pileup environment. In addition, since the muons crossing the forward region of the detector have a higher probability to shower (Fig. \ref{fig:showersVsP_dataVsMc_segment}), the results are then further split between events where at least one shower is tagged from events without any showering detected.

For the low-pileup environment and events without tagged showers, the efficiency measured both in simulation and in data is mostly uniform across the momentum spectrum and is almost 100 (99)\% in simulation (data). It starts to show a decreasing trend for higher pileup activity with the efficiency going down to 98 (96)\% for muons with momentum of a few TeV in simulation (data). Although the simulation results show a dependence on the level of pileup, they do not reproduce the data trend when there are more than 15 vertices. When no shower is found, the decreasing trend seen in simulation, and more pronounced in data, is due to pileup.
In the presence of showers, the inefficiency trend is enhanced in both data and simulation, and in particular for events inside the high pileup environment, where the lowest efficiency value is 95 (93)\% for muons of few TeV in simulation (data). The data vs.\ simulation discrepancy is slightly enhanced in the presence of showering for events recorded in both low- and high-pileup environments. We conclude that muon showering and dense track activity interfere within the muon reconstruction, and lead to the momentum dependence of up to 5\% in the inefficiency when both effects are combined.

The scrutiny of DY events from simulation shows that approximately half of the events responsible for the reconstruction inefficiency do have a standalone muon, but it is not associated with its tracker part. Despite the fact that the tracker part and the standalone muon share common segments, the extrapolation of the standalone muon to the tracker volume is not succeeding. Hence the standalone muon and the global muon formed from it (if any) both exist, but the momentum assignment is wrong. The other half of the events are again good tracker muons, with associated muon segments in several muon chambers, but in these cases no standalone muon is reconstructed. Still, several segments are found across the entire muon system (over the 4 stations) and they match the tracker part. This observation indicates a reconstruction failure at the muon system level, namely the inability to reconstruct the standalone trajectory out of the detected segments.

\begin{figure}[ht]
\begin{center}
\includegraphics[width=0.48\textwidth]{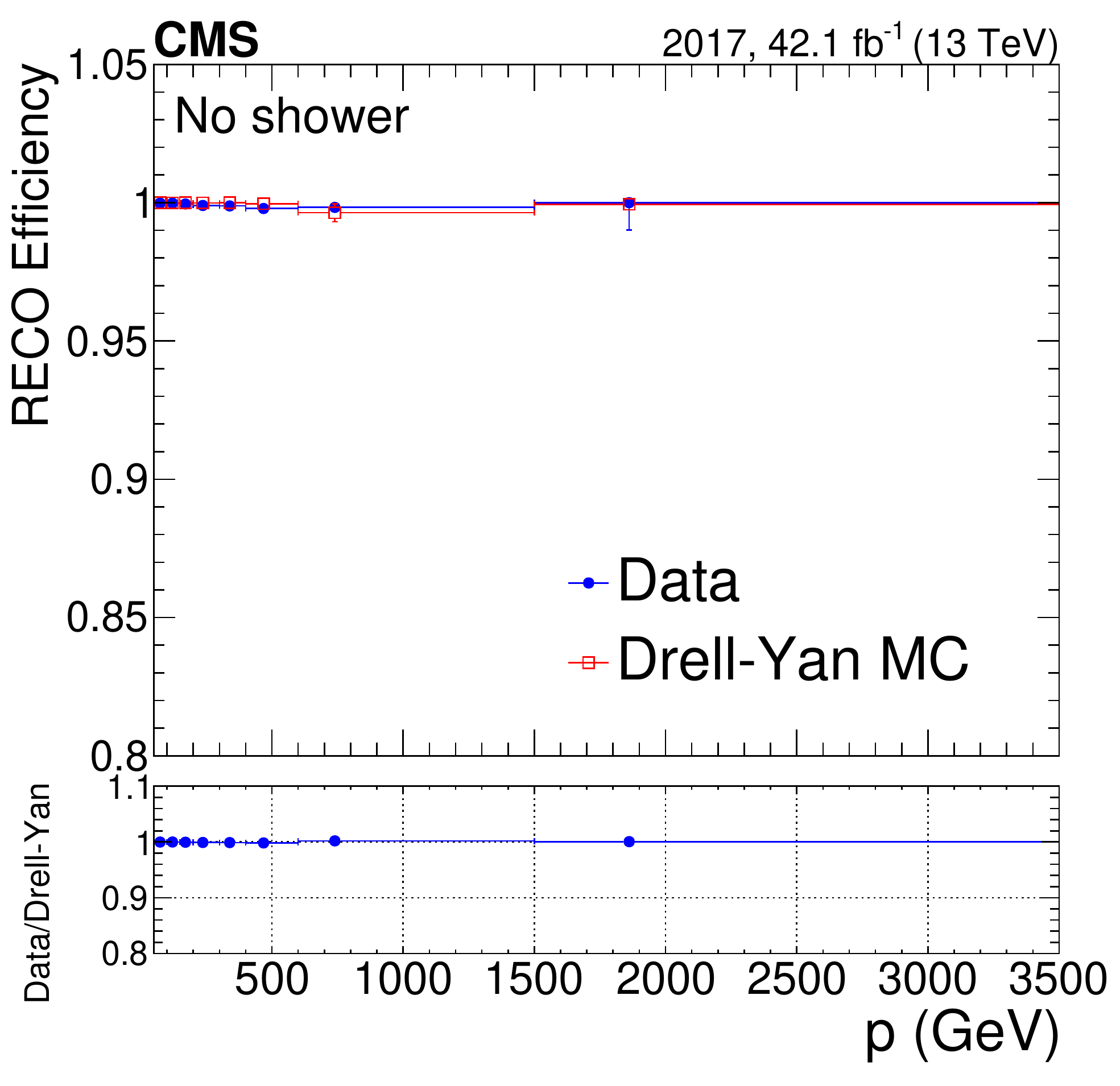}
\includegraphics[width=0.48\textwidth]{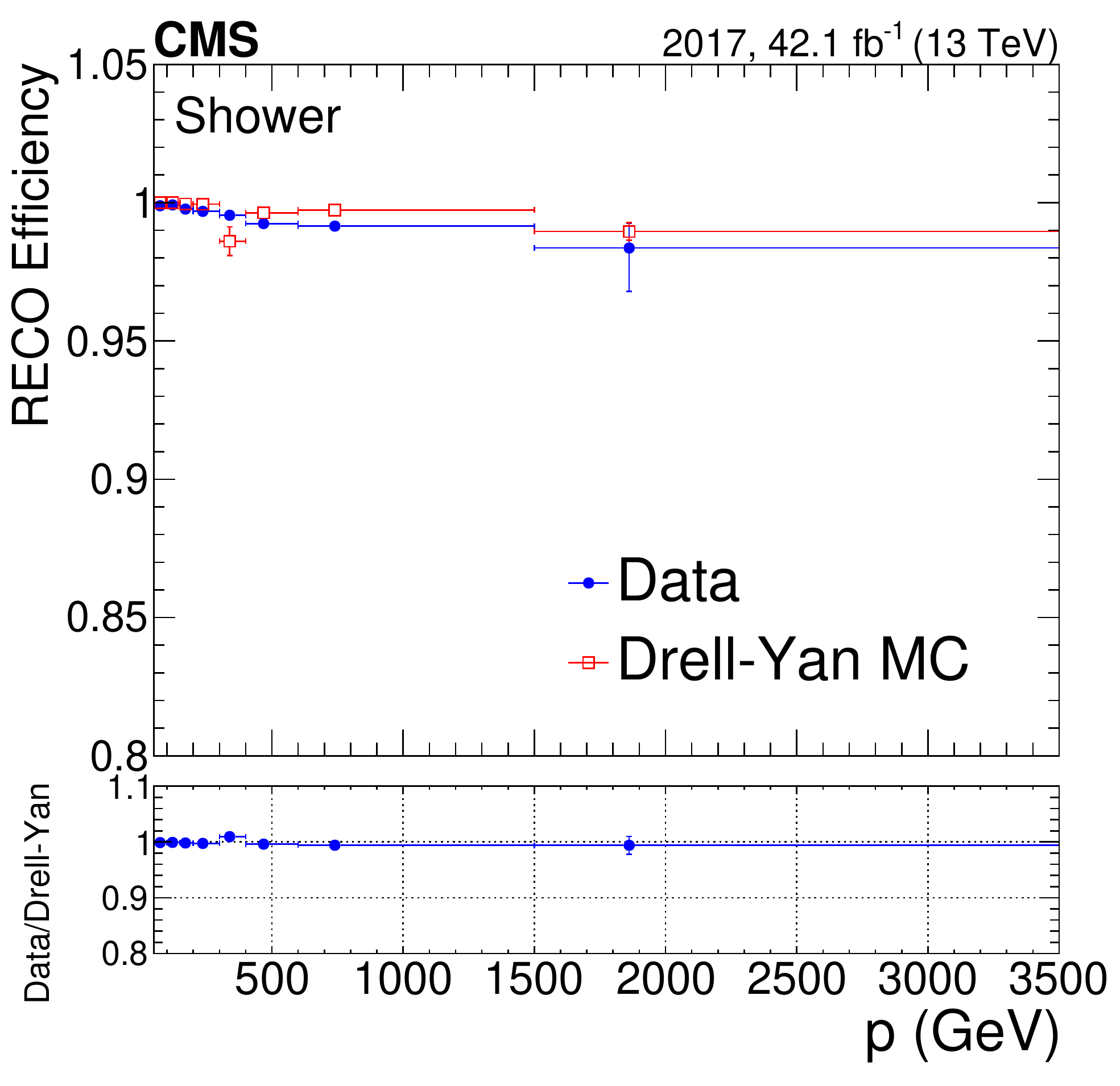}
\end{center}
\caption{Global muon reconstruction efficiency as a function of muon momentum. The left plot is obtained with events without any showers, while the right one contains events with at least one shower. The blue points represent data and the red empty squares represent simulation. The lower panels of the plots show the ratio of data to simulation. The central value in each bin is obtained from the average of the distribution within the bin.}
\label{fig:fig_reco_global}
\end{figure}

The global reconstruction efficiency is computed for probe muons that are also standalone muons and is displayed as a function of the muon momentum in Fig.~\ref{fig:fig_reco_global}. The results are integrated over muon $\eta$ but split according to the (left) absence or (right) presence of showers. The efficiency is almost 100\% over the full momentum spectrum when the events do not contain showering muons. A slight decreasing trend is observed in the presence of muon showering, although the global reconstruction efficiency remains greater than 99\%.

\subsection{Combined L1 and HLT efficiency}

\label{sec:hltTriggerEff}

  \begin{figure}[ht]
  \centering
  {\includegraphics[width=0.49\textwidth]{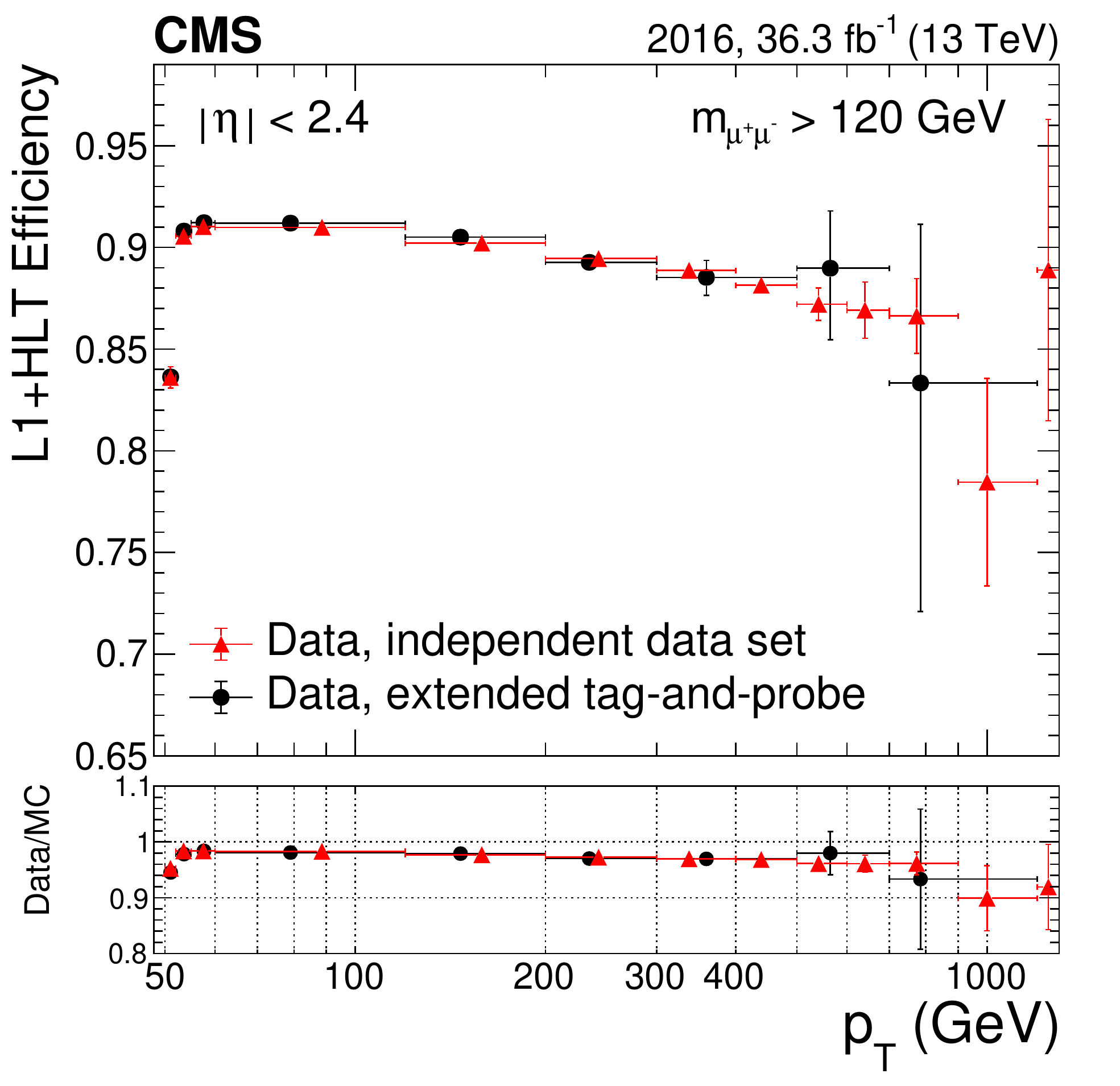}}
  {\includegraphics[width=0.49\textwidth]{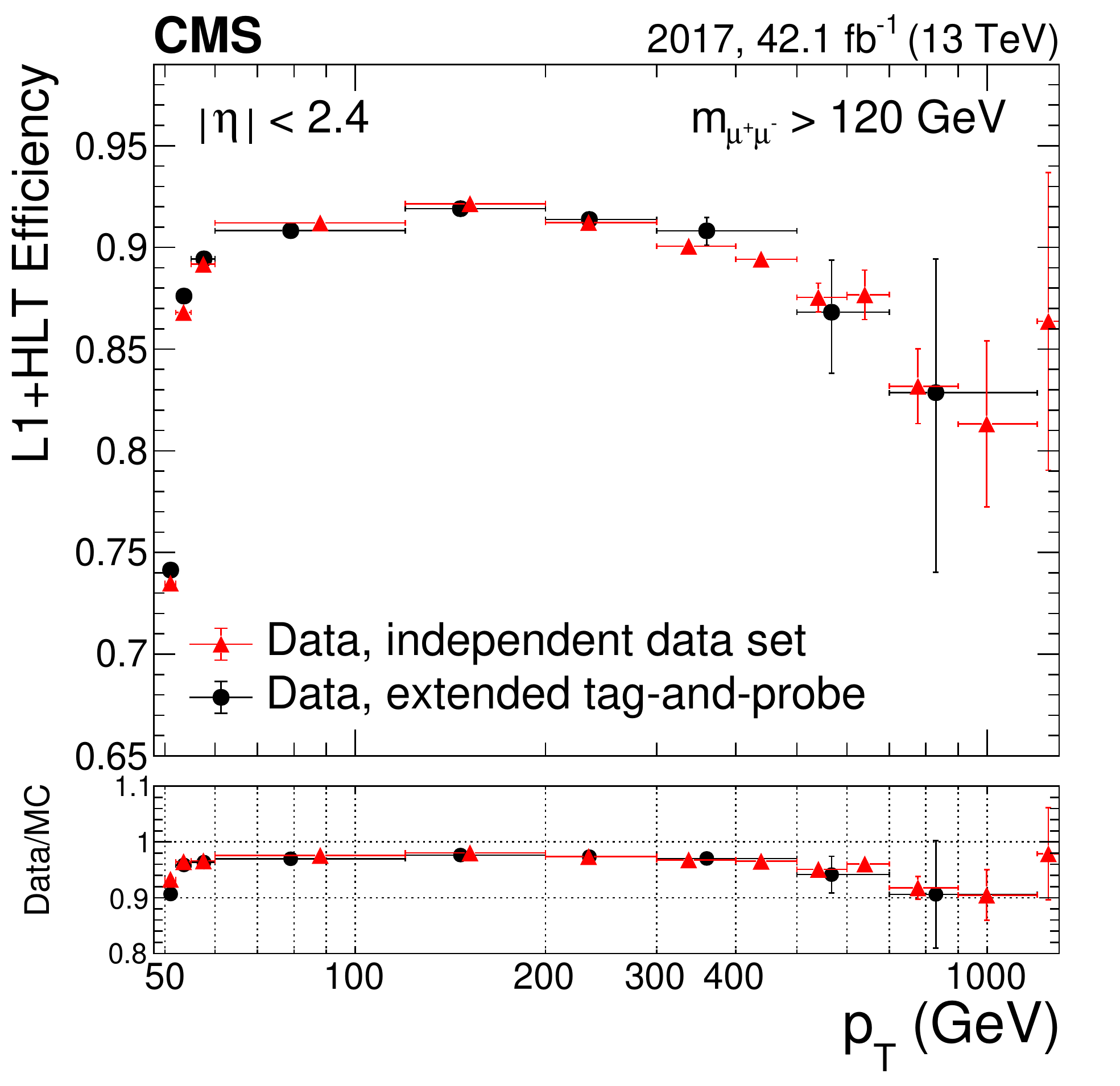}}
  \caption{The combined HLT+L1 efficiency with respect to the offline selection, and the ratio of data to simulation for different methods, as functions of \pt, for (left) 2016 data and (right) 2017 data. The red triangles are measured using an independent data set collected with a \ptmiss trigger; the black circles are measured by the extended \TnP method in which selected events have $m_{\mu\mu} > 120\GeV$. The central value in each bin is obtained from the average of the distribution within the bin.}
  \label{fig:FinalComp}
  \end{figure}

The overall trigger efficiency (combined L1 and HLT) is measured using the extended \TnP method, as well as using events selected
by a set of triggers without muon requirements. The events selected in these independent data sets contain a high-energy electron or large \ptmiss. This second approach leads to a sample enriched in \PW+jets and $\ttbar$ events that could be used to probe muon triggers.

Figure~\ref{fig:FinalComp} shows the trigger efficiency measurement using the extended \TnP (black), and independent data set (red) methods as a function of the muon \pt for 2016 and 2017 data. The two methods are compatible with each other, reinforcing the robustness of the results.
The measured trigger efficiency in 2016 and 2017 data shows a slight decreasing trend as a function of the muon \pt with a value of 90 (85)\% at 60\GeV (1\TeV). The SF between the trigger efficiencies in data and simulation ranges between 0.95 and 0.9.

\begin{figure}[hpt!]
  \centering
  {\includegraphics[width=0.49\textwidth]{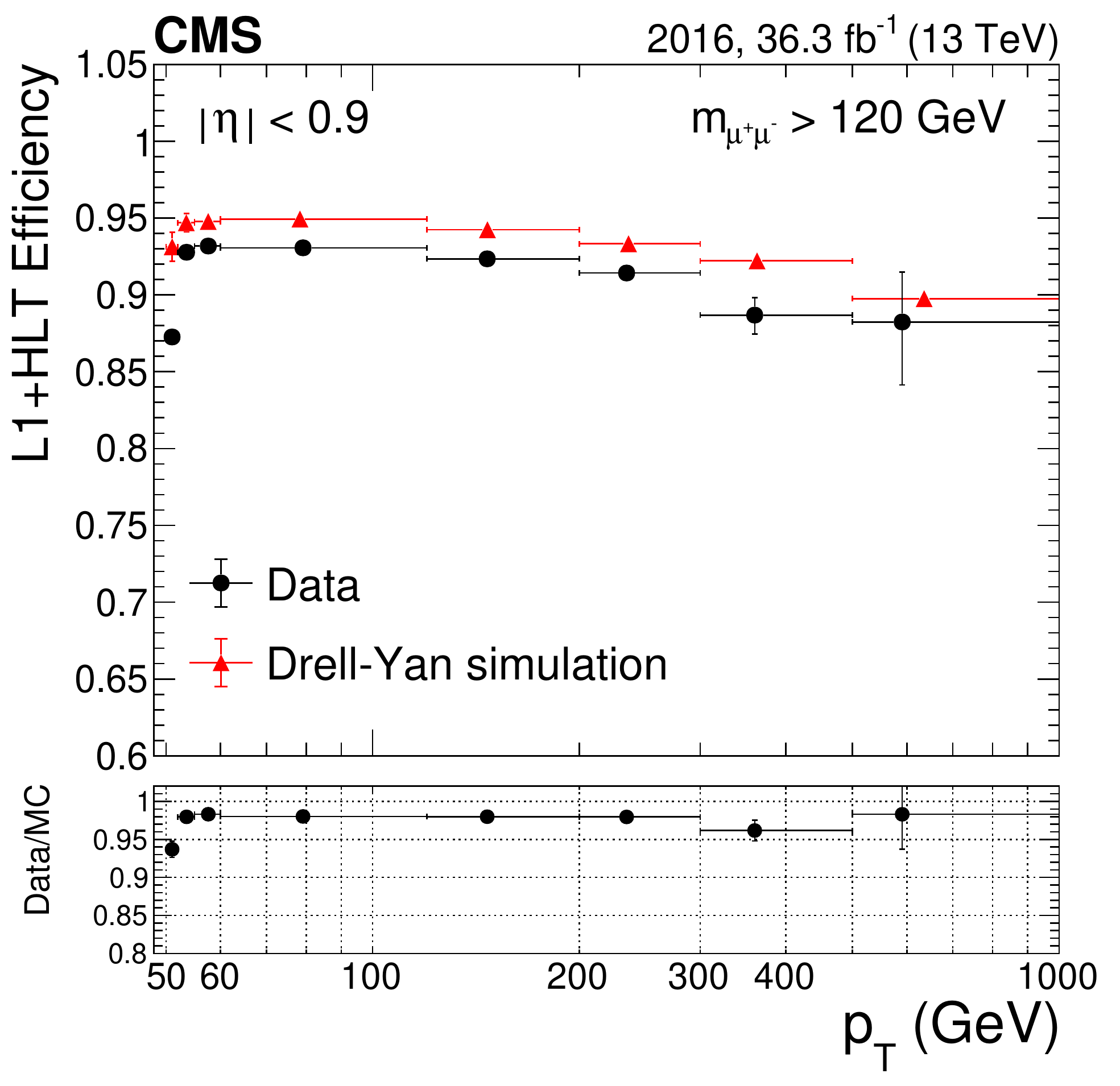}}
  {\includegraphics[width=0.49\textwidth]{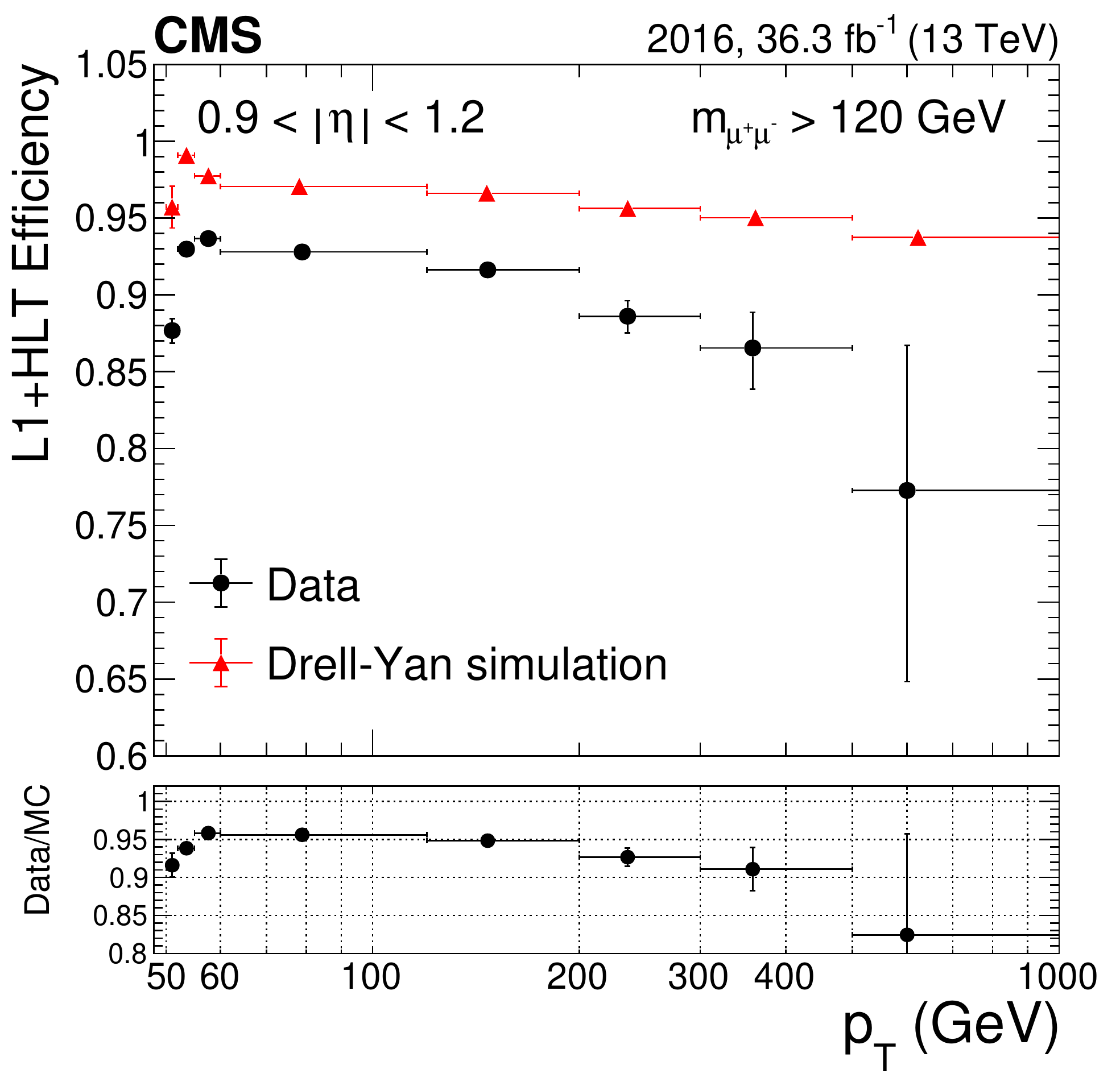}}
  {\includegraphics[width=0.49\textwidth]{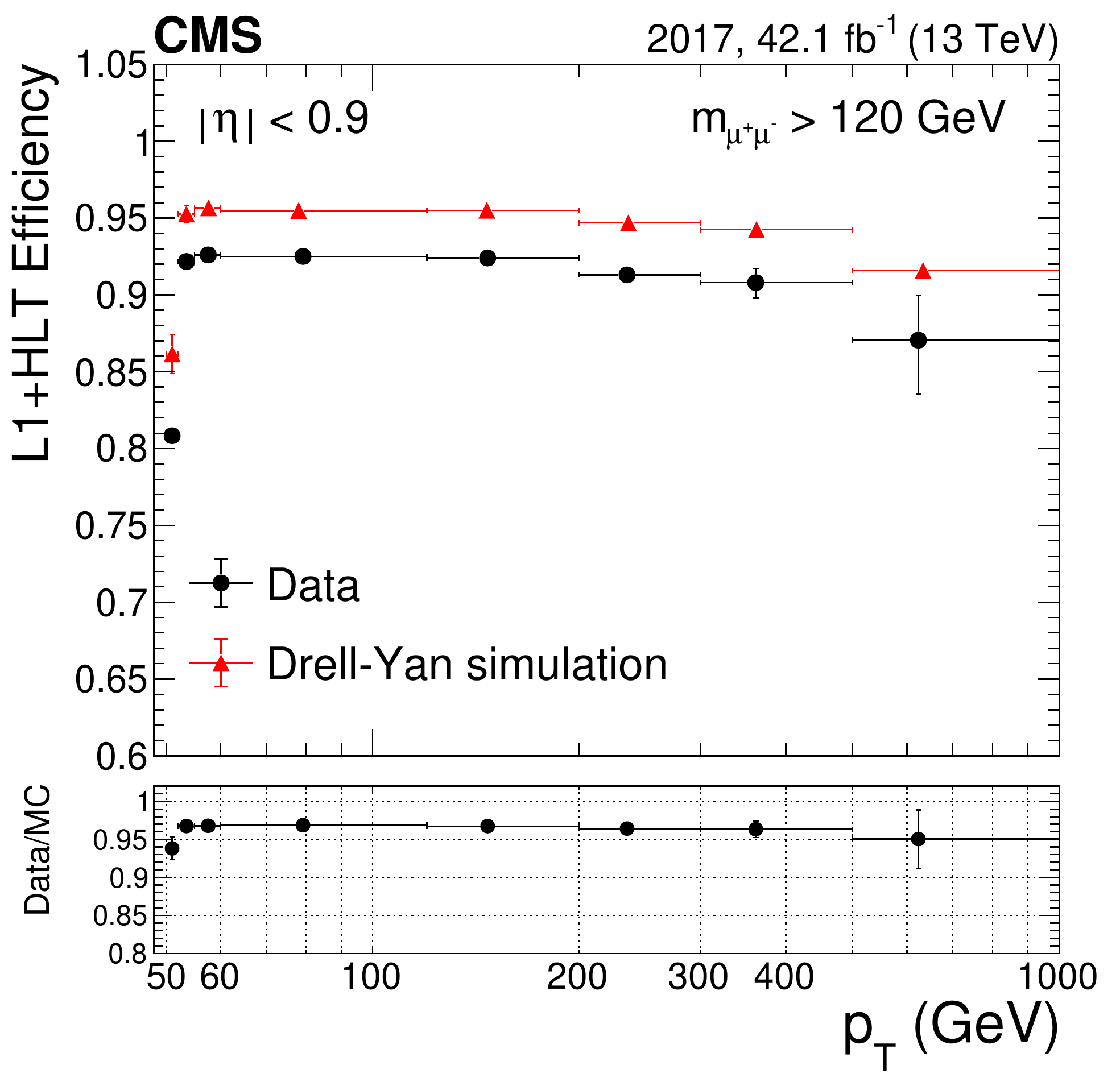}}
  {\includegraphics[width=0.49\textwidth]{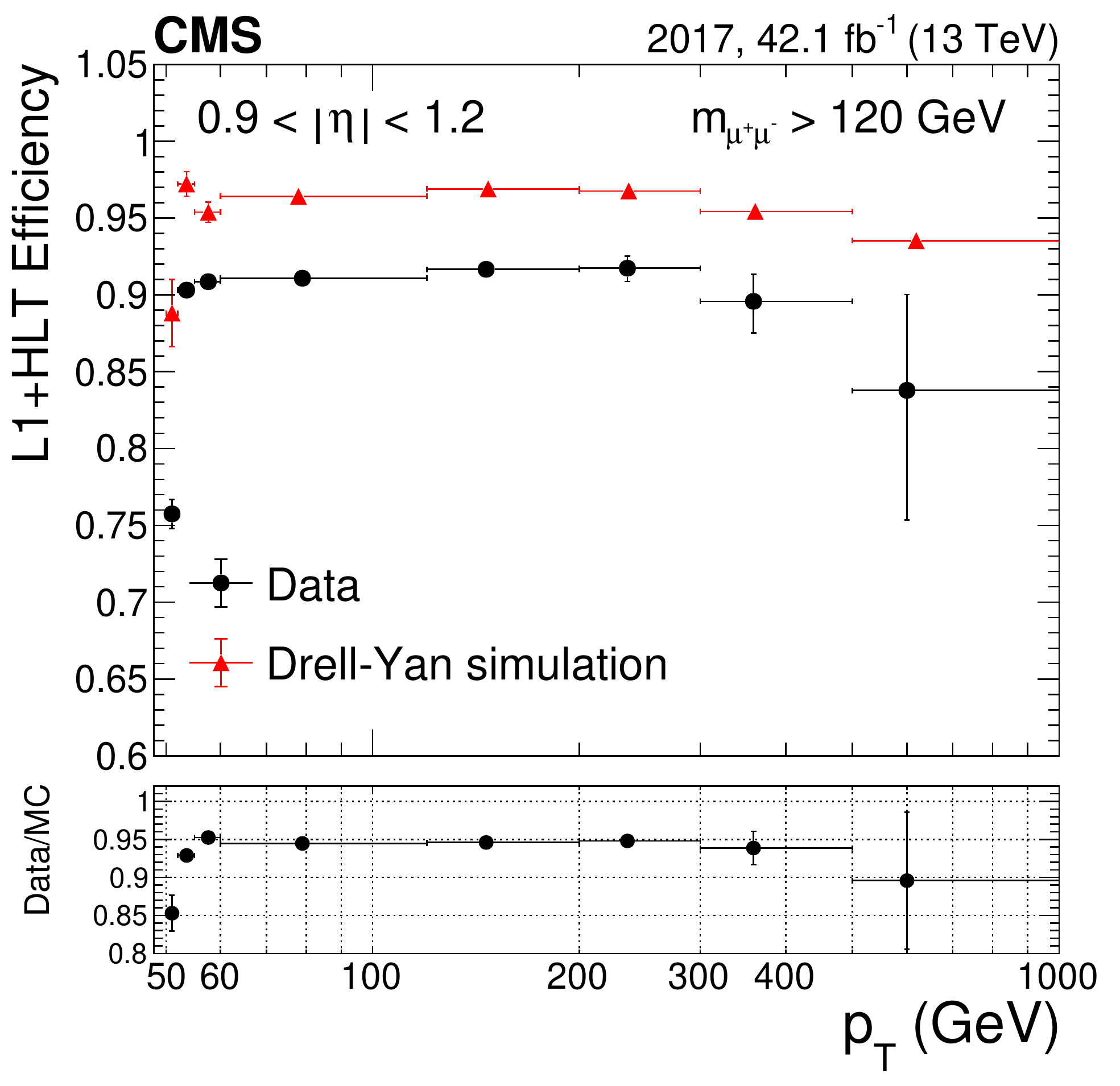}}
  \caption{The combined HLT+L1 efficiency with respect to the offline selection, and the ratio of data to simulation, as a function of \pt, for (upper) 2016 data and (lower) 2017 data and simulation. The left plots are for the barrel region and the right plots are for the overlap region. The red triangles represent the simulation while the black dots are the data. The lower panels display the ratio of efficiencies in data and simulation. The central value in each bin is obtained from the average of the distribution within the bin.}
  \label{fig:pTTrend}
  \end{figure}

The 2016 and 2017 trigger efficiencies obtained with the extended \TnP method are computed separately for the barrel and overlap regions, and compared to simulation in Fig.~\ref{fig:pTTrend}. In both data sets, the efficiency trend as a function of \pt is seen in the barrel but even more pronounced in the overlap region. In the barrel, the ratio of data to simulation is 0.98 (0.97) for 2016 (2017) data and is uniform with \pt in both data sets. The residual efficiency dependence of the results is caused by the L1 component, due to the lower efficiency of the L1 muon trigger for muons with shower tags, as discussed in Section~\ref{sec:l1trigger}. In the overlap region, the inefficiency trend is much more severe in data than in simulation, and the SF are increasing with \pt. They range from 0.95 at 60\GeV and down to 0.85\GeV in the highest bin in 2016 data (and 0.9 in 2017 data). Hence, though the efficiency trend is visible in both the barrel and overlap regions, the \pt dependence of the SF is coming exclusively from the overlap region. This effect has been tracked down to the L1 trigger and the causes are attributed to a nonoptimal arbitration between the DT and CSC segments that are both present in the overlap region. Even though the muon identification relies equally on CSC and on DT segments, the momentum assignment will be more accurate if the estimated value comes from the DT. A fix was implemented in 2018 so that the DT estimated muon assignment is used in these cases.

\subsection{The L1 trigger efficiency}
\label{sec:l1trigger}

The L1 component of the overall muon trigger efficiency at high \pt is
parameterized separately for the two cases when an associated shower is, or is not, tagged.
From Fig.~\ref{fig:FinalComp}, it can be seen that above the initial turn-on
curve, the trigger efficiency is mostly uniform, but appears to be slowly deteriorating as the muon momentum increases.
It is important to quantify the size of this effect from L1, because it can impact all
high-\pt physics measurements.

The approach used here relies on
assuming that the inefficiency appearing at high \pt is due to showering in the muon detectors,
and that the momentum dependence arises because the probability of showering in a station increases with increasing momentum.
That is, the efficiency under study can be
parameterized as a function of the number of showers $\epsilon (\Nshower)$,
which should be independent of the momentum.

The validity of the shower-based approach is verified by studying the L1 muon efficiency as a function of the number of showers
for different muon momentum slices. We observe
that, within a momentum slice,
the trigger efficiency does correlate with the number of showers.
Furthermore, the dependence on \Nshower is the same or similar across the compared $p$ ranges.

The shower probability shown in Fig.~\ref{fig:showersVsP_dataVsMc_segment} is parameterized as a function of
muon momentum as $P_ \mathrm{shower}(p)$. The parameterization is performed by fitting the distribution in the region up to 1\TeV,
separately for the data and for the DY simulation, with a linear function. The upper end of the range is dictated by the lack
of a sufficient number of muons in data above $p\approx 1\TeV$.

The L1 efficiency can thus be calculated as a function of $p$ according to:
\begin{equation}
\epsilon_{\mathrm{L1}} (p) = \sum\limits_{\Nshower=0}^4 \epsilon(\Nshower) P_{\mathrm{Nshower}}(p),
\end{equation}
where $P_{\mathrm{Nshower}}(p)$ is the probability for a muon of momentum $p$ to produce the number
of showers given by $\Nshower$, which can be calculated from $P_{\mathrm{shower}}(p)$ using standard
combinatorial formulas. The  maximum number of showers is 4 since there
are 4 muon stations.

We extract $\epsilon(\Nshower)$ from simulated DY events and from
the 2016 and 2017 data sets recorded with the \ptmiss trigger.
An event selection is applied to remove cosmic ray muons from the data
and to select only well-reconstructed isolated muons passing the high-\pt identification criteria.
Regions in the barrel ($\abs{\eta}<0.9$) and endcap ($\abs{\eta}>1.2$) are analyzed separately.
The overlap region where muons can have hits in both DT and CSC is not considered in this study.

For each muon reconstructed offline, the L1 muon candidates close to the extrapolated muon
trajectory are stored. The candidate with the highest \pt and in time
with the collision
is taken as the L1 candidate assigned to this muon.
The L1 efficiency for the muon is defined based on whether
an L1 candidate with \pt above the L1 threshold (22\GeV) is found or not.

The final efficiency measurement is extracted from a combination
of 2016 and 2017 data sets, which maximizes the sample size.
The resulting L1 efficiency for muons with different numbers of showers
is shown in Table~\ref{tab:l1eff_vs_showers}.

\begin{figure}[ht]
  \begin{center}
    \includegraphics[width=.49\textwidth]{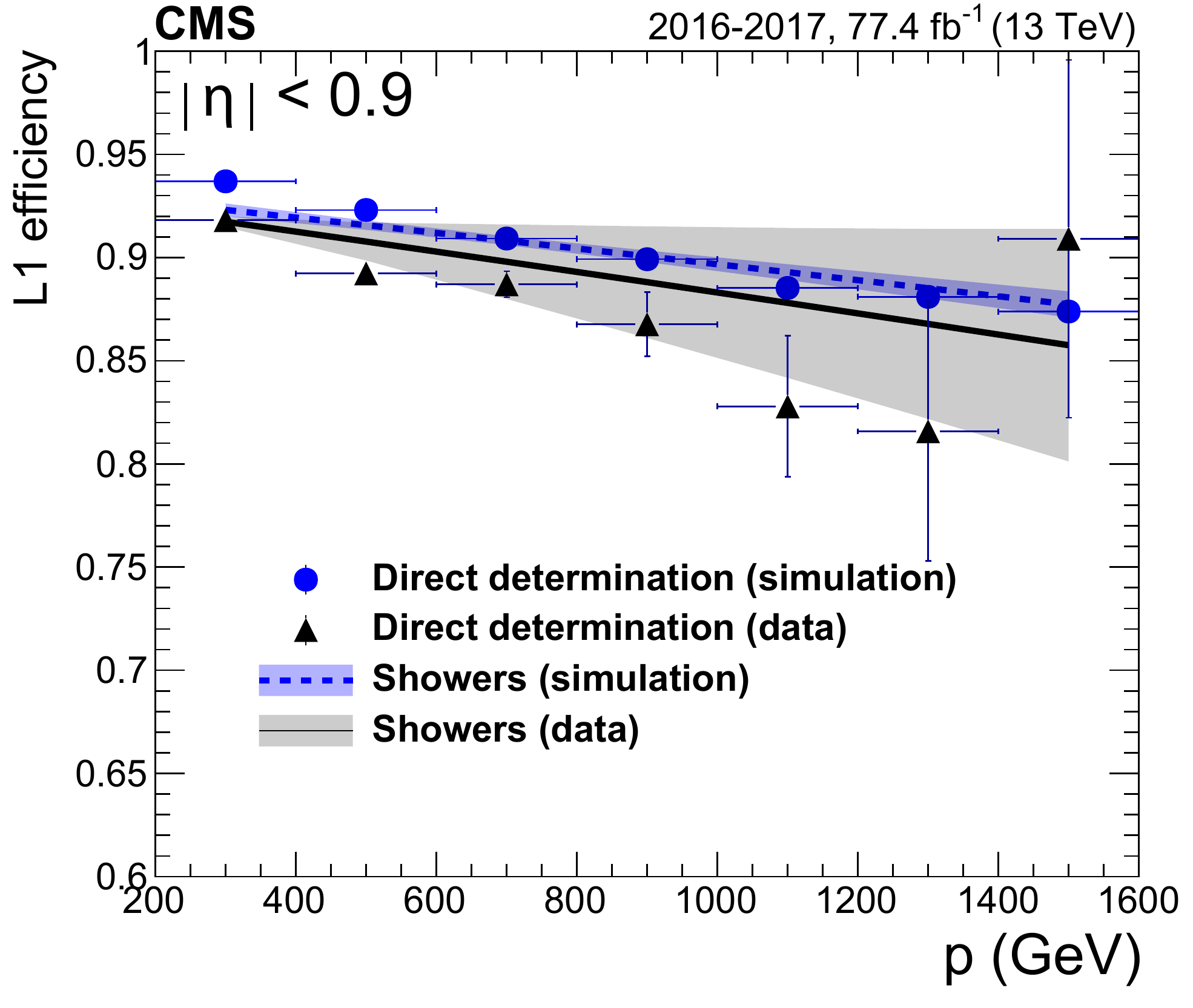}
    \includegraphics[width=.49\textwidth]{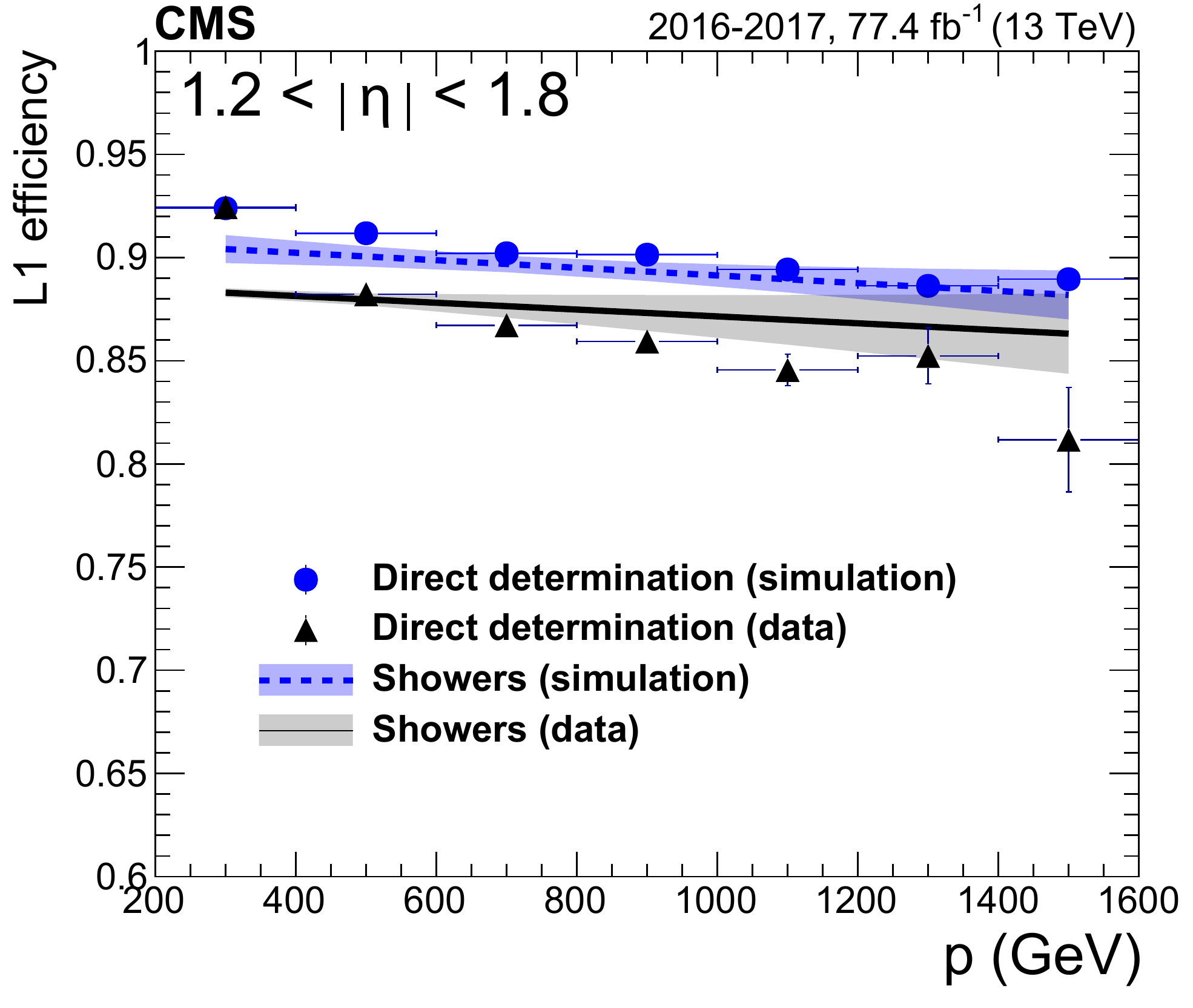}
    \includegraphics[width=.49\textwidth]{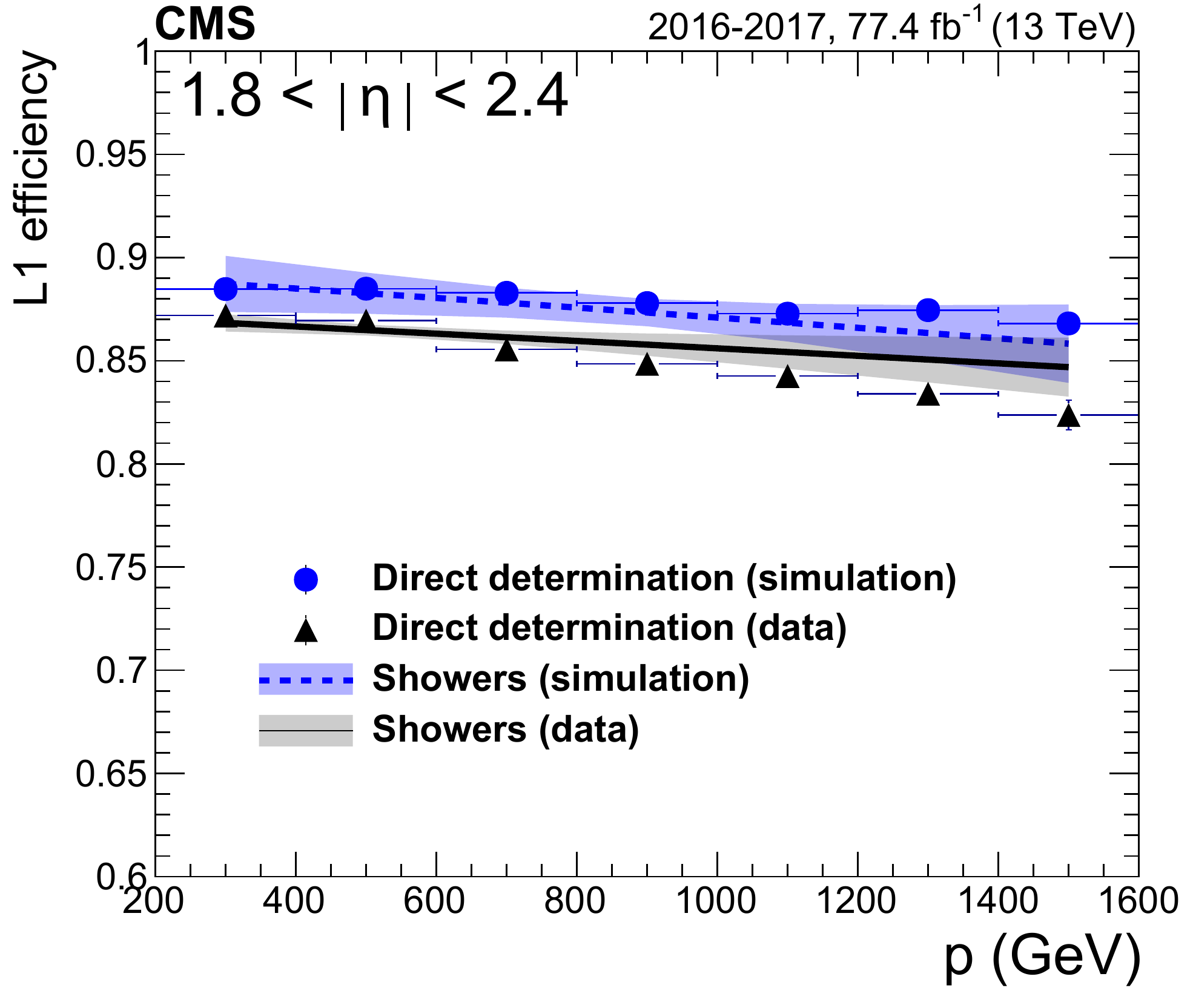}
  \end{center}
  \caption{The L1 efficiency in three $\eta$ regions:
  (upper left) barrel; (upper right) for muon with $1.2 < \abs{\eta} < 1.8$; and (lower) endcap with muon $\abs{\eta} > 1.8$.
  The plots show a comparison between directly determining the
  efficiency from simulation (blue dots) and with data (black
  triangles) with respect to calculating it from shower multiplicity,
  both in 2016+2017 combined data (black line) and 2017 simulation (dashed blue line).
  The shaded bands include the statistical uncertainties of the
  measurements and the systematic uncertainty of the showering probability determination.}
  \label{fig:l1eff}
\end{figure}

\begin{table}[ht]
\renewcommand{\arraystretch}{1.3}
\caption{The L1 trigger efficiency for barrel and endcap muons measured as a function of the number of showers in the muon stations.
The endcap was split into near ($\abs{\eta}<1.8$) and far ($\abs{\eta}>1.8$) sections.}
\label{tab:l1eff_vs_showers}
\centering

\begin{tabular}{ccccccc}
\Nshower & \multicolumn{2}{c}{Barrel L1 efficiency}
& \multicolumn{2}{c}{Near endcap L1 efficiency} & \multicolumn{2}{c}{Far endcap L1 efficiency} \\ \hline
& Data & Simulation & Data & Simulation & Data & Simulation\\
0 & 93.9$\pm$0.1\%& 94.4$\pm$0.1\%& 89.3$\pm$0.1\%& 91.4$\pm$0.1\%& 88.3$\pm$0.1\%& 90.0$\pm$0.1\%\\
1 & 82.2$\pm$0.1\%& 82.7$\pm$0.1\%& 84.9$\pm$0.2\%& 87.2$\pm$0.1\%& 83.5$\pm$0.1\%& 85.9$\pm$0.1\%\\
2 & 67.1$\pm$0.7\%& 67.3$\pm$0.3\%& 78.9$\pm$0.5\%& 81.9$\pm$0.3\%& 77.0$\pm$0.2\%& 79.8$\pm$0.3\%\\
3 & 49.8$\pm$3.4\%& 50.1$\pm$1.4\%& 76.9$\pm$2.0\%& 76.0$\pm$1.0\%& 70.2$\pm$0.8\%& 72.7$\pm$0.9\%\\
4 & 40$\pm$15\%& 36$\pm$9\%& 68$\pm$11\%& 75$\pm$5\%& 63$\pm$3\%& 60$\pm$4\%\\
\end{tabular}
\end{table}

These numbers are combined with the parameterization of the number of showers versus $p$, as described above, yielding
the L1 efficiency, as a function of $p$, shown in Fig.~\ref{fig:l1eff}. The results shown as a black line in the plot
were derived using the shower-based approach described above, taking
both the shower probability and the L1 efficiency from data. The shaded
bands represent the statistical and systematic uncertainties of the
shower probability determination. They are dominated by the small
number of events at high momentum, particularly in the barrel region (cf.\ Fig.~\ref{fig:showersVsP_dataVsMc_segment}).
The efficiency calculated directly from the data events is shown as black points. The two methods give comparable results, indicating that
the presence of showers contributes to the L1 inefficiency at high momentum.
The L1 efficiency measured in the simulated DY sample is shown for comparison, as blue points and lines. The two methods agree well, with
a decreasing efficiency trend similar to that observed in data.

\section{Momentum assignment performance} \label{sec:Momentum}

At low and intermediate momenta, below 100\GeV, the muon \pt resolution is dominated by the hits measured in
the silicon tracker. In contrast, hits and segments measured in the muon chambers are significantly affected by multiple scattering
of the muon trajectory while passing through the calorimeters and the flux-return yoke. This multiple scattering is reduced with increasing momentum, and
above 200\GeV the muon chamber measurements start to improve on the measured \pt.
The ultimate performance at high \pt is then determined by the
precision of the muon chamber measurements and by the alignment of muon
chambers relative to each other and to the inner tracker.

The alignment of the silicon tracker is a challenging task, achieving
a statistical accuracy better than 10\mum on the position of
individual detector modules \cite{trkalign-craft,trkalign-pp}.
To these small remaining alignment uncertainties one has to add the
intrinsic resolution of silicon hits (typical 10--30\mum). The intrinsic precision of the muon DT
chambers in the barrel region and the CSC in the endcap region, is of the order of 100--200\mum,
to which the possible chamber misalignment is added in
quadrature \cite{Chatrchyan:2013sba}.

Until 2015, the CMS muon reconstruction neglected the alignment uncertainties of the muon chambers,
referred to as alignment position errors (APE).
Resultant shortcomings were evident, as observed deviations were larger than expected uncertainties for the muon segment parameters
with respect to the extrapolated track from the inner tracker.
The best possible reconstruction for a high-\pt muon track can be reached by a correct relative weighting
of tracker and muon detector hits. In the high-momentum regime this balance requires including appropriate muon alignment uncertainties
in the Kalman filter.
From the beginning of Run 2, the muon reconstruction has been using nonzero muon APEs~\cite{DPS-POG}.
Muon APEs have been introduced for local reconstructed segments in each station for all six segment degrees of freedom (three local positions $x$, $y$, and $z$; and three local angles
$\phi_x$, $\phi_y$, and $\phi_z$), chamber-by-chamber, for both DT and CSC chambers;
they are taken as uncorrelated, as a first approximation.

The muon momentum resolution and the closely related charge
assignment are studied in detail using simulation in Section \ref{sec:Simreso}.
These studies span the entire momentum spectrum with high precision and provide estimates
of the impact of different detector alignment conditions, with and without
the APEs. The performance of the momentum resolution and scale measurements are then
assessed in data from both cosmic ray muons and collisions, and
compared to simulation, in Sections \ref{sec:Datareso} and \ref{sec:DataScale}, respectively.

\subsection{Momentum performance in simulation} \label{sec:Simreso}

\begin{figure}[ht]
\includegraphics[width =3in]{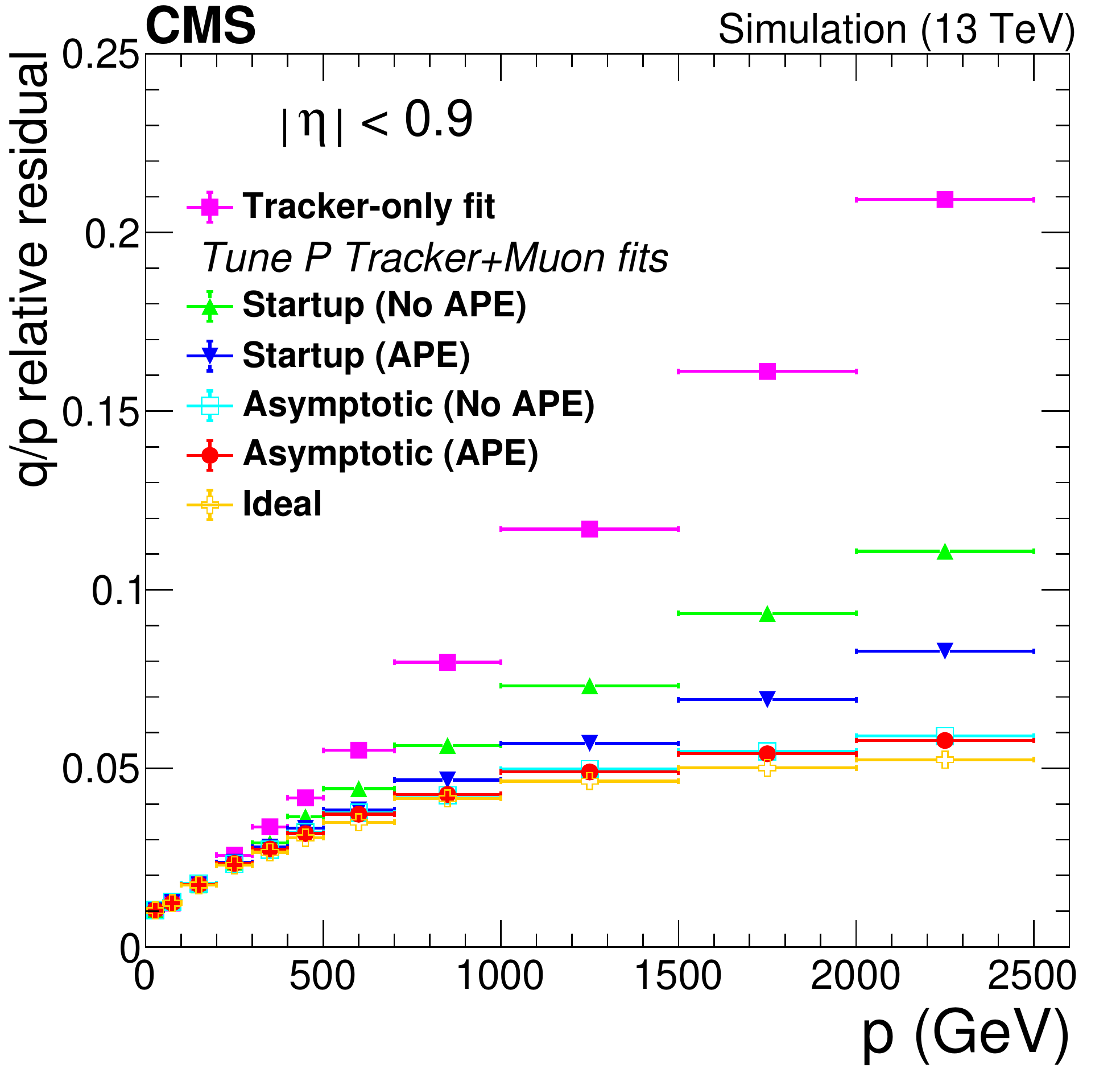}
\includegraphics[width=3in]{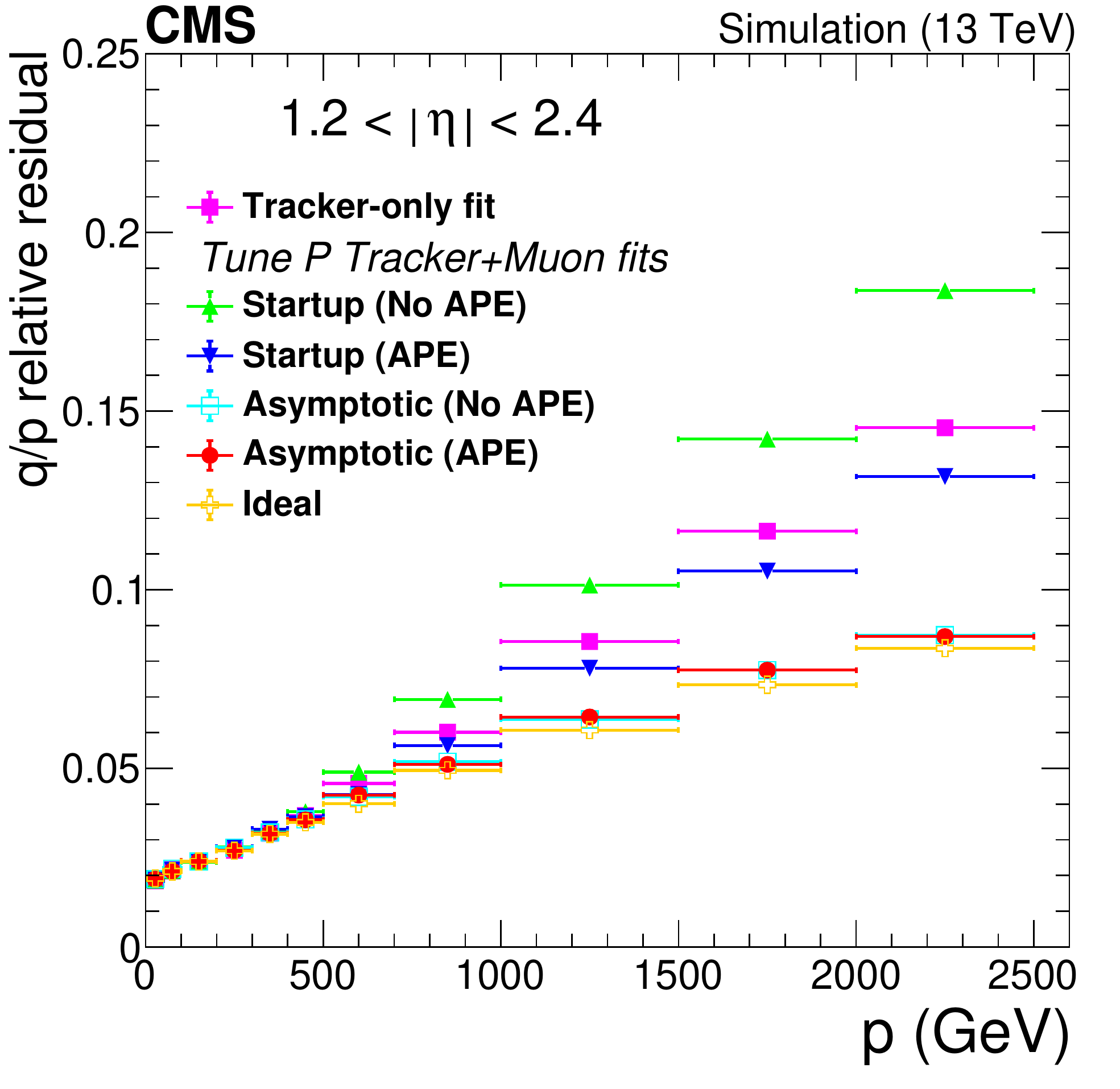}
\caption{Muon momentum resolution (standard deviation $\sigma$
of the fit of the core distributions to a Gaussian function)
in (left) the barrel region $\abs{\eta}<0.9$ and (right) the endcap region $1.2<\abs{\eta}<2.4$,
for the \TuneP algorithm, as a function of muon momentum, for the various misalignment scenarios with and without APEs.
A comparison with the ideal scenario is also given. The performance of the tracker-only fit is shown for comparison.
}
\label{fig_tunePsigmaRelRes}
\end{figure}

The momentum resolution of highly energetic muons can be measured in
simulated events, where the true muon momentum is known. The resolution can be
extracted from the distribution of the relative residual in $q/p$:
\begin{equation}
\residgen =
\frac{(q/p)^{\mathrm{reco}} -
  (q/p)^{\mathrm{gen}}}{(q/p)^{\mathrm{gen}}},
 \label{MCTruth}
\end{equation}
where $q/p$ is the charge sign divided by the momentum of the muon. The expectations for various alignment scenarios have been tested in
simulation on back-to-back dimuons with distributions uniform in
$\eta$, $\phi$, and $p$, within the range $5\GeV < p < 2.5\TeV$. For smaller intervals of momentum within that range, the standard deviation $\sigma$ of a fit to a Gaussian function of the distribution for the \TuneP algorithm is shown in Fig.~\ref{fig_tunePsigmaRelRes}, as a function of the momentum.
The performance of the tracker-only fit is also given for comparison.

\begin{figure}[b]
\begin{center}
\includegraphics[width =3in]{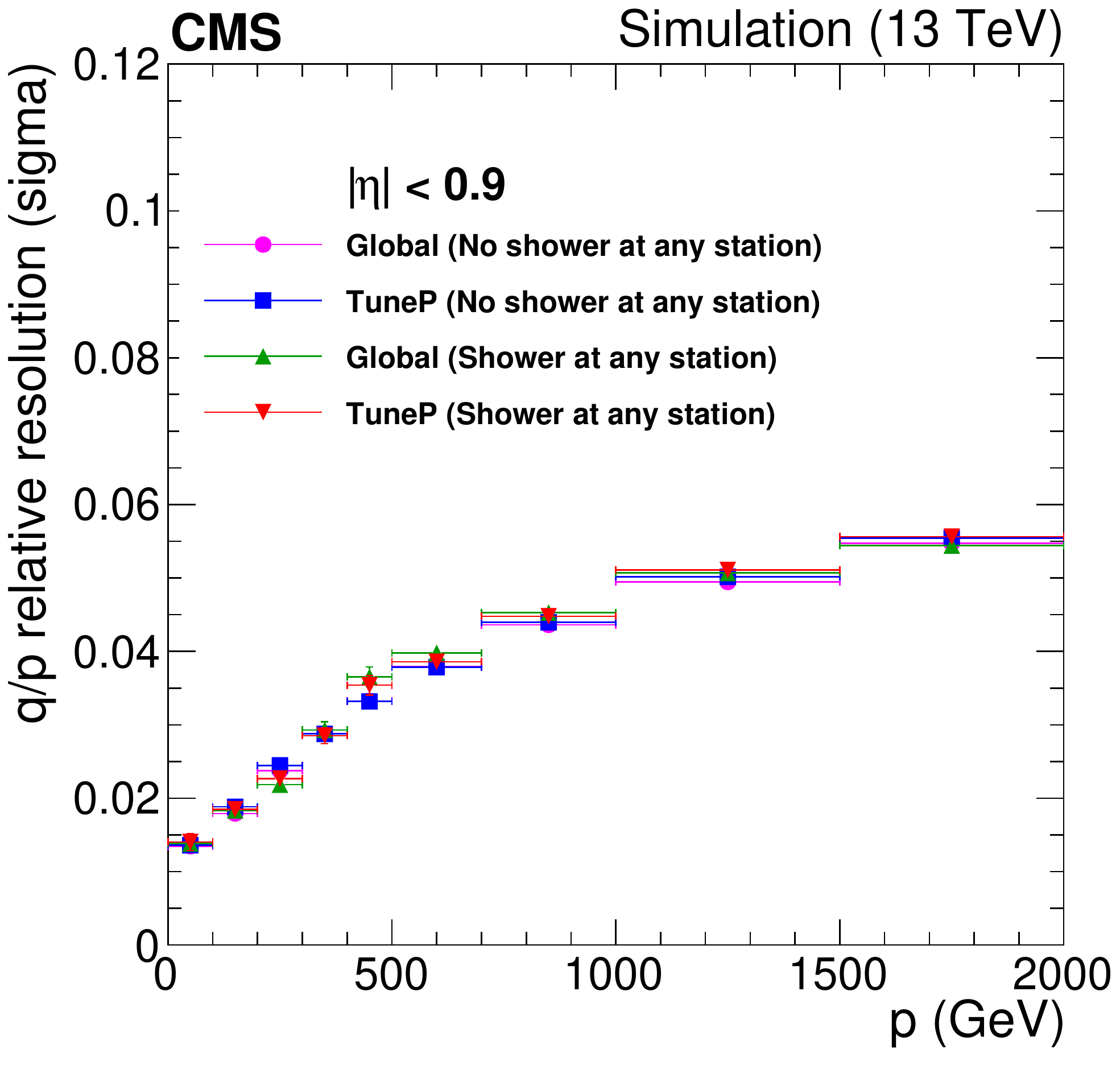}
\includegraphics[width =3in]{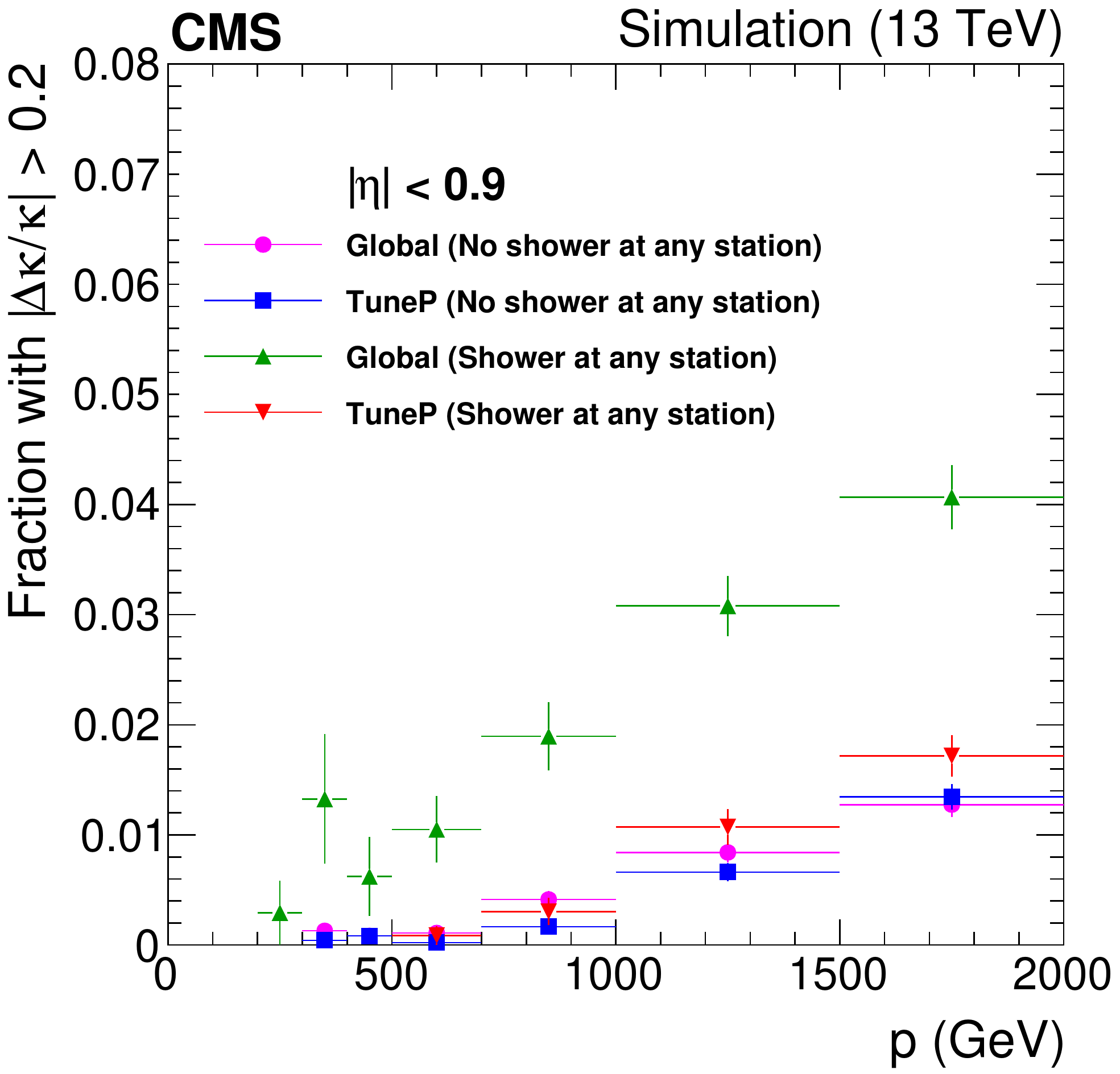}
\includegraphics[width =3in]{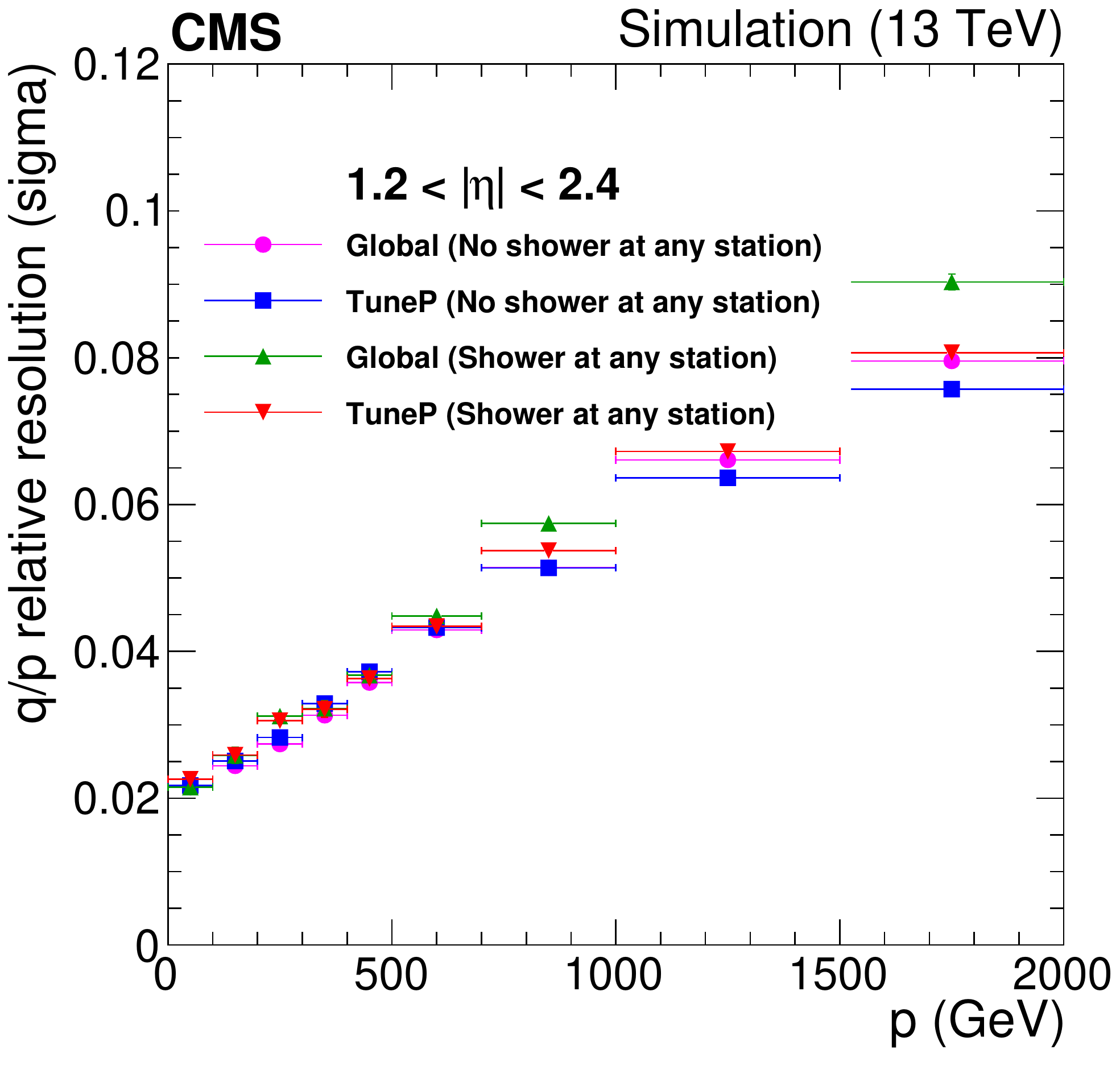}
\includegraphics[width =3in]{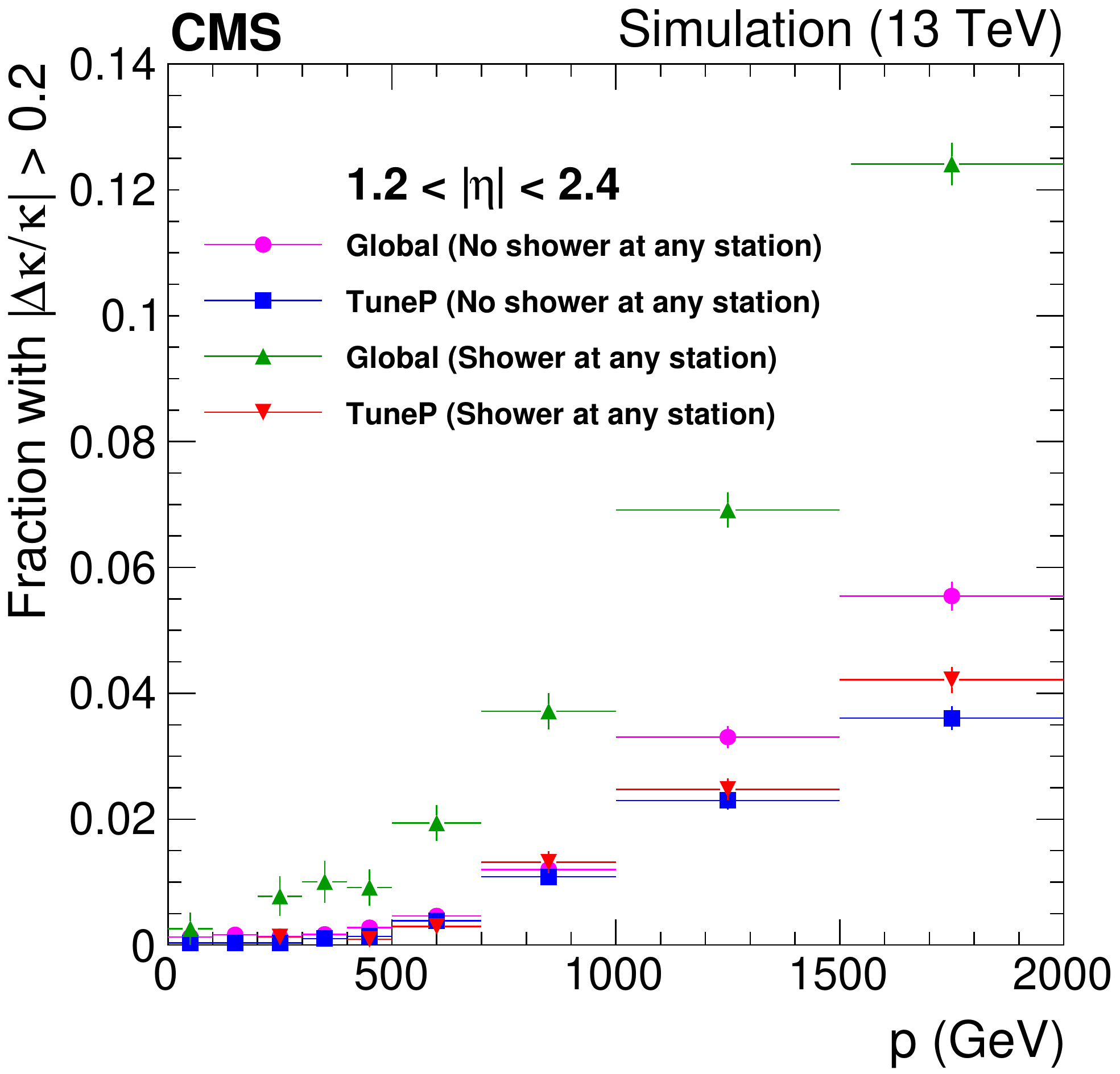}
\end{center}
\caption{
Comparison of the \TuneP and global reconstruction algorithms for simulated muons in the
(upper) barrel and (lower) endcap,
for the cases with and without the presence of tagged showers in any muon station.
The left plots show the momentum resolution (Gaussian $\sigma$); the right plots show the tail fraction with $\abs{\delta k /k} > 20\%$, as a function of muon momentum.
}
\label{fig:showerSeg}
\end{figure}

Startup and asymptotic scenarios with and without the corresponding APEs have been simulated,
together with the ideal scenario with no misalignment (APEs set to zero).
The startup scenario corresponds to the preliminary alignment at the beginning of a data-taking
period. The startup
performance is expected to be suboptimal, in particular because of
alignment after the opening and closure of the detector during the
LHC shutdown periods.
The final
alignment of the individual muon chambers (both DT and CSC), also
called asymptotic, is
determined starting from the aligned silicon tracker geometry, by
extrapolating selected muon tracks from the inner tracker to the muon
chambers~\cite{mualign-TB}. The alignment algorithm can use both cosmic ray muons and muons from \pp collisions, selected with high purity and $\pt$ above
a minimum threshold to limit the multiple scattering ($\pt > 30\GeV$
for collision muons in 2016 data taking).
Significant improvements are found with the inclusion of APEs in the startup scenario. In the endcaps, the startup performance
is worse than tracker-only, but gets recovered with APEs.
In the barrel region, the performance gets closer to asymptotic by including the APEs.
Overall, there are also small improvements for the asymptotic scenario due to the inclusion of APEs.

To further assess the performance of the \TuneP algorithm, it is important
to study not only the Gaussian core resolution, but also the tails of
the residual distribution that are sensitive to muon showering.  We characterize the tails by
the fraction of muons with relative momentum residual
$ \abs{\delta k /k} > 20\%$ (with $k = q/p$), as a function of the muon momentum.
The comparisons of the momentum resolution and the tails between the global muon fit and
the \TuneP choice are shown in Fig.~\ref{fig:showerSeg} for the asymptotic conditions of alignment and APEs. Two cases are
defined by whether or not at least one shower was found in the muon
system.

With \TuneP, the momentum resolution $\sigma$ is about 2\% for muons with $p<200\GeV$ in the barrel, whereas it is slightly above that value in
the endcap. At 2\TeV, the resolution
reaches about 6\% in the barrel and 8\% in the endcap.
A clear advantage of the strategy to remove contaminated muon stations from the trajectory fit
is seen by comparing the resolution tails of the global muon fit and \TuneP.
The \TuneP \pt assignment is mostly independent of showering. This does not
come at the
expense of the core resolution, which does not degrade with respect to
the global muon fit, but rather is also slightly improved.

Finally, the TuneP momentum assignment provides a reliable
determination of the muon charge sign up to very high momenta. Several
studies made on DY and single muon simulations predict a charge misassignment
probability varying from
$10^{-5}$  to $10^{-4}$ for muon momenta from 100\GeV to 2\TeV.
Cosmic ray muons can be used to partially validate these probabilities
\cite{Chatrchyan:2009ae}. During Run 2, we collected  20\,000 cosmic ray muons crossing the tracker volume of CMS with $\pt > 30\GeV$. Only one cosmic ray muon appears to have a wrong charge assignment, with apparent $\pt =  640\GeV$ (estimated from the lower CMS hemisphere).

\subsection{Momentum resolution from cosmic ray muons and collision events} \label{sec:Datareso}

In addition to the muons produced from heavy-boson decays, high-\pt muons from cosmic ray interactions and decays in the atmosphere \cite{Agafonova:2013wya} provide an excellent source of clean events that the CMS detector can measure.
As the muons traverse the CMS detector close to vertically, two reconstructed legs
(upper and lower) provide independent measurements of the
momentum for a single physical muon. The muon momentum scale and
resolution can then be assessed. For each selected event, we require two global
muons, one in each hemisphere of the detector, with good tracker track
quality to further ensure that the track is crossing well within the tracker volume,
similarly to muons produced in \pp collisions.

The two global muon tracks belong to the same
cosmic ray muon trajectory and should then have similar momentum. It is
then possible to extract the relative $q/\pt$ residual, $\residcosmic(q/\pt)$, defined as:
\begin{equation}
 \residcosmic (q/\pt) = \frac{{(q/\pt)}^\text{Upper} - {(q/\pt)}^\text{Lower}}{\sqrt{2}(q/\pt)^\text{Lower}},
\end{equation}
where ${(q/\pt)}^\text{Upper}$ and ${(q/\pt)}^\text{Lower}$
are the charge sign divided by $\pt$ for the
upper and the lower muon tracks, respectively. The factor of
$\sqrt{2}$ accounts for the fact that the $q/\pt$ measurements of the
two tracks are independent.

\begin{figure}[t]
\begin{center}
\includegraphics[width=0.48\textwidth]{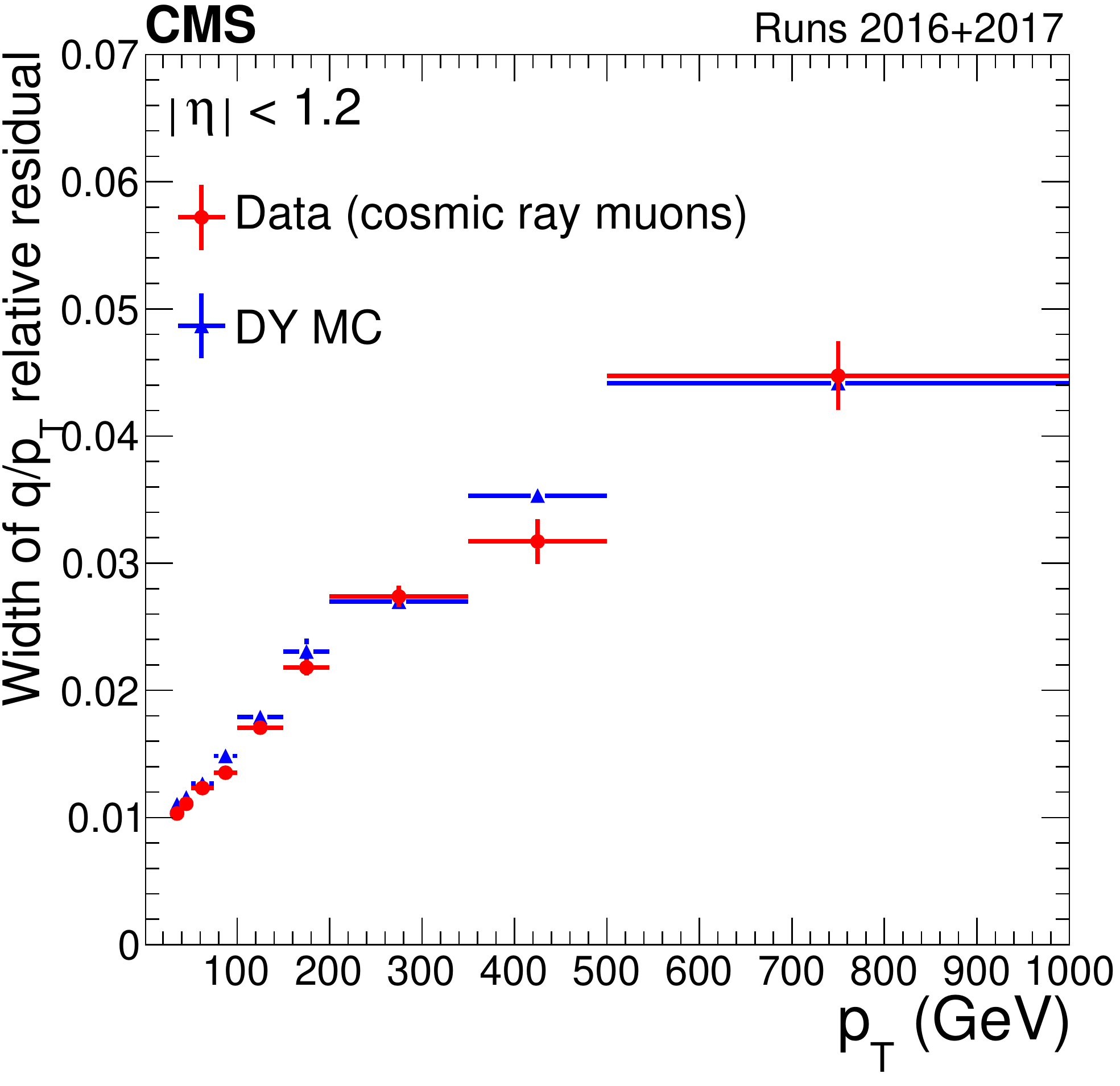}
\includegraphics[width=0.48\textwidth]{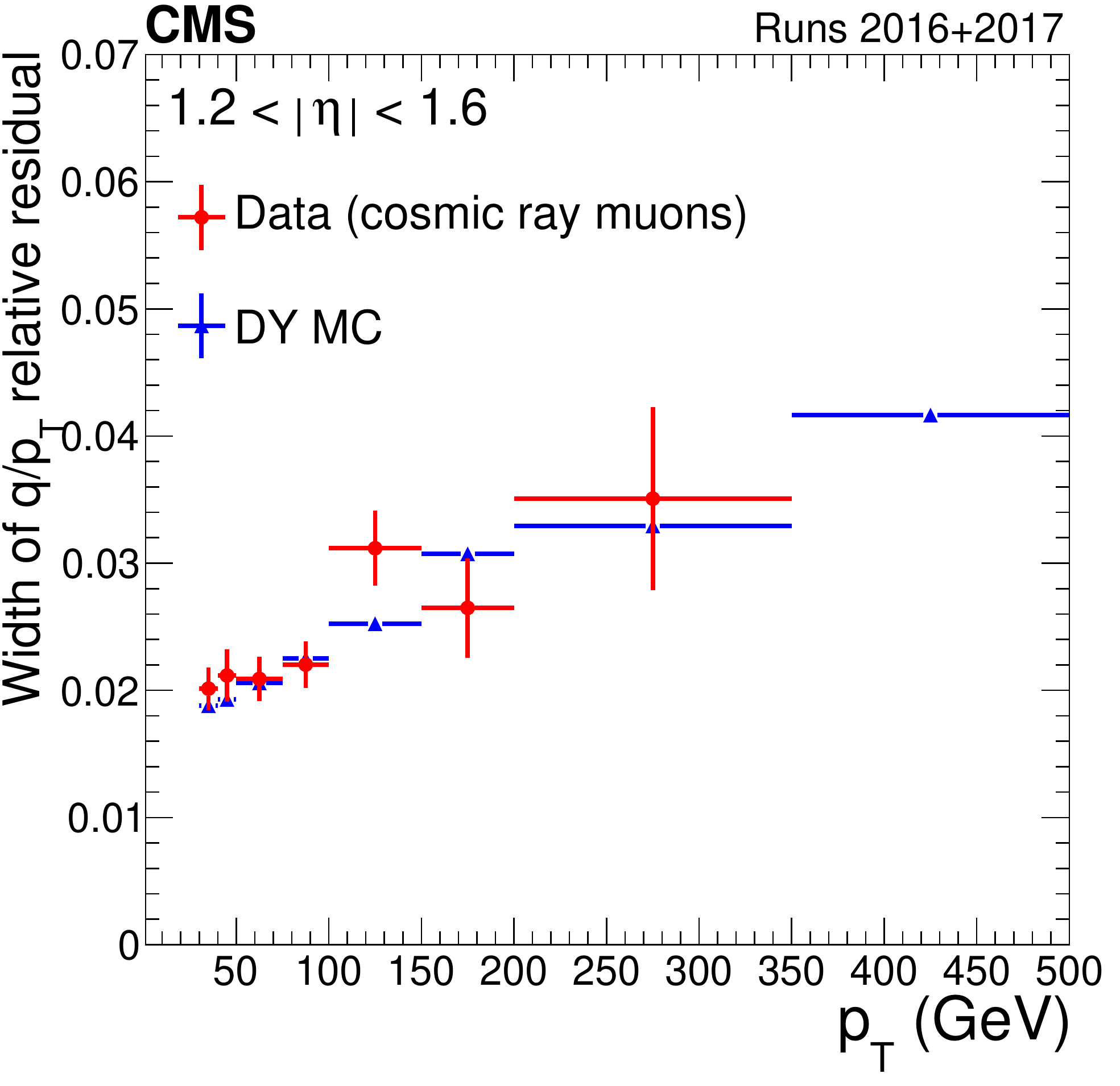}
\caption{Gaussian $\sigma$ of fits to $q/\pt$ relative residuals for
  \TuneP cosmic ray muons collected in 2016 and 2017 for (left) the barrel
  ($\abs{\eta}<1.2$) and (right) the endcap ($1.2<\abs{\eta}<1.6$) regions,
compared to the resolution extracted from DY simulation.}
\label{fig:cosmic_width_barrel_dy}
\end{center}
\end{figure}

Figure~\ref{fig:cosmic_width_barrel_dy} compares the $\pt$ resolution from $\residcosmic$
measured with the cosmic ray muons collected in 2016 and 2017,
crossing the barrel \mbox{($\abs{\eta}<1.2$)} and the endcap
($1.2<\abs{\eta}<1.6$) regions. The fits use the
\TuneP algorithm and the resolution obtained from simulated DY events,
$\residgen$, is defined in Eq.~(\ref{MCTruth}).
One third of the cosmic ray muon sample was collected during collisions using
the same single-muon trigger used to record high-momentum muons from
heavy-boson decays in order to guarantee the same detection environment;
the remainder was collected during dedicated cosmic ray muon runs with no LHC beams. The full cosmic ray muon sample has the
same reconstruction procedure as that used for \pp collisions. Good agreement is found between the cosmic ray muon data and the simulated DY
events. The uncertainties in the highest bins are dominated
by the small number of cosmic ray muons recorded (only 247 events with $\pt > 500\GeV$).

\begin{figure}[ht]
\begin{center}
\includegraphics[width=0.48\textwidth]{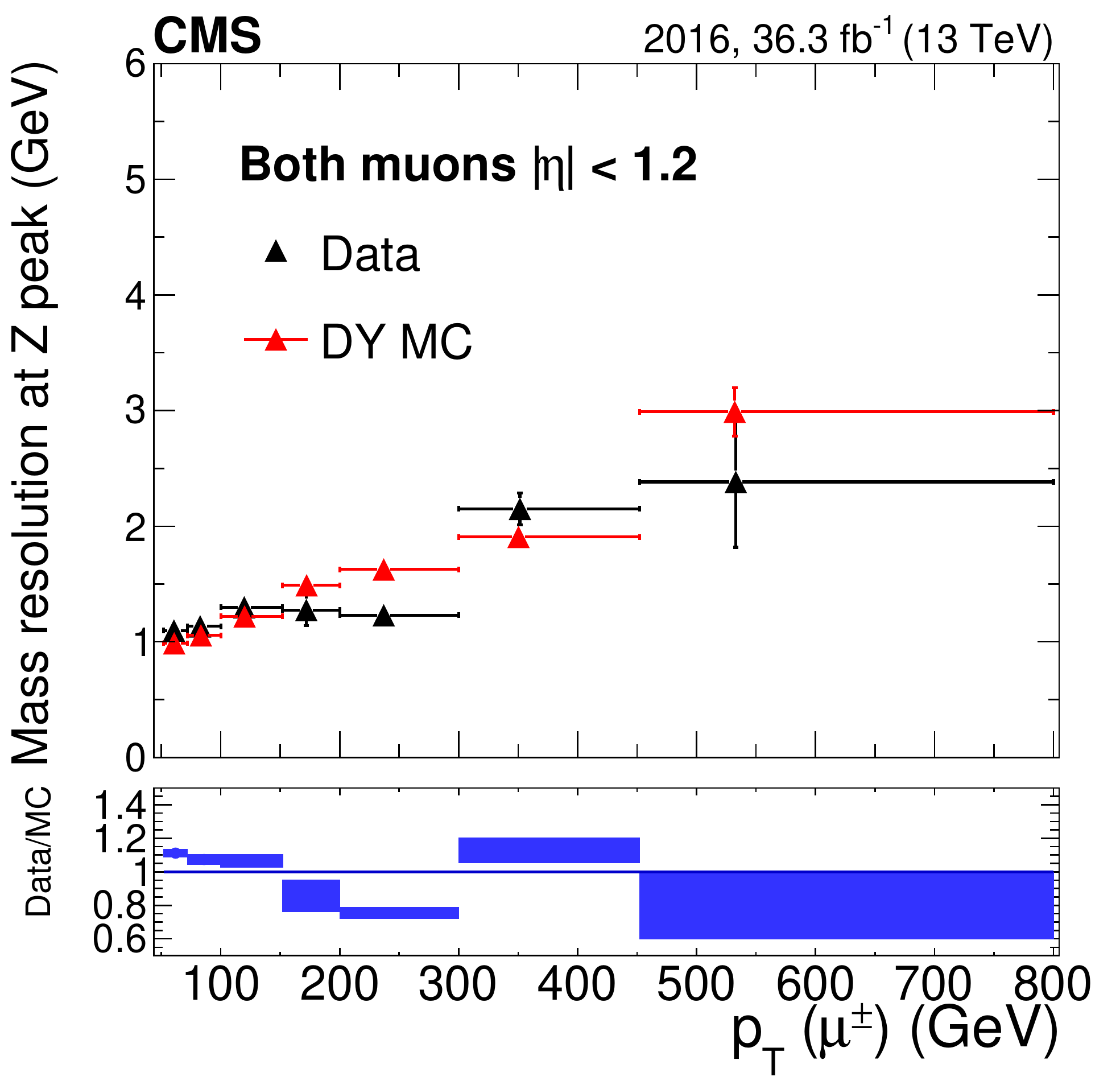}
\includegraphics[width=0.48\textwidth]{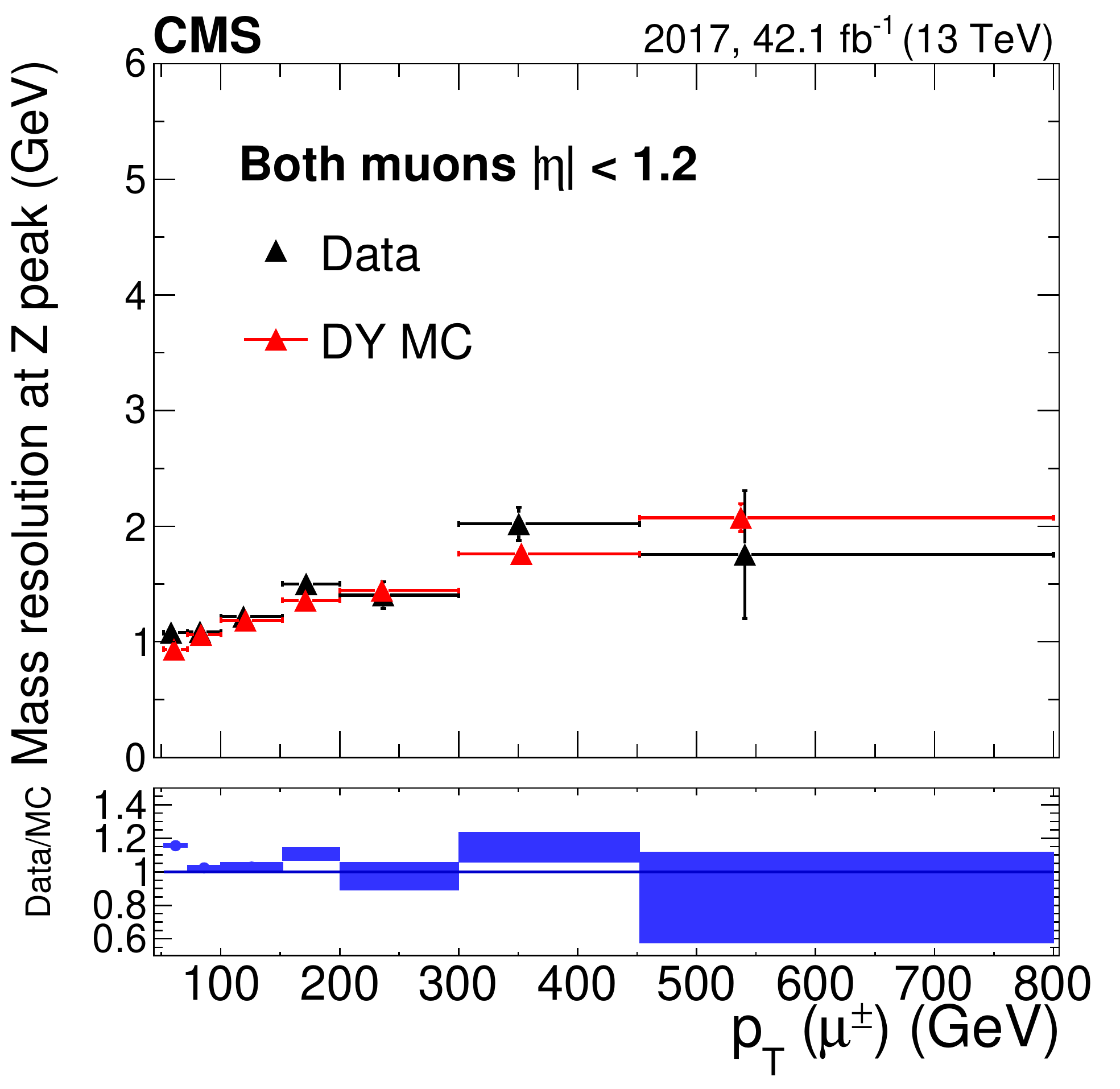}
\end{center}
\caption{Dimuon mass resolution (Gaussian $\sigma$), as a function of
  muon \TuneP \pt in the BB category. Each dimuon event is counted
  twice since each muon ($\mu^+$ and $\mu^-$) in the event is filling the histograms. Results for (left) 2016 and
  (right) 2017 are shown. Data are shown in black while the resolution
  obtained from simulation is shown in red. The lower panels of the
  plots show the ratio of data to simulation; the blue boxes represent
  the statistical uncertainties. The central value in each bin is obtained from the average of the distribution within the bin.}
\label{fig:MassResBB}
\end{figure}

\begin{figure}[htp]
\begin{center}
\includegraphics[width=0.48\textwidth]{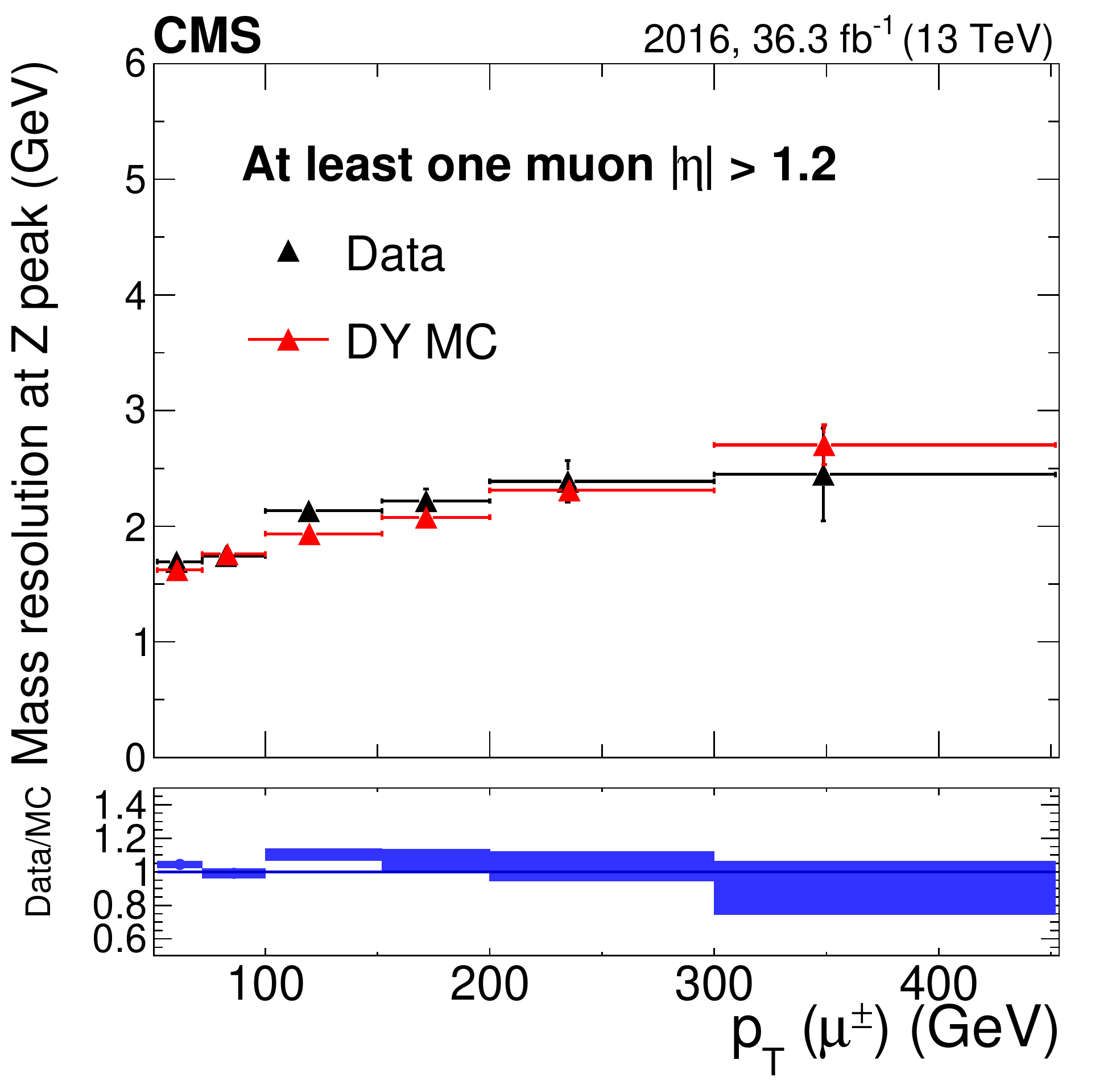}
\includegraphics[width=0.48\textwidth]{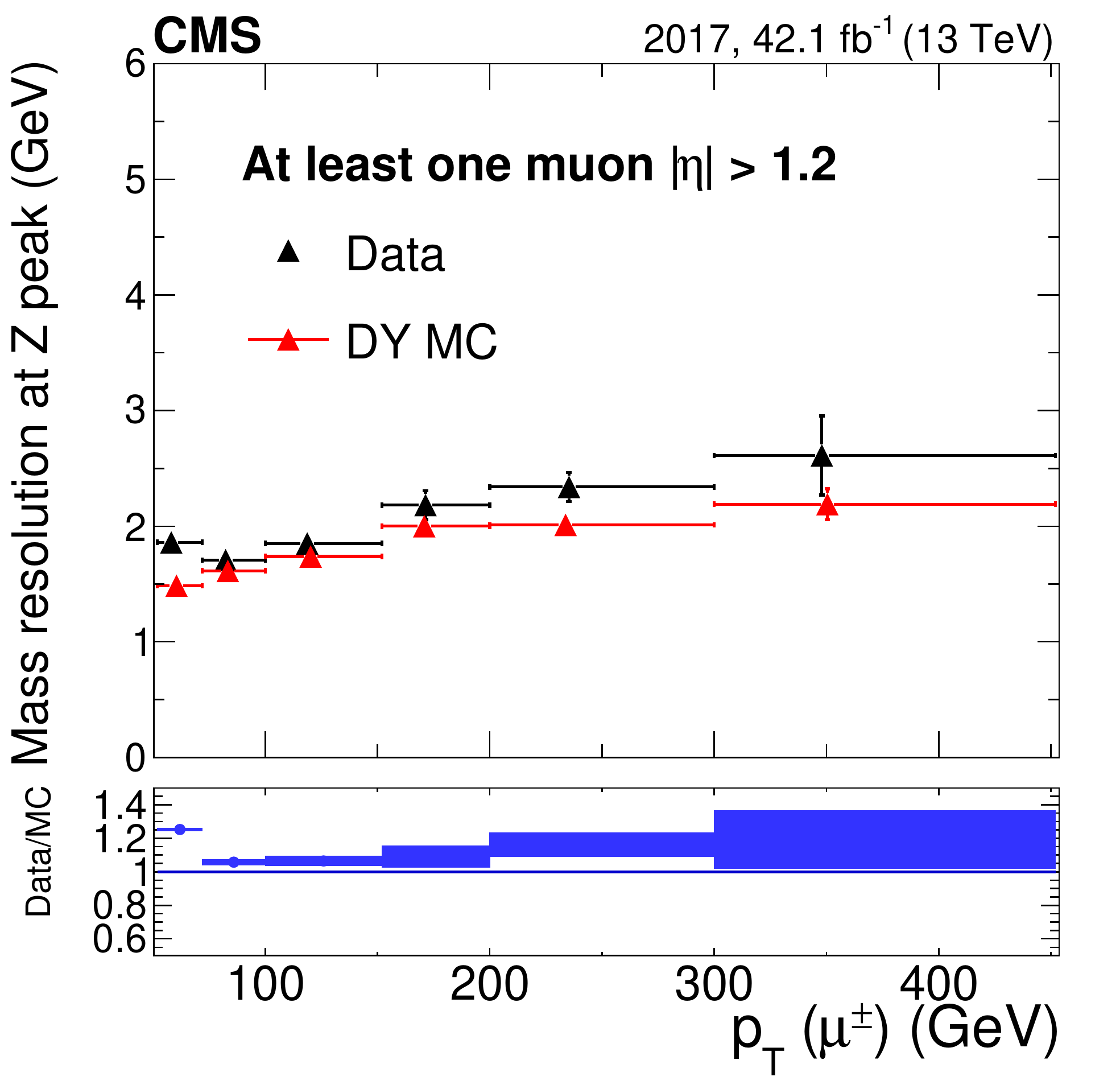}
\includegraphics[width=0.48\textwidth]{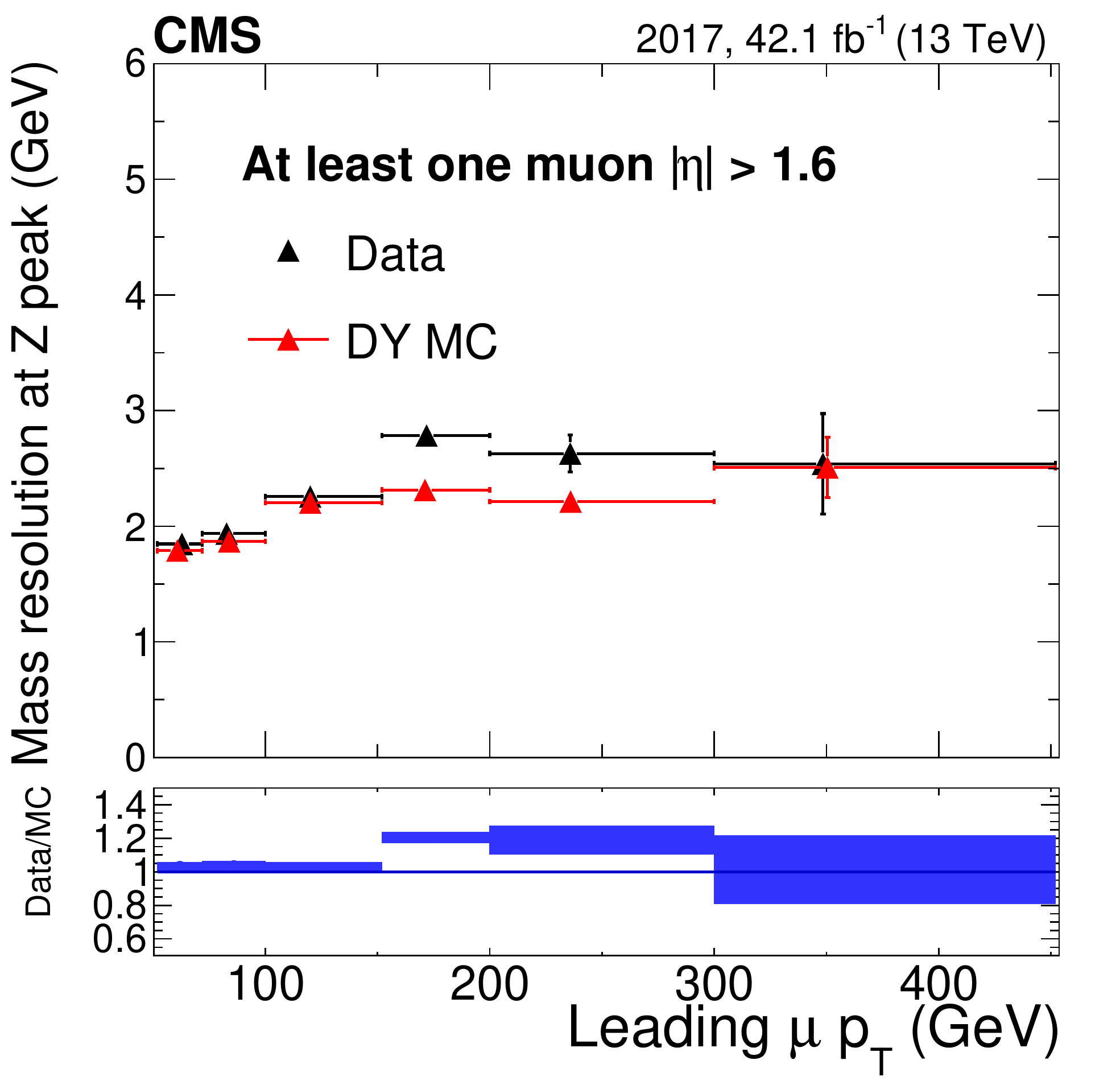}
\end{center}
\caption{Dimuon mass resolution (Gaussian $\sigma$), as a function of
  muon \TuneP \pt in the BE+EE category. Each dimuon event is counted
  twice since each muon ($\mu^+$ and $\mu^-$) in the event is filling the histograms. Results for (upper left) 2016
  and (upper right) 2017 are shown. The lower plot is for 2017 data
  where the BE+EE category is defined with at least one of the two
muons with $\abs{\eta}>1.6$. Data are shown in black while the resolution
obtained from simulation is shown in red. The central value in each bin is obtained from the average of the distribution within the bin.}
\label{fig:MassResBE}
\end{figure}

The coverage in $\eta$ is limited with cosmic ray muons, which
are predominantly close to vertical. Hence the
momentum resolution performance is measured best for $ \abs{\eta}<1.6$. To
overcome this limitation, events in the \PZ boson peak from \pp collisions can be used to assess
the dimuon mass resolution, as a function of the \pt of the individual
muons in a dimuon pair. The mass resolution function is the
convolution of a  Breit--Wigner distribution that models the intrinsic
decay width of the \PZ boson (both mean and width set to the PDG values~\cite{PDG2018})
with a double Crystal Ball function~\cite{Oreglia_CB1,Gaiser_CB2} that models the detector
effects. The \PZ boson peak is fit
in a mass range $75 < m_{\mu\mu} < 105\GeV$.
Each muon in the event is counted separately when filling the histograms, according to muon \pt, so that each event is counted twice.
The resulting dimuon mass resolution as a function of \pt is
shown in Figs.~\ref{fig:MassResBB} and~\ref{fig:MassResBE} for
events having both muons in the barrel (BB) or at least one of the two
muons in the endcap (BE+EE), respectively. The BE+EE results are
further split to isolate the forward endcap part (at least one of the two
muons with $\abs{\eta}>1.6$) in the lower plot in
Fig.~\ref{fig:MassResBE}.

For the BB events, the mass resolution in data agrees with
simulation for both 2016 and 2017 data. These results confirm what
is observed with cosmic ray muons. Above 300\GeV the 2017 resolution
is slightly better than that in 2016. This is due to
changes in the muon system alignment and improved values of APEs. For the BE+EE events, an offset of about 15\% can be seen
over the entire \pt range in the 2017 data. The discrepancy is
localized in the forward endcap region, as can be seen in the bottom
plot of Fig.~\ref{fig:MassResBE} that restricts the
BE+EE category to events with only one of the two
muons with $\abs{\eta}>1.6$. The results are presented as a function of
the leading muon \pt; the shift in the
resolution between data and simulation is seen only when the events
have at least one high-\pt muon.  This endcap region is known to have a tracker
alignment bias, as can also be seen in the scale
results in Fig.~\ref{fig:result_2017}.

\subsection{Momentum scale from collision events} \label{sec:DataScale}

The scale of the muon $\pt$ is sensitive to three
effects that can potentially introduce biases:
muon energy losses, detector misalignments, and magnetic
field variations. The calibration of the momentum scale is performed
by modifying the curvature of the muon, while taking into account these
three physics effects.
The detector alignment biases result
in an additive correction $k_{\mathrm{b}}$ (to the (signed) muon curvature
$\kappa$) that has the same sign for both positively and negatively charged muons,
resulting in an increase in the measured $p$ for one sign charge, and
a decrease in measured $p$ for the other sign.
The variations from the magnetic field lead
to a multiplicative correction factor to the curvature. The energy
loss is taken into account with an additive term, that increases the
muon momentum independent of the
muon charge.

For intermediate- and low-\pt muons, two different methods are used for
Run 2 data to estimate the additive and multiplicative correction factors to the muon curvature. The first method selects muons from
\PZ boson decays and derives the corrections from the mean value of the
distribution of $q/\pt$, with further tuning performed using the mean
of the dimuon invariant mass spectrum~\cite{Bodek:2012id}. The second
method selects muons from \PZ boson, \PJGy, and \PGUP{1S} resonances, and
determines corrections using a Kalman filter. The
corrections are provided as a function of the muon $\eta$ and $\phi$
in both methods and as a function of the muon \pt in the second method.
In Run 2 data, the dominant source of scale bias is coming from the
detector alignment.

\begin{figure}[th]
\centering
\includegraphics[width=.49\linewidth]{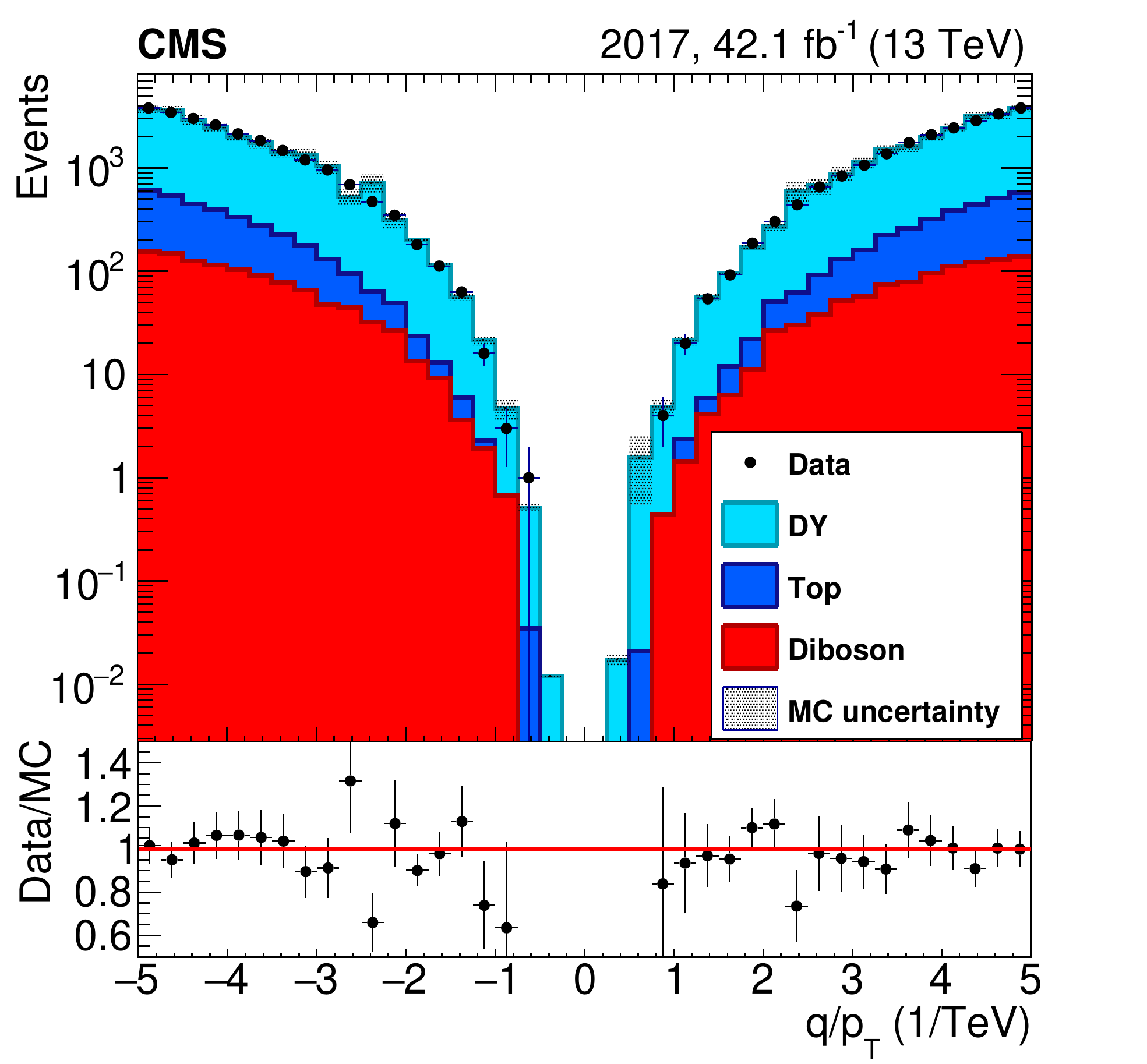}
\caption{Data to simulation comparison of the curvature
  distribution in $\PZ \to \mu^+\mu^-$ events, for 2017 data with muon $\pt > 200\GeV$.}
\label{fig:k}
\end{figure}

For high-\pt muons, the reconstruction of the muon \pt relies both on
the tracker and on the muon system inputs. Thus, the derived corrections
from the two previous methods that focus only on the tracker
information, are not directly applicable. In addition, the intrinsic
alignment of the muon chambers within the muon system, and their
alignment respectively to the tracker, are sources of
potential scale bias. The generalized endpoint (GE) method~\cite{Sirunyan:2018fpa} quantifies biases in the \pt determination relying on muons
produced from DY events. The method consists of comparing the muon curvature distribution between data and
simulated events, modifying the simulated values by a constant additive bias
term ($k_{\mathrm{b}}$), such that the distribution is distorted as $\kappa
\to \kappa + k_{\mathrm{b}}$. A $\chi^2$ test is performed between the curvature distribution in data
and in simulation, as a function of the injected bias $k_{\mathrm{b}}$, in order
to find the minimum of the distribution.
Such a distortion reproduces a potential detector alignment bias that
changes $p$ in opposite directions for positively and negatively
charged muons. The muon curvature without any
additive bias in simulation is shown in Fig.~\ref{fig:k}.

\begin{figure}[h]
\centering
\includegraphics[width=.49\linewidth]{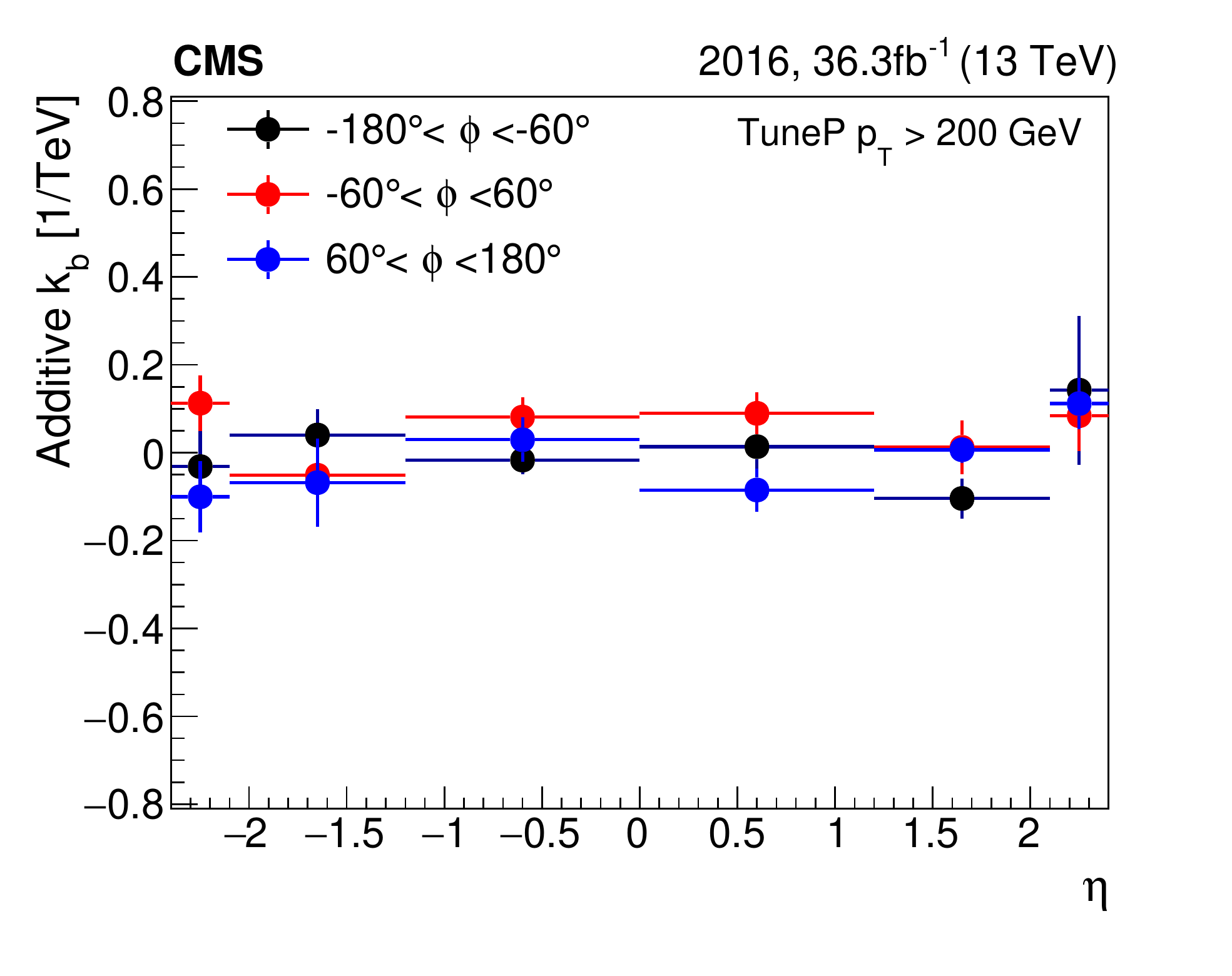}
\includegraphics[width=.49\linewidth]{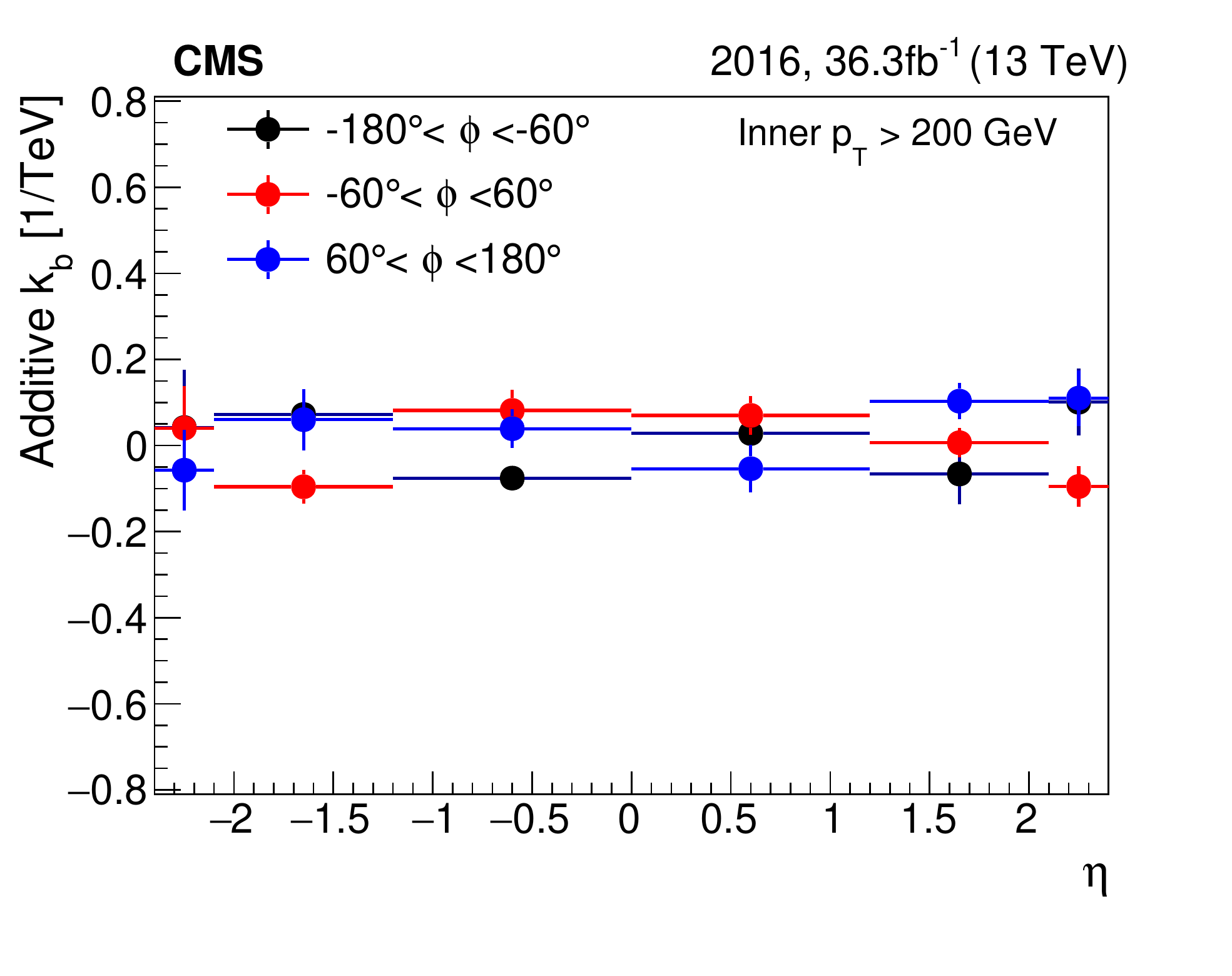}
\caption{Measurement of the scale bias for muons above 200\GeV with 2016 data. On the left the \pt corresponds to \TuneP, while on the right
  it corresponds to the tracker-only assignment. }
\label{fig:result_2016}
\end{figure}

\begin{figure}[h]
\centering
\includegraphics[width=.49\linewidth]{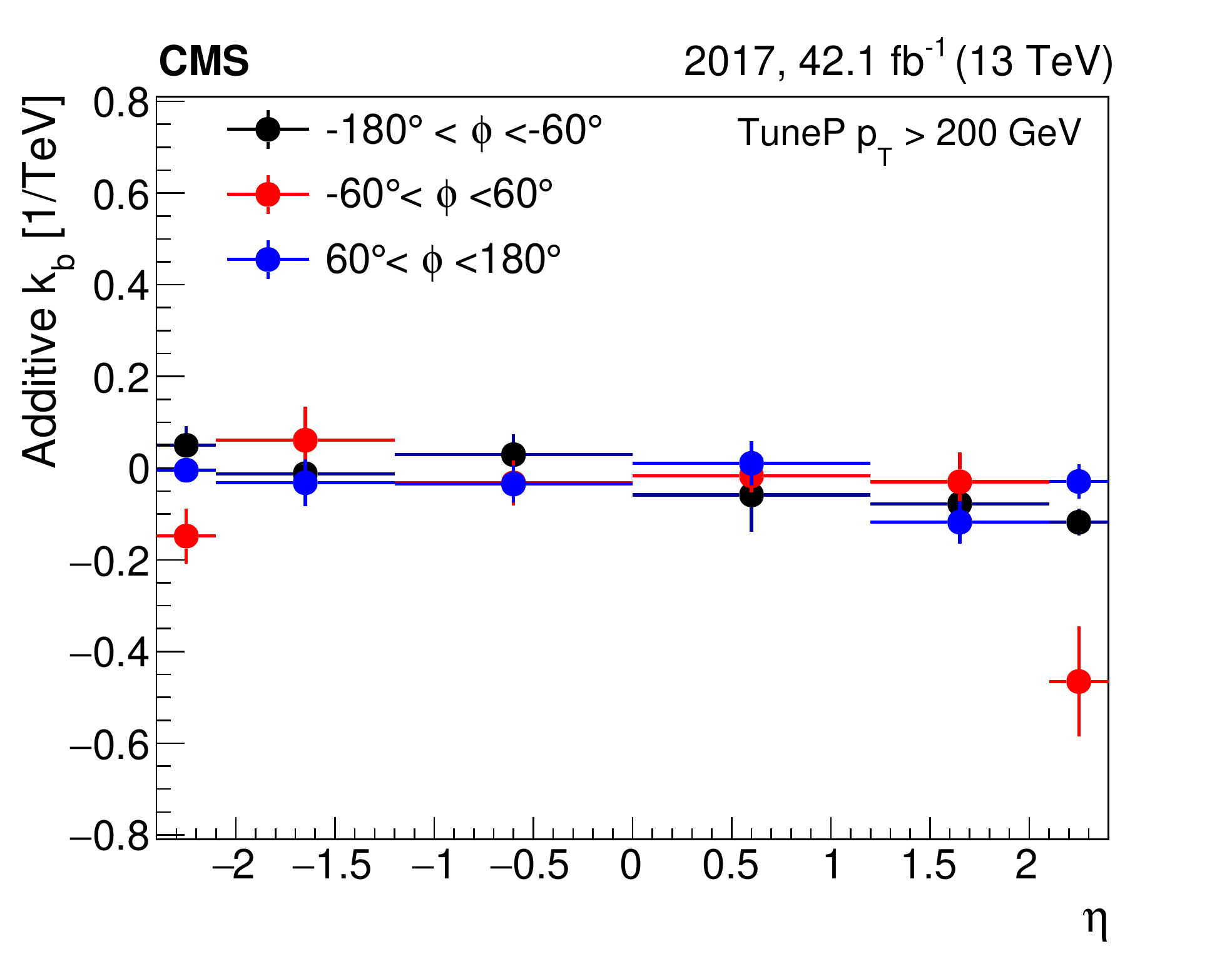}
\includegraphics[width=.49\linewidth]{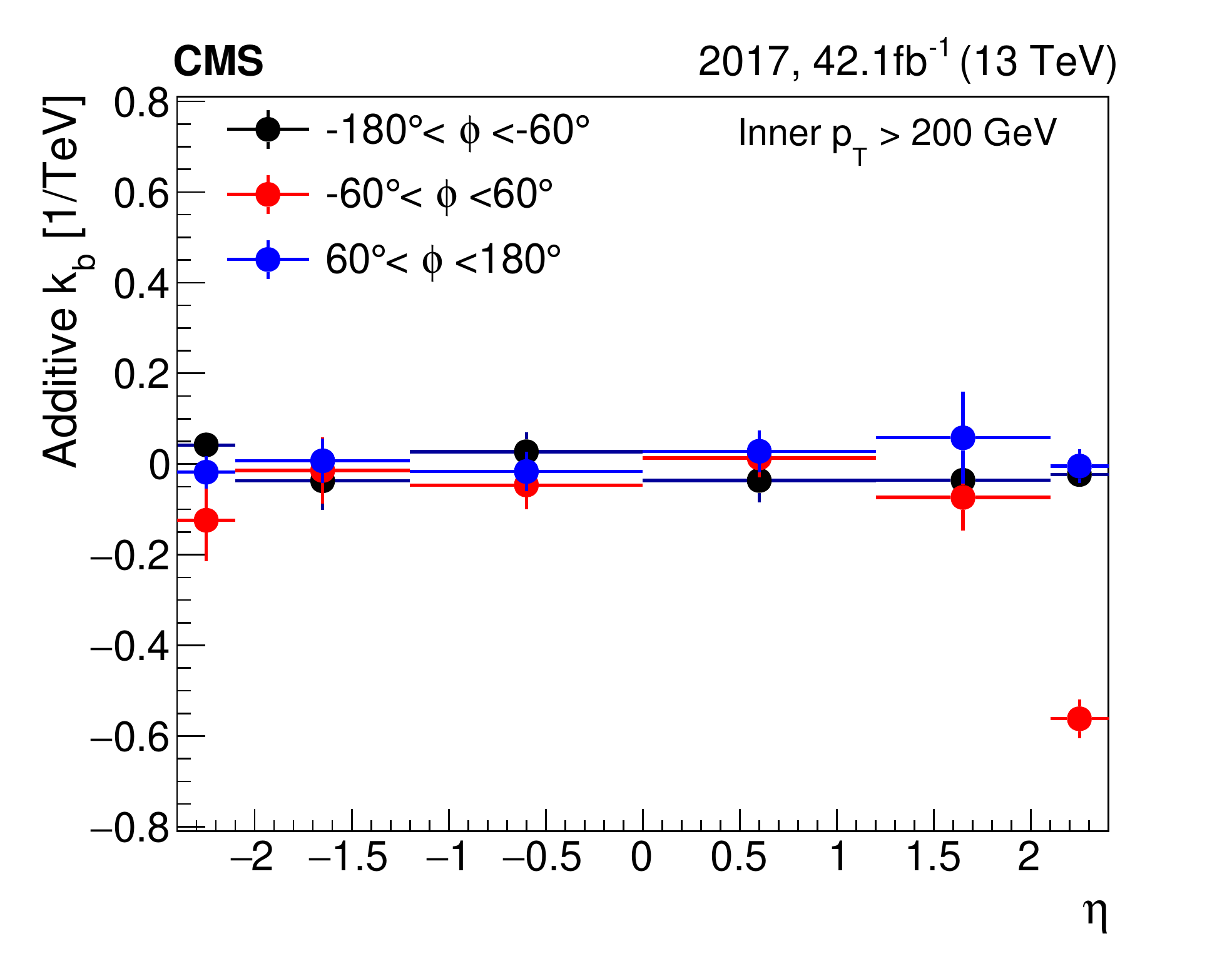}
\caption{Measurement of the scale bias for muons above 200\GeV with 2017
  data. On the left the \pt corresponds to the \TuneP, while on the right
  it corresponds to the tracker-only assignment.}
\label{fig:result_2017}
\end{figure}

The estimated additive biases measured with GE are presented as a function of $\eta$ and $\phi$, for the
2016 data in Fig.~\ref{fig:result_2016} and for the 2017 data in
Fig.~\ref{fig:result_2017}. For each year, the results are obtained
using both the tracker and the \TuneP \pt assignment. The sample size of high-momentum muons is limited, but it
is visible that the detector parts that are most affected by the
misalignment are the endcaps, in both years, with an estimated bias
$k_{\mathrm{b}} \approx 0.15/\TeV$ in 2016 data and a maximum
$k_{\mathrm{b}} \approx 0.5/\TeV$ in 2017 data localized in the forward
positive endcap and in a given $\phi$ sector ($-60^\circ<\phi<60^\circ$). No significant differences are found when comparing
the bias values obtained with \TuneP \pt and tracker \pt. Thus the misalignment
is mostly coming from the tracker component while the muon system alignment
does not contribute significantly. The local tracker misalignment found in 2017 data is suspected to be caused by radiation effects that impact the pixel and strip detector calibration throughout the run period. In addition, for that specific year the alignment procedure was limited by the sample size of muons from Z decays and from cosmic rays, which are needed to ensure the right muon mixture input when performing the detector calibration. These results are in agreement with those from Ref.~\cite{Bodek:2012id}.

\section{Summary}
The performance of muon reconstruction, identification, trigger,
and momentum assignment has been studied in a sample enriched in high-momentum muons using proton-proton collisions at $\sqrt{s} = 13\TeV$, collected by the CMS experiment at the LHC in 2016-2017, and corresponding to the integrated luminosity of 78.4\fbinv.
Depending on the longitudinal component of the momentum,
muons with transverse momentum $\pt>200\GeV$ can have radiative energy losses in steel that are no longer negligible compared to ionization energy losses.
Dedicated methods have been developed to study the performance impact of the detector alignment and electromagnetic showers
along the muon track.
Overall, the measurements are described accurately by the simulation and their reach in momentum is limited by the statistical uncertainties. The largest discrepancy between data and simulation is found at the trigger level with a 10\% efficiency difference for muons with \pt around 1\TeV.
Representative figures of merit that illustrate the muon performance at high momentum are listed below.

\begin{itemize}
\item The identification efficiency measured in data is $>$98\% over the full \pt spectrum, up to $1000\GeV$ for a pseudorapidity magnitude $\abs{\eta}<2.4$. No dependence on the momentum p is observed. The ratio of data to simulation is within 1\% of unity for $0 <\abs{\eta} < 0.9$ and $2.1 <\abs{\eta} < 2.4$, and within 0.5\% for $0.9 <\abs{\eta}< 2.1$.
\item The standalone reconstruction efficiency measured in data is $>$98\% over the full \pt spectrum and up to $1500\GeV$ for $\abs{\eta} < 1.6$. No dependence on the momentum p is observed. The ratio of data to simulation is uniform and equal to 0.99. In the forward detector region ($\abs{\eta} > 1.6$), an inefficiency trend starting at $p = 200\GeV$ is observed both in simulation and in data, although it is slightly more pronounced in the latter. The muon showering and the dense track activities for muon momentum around $1000\GeV$ interfere within the muon reconstruction, and lead to a momentum dependence with up to 5\% inefficiency when both effects are combined.
\item The total trigger efficiency measured in data shows a decreasing trend  from 92\% at $\pt= 100\GeV$ down to 80\% at $\pt= 1000\GeV$,  integrated over muon $\eta$. The simulation does not reproduce the severity of the slope and the ratio of data to simulation deviates from unity at the level of 10\%. This discrepancy is driven by the first level (L1) trigger and is localized in the overlap region ($0.9 <\abs{\eta} < 1.2$) because of a nonideal interplay between DT and CSC signals. This was improved in 2018.
\item The L1 efficiency suffers from showering effects. Direct measurements from data and simulation are compared to a parameterization derived from showering inputs. The trend as a function of $p$ is reproduced by the parameterization within the uncertainties. The biggest impact of showering is seen in the barrel region ($\abs{\eta}<0.9$), both in simulation and in data. The simulation does not fully reproduce the slope seen in data in both barrel and endcap regions, thus indicating a slight underestimation of the showering effect at L1 in simulation.
\item \TuneP momentum assignment and performance is robust against the presence of showers. The simulation reproduces the choice of TuneP among the different TeV refitters in data.
\item The \PZ boson mass resolution is $<$3\GeV for events with the leading muon \pt  up to 450\GeV over the full $\eta$ range. The \PZ boson mass resolution is very similar between simulation and data, except in the endcap region for the 2017 data, where a tracker alignment bias degrades the resolution by 20\% for events with the leading muon \pt  above 150\GeV.
\item The trajectory curvature bias $k_{\mathrm{b}}$ is compatible with zero in the barrel region, but is as large as $0.5 \TeV^{-1}$ for $\abs{\eta} > 2.1$ in 2017 data.
\end{itemize}
These results show that the performance of the CMS detector is outstanding for high energy muons and is largely well described by simulation.

\begin{acknowledgments}
\hyphenation{Bundes-ministerium Forschungs-gemeinschaft Forschungs-zentren Rachada-pisek} We congratulate our colleagues in the CERN accelerator departments for the excellent performance of the LHC and thank the technical and administrative staffs at CERN and at other CMS institutes for their contributions to the success of the CMS effort. In addition, we gratefully acknowledge the computing centers and personnel of the Worldwide LHC Computing Grid for delivering so effectively the computing infrastructure essential to our analyses. Finally, we acknowledge the enduring support for the construction and operation of the LHC and the CMS detector provided by the following funding agencies: the Austrian Federal Ministry of Education, Science and Research and the Austrian Science Fund; the Belgian Fonds de la Recherche Scientifique, and Fonds voor Wetenschappelijk Onderzoek; the Brazilian Funding Agencies (CNPq, CAPES, FAPERJ, FAPERGS, and FAPESP); the Bulgarian Ministry of Education and Science; CERN; the Chinese Academy of Sciences, Ministry of Science and Technology, and National Natural Science Foundation of China; the Colombian Funding Agency (COLCIENCIAS); the Croatian Ministry of Science, Education and Sport, and the Croatian Science Foundation; the Research Promotion Foundation, Cyprus; the Secretariat for Higher Education, Science, Technology and Innovation, Ecuador; the Ministry of Education and Research, Estonian Research Council via IUT23-4, IUT23-6 and PRG445 and European Regional Development Fund, Estonia; the Academy of Finland, Finnish Ministry of Education and Culture, and Helsinki Institute of Physics; the Institut National de Physique Nucl\'eaire et de Physique des Particules~/~CNRS, and Commissariat \`a l'\'Energie Atomique et aux \'Energies Alternatives~/~CEA, France; the Bundesministerium f\"ur Bildung und Forschung, the Deutsche Forschungsgemeinschaft (DFG) under Germany’s Excellence Strategy -- EXC 2121 ``Quantum Universe" -- 390833306, and Helmholtz-Gemeinschaft Deutscher Forschungszentren, Germany; the General Secretariat for Research and Technology, Greece; the National Research, Development and Innovation Fund, Hungary; the Department of Atomic Energy and the Department of Science and Technology, India; the Institute for Studies in Theoretical Physics and Mathematics, Iran; the Science Foundation, Ireland; the Istituto Nazionale di Fisica Nucleare, Italy; the Ministry of Science, ICT and Future Planning, and National Research Foundation (NRF), Republic of Korea; the Ministry of Education and Science of the Republic of Latvia; the Lithuanian Academy of Sciences; the Ministry of Education, and University of Malaya (Malaysia); the Ministry of Science of Montenegro; the Mexican Funding Agencies (BUAP, CINVESTAV, CONACYT, LNS, SEP, and UASLP-FAI); the Ministry of Business, Innovation and Employment, New Zealand; the Pakistan Atomic Energy Commission; the Ministry of Science and Higher Education and the National Science Centre, Poland; the Funda\c{c}\~ao para a Ci\^encia e a Tecnologia, Portugal; JINR, Dubna; the Ministry of Education and Science of the Russian Federation, the Federal Agency of Atomic Energy of the Russian Federation, Russian Academy of Sciences, the Russian Foundation for Basic Research, and the National Research Center ``Kurchatov Institute"; the Ministry of Education, Science and Technological Development of Serbia; the Secretar\'{\i}a de Estado de Investigaci\'on, Desarrollo e Innovaci\'on, Programa Consolider-Ingenio 2010, Plan Estatal de Investigaci\'on Cient\'{\i}fica y T\'ecnica y de Innovaci\'on 2017--2020, research project IDI-2018-000174 del Principado de Asturias, and Fondo Europeo de Desarrollo Regional, Spain; the Ministry of Science, Technology and Research, Sri Lanka; the Swiss Funding Agencies (ETH Board, ETH Zurich, PSI, SNF, UniZH, Canton Zurich, and SER); the Ministry of Science and Technology, Taipei; the Thailand Center of Excellence in Physics, the Institute for the Promotion of Teaching Science and Technology of Thailand, Special Task Force for Activating Research and the National Science and Technology Development Agency of Thailand; the Scientific and Technical Research Council of Turkey, and Turkish Atomic Energy Authority; the National Academy of Sciences of Ukraine; the Science and Technology Facilities Council, UK; the US Department of Energy, and the US National Science Foundation.

Individuals have received support from the Marie-Curie program and the European Research Council and Horizon 2020 Grant, contract Nos.\ 675440, 752730, and 765710 (European Union); the Leventis Foundation; the A.P.\ Sloan Foundation; the Alexander von Humboldt Foundation; the Belgian Federal Science Policy Office; the Fonds pour la Formation \`a la Recherche dans l'Industrie et dans l'Agriculture (FRIA-Belgium); the Agentschap voor Innovatie door Wetenschap en Technologie (IWT-Belgium); the F.R.S.-FNRS and FWO (Belgium) under the ``Excellence of Science -- EOS" -- be.h project n.\ 30820817; the Beijing Municipal Science \& Technology Commission, No. Z181100004218003; the Ministry of Education, Youth and Sports (MEYS) of the Czech Republic; the Lend\"ulet (``Momentum") Program and the J\'anos Bolyai Research Scholarship of the Hungarian Academy of Sciences, the New National Excellence Program \'UNKP, the NKFIA research grants 123842, 123959, 124845, 124850, 125105, 128713, 128786, and 129058 (Hungary); the Council of Scientific and Industrial Research, India; the HOMING PLUS program of the Foundation for Polish Science, cofinanced from European Union, Regional Development Fund, the Mobility Plus program of the Ministry of Science and Higher Education, the National Science Center (Poland), contracts Harmonia 2014/14/M/ST2/00428, Opus 2014/13/B/ST2/02543, 2014/15/B/ST2/03998, and 2015/19/B/ST2/02861, Sonata-bis 2012/07/E/ST2/01406; the National Priorities Research Program by Qatar National Research Fund; the Ministry of Science and Education, grant no. 3.2989.2017 (Russia); the Programa de Excelencia Mar\'{i}a de Maeztu, and the Programa Severo Ochoa del Principado de Asturias; the Thalis and Aristeia programs cofinanced by EU-ESF, and the Greek NSRF; the Rachadapisek Sompot Fund for Postdoctoral Fellowship, Chulalongkorn University, and the Chulalongkorn Academic into Its 2nd Century Project Advancement Project (Thailand); the Nvidia Corporation; the Welch Foundation, contract C-1845; and the Weston Havens Foundation (USA). \end{acknowledgments}

\bibliography{auto_generated} 
\cleardoublepage \appendix\section{The CMS Collaboration \label{app:collab}}\begin{sloppypar}\hyphenpenalty=5000\widowpenalty=500\clubpenalty=5000\vskip\cmsinstskip
\textbf{Yerevan Physics Institute, Yerevan, Armenia}\\*[0pt]
A.M.~Sirunyan$^{\textrm{\dag}}$, A.~Tumasyan
\vskip\cmsinstskip
\textbf{Institut f\"{u}r Hochenergiephysik, Wien, Austria}\\*[0pt]
W.~Adam, F.~Ambrogi, T.~Bergauer, M.~Dragicevic, J.~Er\"{o}, A.~Escalante~Del~Valle, M.~Flechl, R.~Fr\"{u}hwirth\cmsAuthorMark{1}, M.~Jeitler\cmsAuthorMark{1}, N.~Krammer, I.~Kr\"{a}tschmer, D.~Liko, T.~Madlener, I.~Mikulec, N.~Rad, J.~Schieck\cmsAuthorMark{1}, R.~Sch\"{o}fbeck, M.~Spanring, D.~Spitzbart, W.~Waltenberger, C.-E.~Wulz\cmsAuthorMark{1}, M.~Zarucki
% ...
\vskip\cmsinstskip
\dag: Deceased\\
1:  Also at Vienna University of Technology, Vienna, Austria\\
% ...
\end{sloppypar}
\end{document}